\newcommand*{\ATLASLATEXPATH}{latex/}
\documentclass[UKenglish,cernpreprint,texlive=2016,txfonts]{\ATLASLATEXPATH atlasdoc}

\usepackage[subfigure]{\ATLASLATEXPATH atlaspackage}
\usepackage{a4wide}
\usepackage{\ATLASLATEXPATH atlasbiblatex}

\usepackage{\ATLASLATEXPATH atlascontribute}

\usepackage{\ATLASLATEXPATH atlasphysics}
\usepackage{\ATLASLATEXPATH atlasprocess}
\usepackage{\ATLASLATEXPATH atlasmisc}
\usepackage{relsize}
\usepackage{multirow}

\graphicspath{{logos/}{figures/}}

\usepackage{atlasmetdefs}

\AtlasTitle{Performance of algorithms that reconstruct missing transverse momentum in $\sqrt{s}$ $=$ 8 TeV proton--proton collisions in the ATLAS detector}

\author{The ATLAS Collaboration}

\AtlasRefCode{PERF-2014-04}

\PreprintIdNumber{CERN-EP-2016-134}

\AtlasJournal{Eur.\ Phys.\ J.\ C}
\AtlasJournalRef{Eur. Phys. J. C77 (2017) 241}
\AtlasDOI{10.1140/epjc/s10052-017-4780-2}

\AtlasAbstract{%
The reconstruction and calibration
  algorithms used to calculate missing transverse momentum (\met{}) with
  the ATLAS detector exploit energy deposits in the calorimeter and tracks
  reconstructed in the inner detector as well as the muon spectrometer. 
  Various strategies are used to suppress effects arising from additional 
  proton--proton interactions, called pileup, concurrent with the hard-scatter
  processes. Tracking information is used to
  distinguish contributions from the pileup interactions using their
  vertex separation along the beam axis. The performance of the \met{}
  reconstruction algorithms, especially with respect to the amount of pileup, is evaluated using data
  collected in proton--proton collisions at a centre-of-mass energy of 
  $8$ \TeV{} during 2012, and 
  results are shown for a data sample corresponding to an integrated 
luminosity of $20.3\, \ifb$. The simulation and modelling of \met~in events containing a $Z$ boson
decaying to two charged leptons (electrons or muons) or a $W$ boson decaying to
a charged lepton and a neutrino are compared to data. The
acceptance for different event topologies, with and without high
transverse momentum neutrinos, is shown for a range of threshold
criteria for \met{}, and estimates of the systematic
uncertainties in the \met~measurements are presented.

}

\hypersetup{pdftitle={ATLAS draft},pdfauthor={The ATLAS Collaboration}}

\begin{document}

\maketitle

\tableofcontents


\section{Introduction}
\label{intro}

The Large Hadron Collider (LHC) provided proton--proton ($pp$) collisions at
a centre-of-mass energy of 8~\TeV{} during 2012. Momentum
conservation transverse to the beam axis\footnote{ATLAS uses a right-handed coordinate system with its origin at the nominal interaction point (IP) in the centre of the detector and the $z$-axis along the beam pipe. The $x$-axis points from the IP to the centre of the LHC ring, and the $y$-axis points upward. Cylindrical coordinates $(r,\phi)$ are used in the transverse plane, $\phi$ being the azimuthal angle around the beam pipe. The pseudorapidity is defined in terms of the polar angle $\theta$ as $\eta=-\ln\tan(\theta/2)$.} implies that the transverse
momenta of all particles in the final state should sum to zero. 
Any imbalance 
 may indicate the presence
of undetectable particles such as neutrinos or new, stable particles escaping detection.

The missing transverse momentum (\metvec) is reconstructed as the negative vector 
sum of the transverse momenta (\ptvec{}) of all
detected particles, and its magnitude is represented by the symbol \metmag{}. 
The measurement of \met~strongly depends on the energy scale and
resolution of the reconstructed ``physics objects''. The physics objects considered in the
\met~calculation are electrons,
photons, muons, $\tau$-leptons, and jets. Momentum contributions not attributed to any of the physics
objects mentioned above are reconstructed as the
\met{} ``soft term''.
Several algorithms for reconstructing the \met~soft term
utilizing a combination of calorimeter signals and tracks in the inner detector are considered.

The \met{} reconstruction algorithms and calibrations
developed by ATLAS for 7~\TeV~data from 2010
are summarized in Ref.~\cite{ATLASMETPaper2011}.
The 2011 and 2012 datasets are more affected by
contributions from additional $pp$ collisions, referred to as
``pileup'', concurrent with the hard-scatter process.
Various techniques have been developed to suppress such
contributions. 
This paper describes the pileup dependence, calibration, and resolution
of the \met{} reconstructed with different algorithms and pileup-mitigation techniques.

The performance of \met~reconstruction algorithms, or
``\met~performance'', refers to the use of derived quantities like the
mean, width, or tail of the \met{} distribution to study pileup dependence
and calibration. The \met{} reconstructed with different algorithms is
studied in both data and Monte Carlo (MC) simulation, and the level of
agreement between the two is compared using datasets in which events with a leptonically decaying $W$ or $Z$ boson dominate. 
The $W$ boson sample provides events with intrinsic
\met{} from non-interacting
particles (e.g. neutrinos). 
Contributions to the \met{} due to mismeasurement are referred to as
fake \met{}. 
Sources of fake \met{}
may include \pT{} mismeasurement, miscalibration, and particles going
through un-instrumented regions of the detector. In MC simulations, the \met{} from each algorithm is compared to the
true \met~(\mettruemag{}), which is defined as the magnitude of the vector sum of
\ptvec~of stable\footnote{ATLAS defines stable particles as those having a mean
 lifetime $>0.3\times10^{-10}$~s.} weakly interacting particles from
the hard-scatter collision. Then the
selection efficiency after a \met-threshold requirement is studied in simulated events with 
high-\pT~neutrinos (such as 
top-quark pair production and 
vector-boson fusion \Htau) or possible new weakly interacting particles that 
escape detection 
(such as the lightest supersymmetric particles).

This paper is organized as follows. Section~\ref{sec:detector} gives a
brief introduction to the ATLAS detector. Section~\ref{sec:Data} 
describes the data and MC simulation used as well as the event selections applied.
Section~\ref{sec:Rec} outlines how the \met~is reconstructed and
calibrated while Section~\ref{sec:Distributions} presents 
the level of agreement between data and MC simulation 
in $W$ and $Z$ boson production events.
Performance studies of the \met~algorithms on data 
and MC simulation are shown for samples with different event topologies in Section~\ref{sec:Performance}. The
choice of jet selection criteria used in the \met{} reconstruction is discussed 
in Section~\ref{sec:jet_opt}.
Finally, the systematic uncertainty in the absolute scale and
resolution of the \met~is discussed in
Section~\ref{sec:systematics}. To provide a reference,
Table~\ref{tab:metsummarytable} summarizes the different \met{} terms
discussed in this paper.

\begin{table}
	\caption{Summary of definitions for \met~terms used in this paper.}
	\footnotesize
	\begin{center}
\begin{tabular*}{\linewidth}{@{\extracolsep{\fill}}|c|l|}
\hline
		Term 	& Brief Description \\ \hline
 \multirow{5}{*}{\parbox{3.2cm}{Intrinsic \met{}} }&
                                                     \multirow{5}{*}{\parbox{11.5cm}{Missing
                                                     transverse momentum
                                                     arising from the presence of
                                                     neutrinos or other
                                                     non-interacting particles in an event. 
                                                     In case of simulated events
                                                     the true \met~(\mettrue) corresponds to
                                                     the \met~in such events defined as the 
                                                     magnitude of the vector sum of 
                                                     \ptvec~of non-interacting particles 
                                                     computed from the generator information.}} \\
& \\
& \\
& \\
& \\
 \multirow{3}{*}{\parbox{3.2cm}{Fake \met{}}} &
                                                \multirow{3}{*}{\parbox{11.5cm}{Missing
                                                     transverse momentum
                                                arising from the
                                                miscalibration or misidentification of
                                                physics objects in the
                                                event. It is typically
                                                studied in
                                                \Zmumu~events where
                                                the intrinsic \met{}
                                                is normally expected to be zero.}}  \\
& \\
&  \\

 \multirow{3}{*}{\parbox{3.2cm}{Hard Terms}} &
                                                \multirow{3}{*}{\parbox{11.5cm}{The
                                               component of the
                                               \met~computed from
                                               high-\pT{} physics
                                               objects, which includes
                                               reconstructed
                                               electrons, photons,
                                               muons, $\tau$-leptons, and
                                               jets. 
                                               }}  \\
& \\
& \\

 \multirow{3}{*}{\parbox{3.2cm}{Soft Terms}} &
                                                \multirow{3}{*}{\parbox{11.5cm}{Typically
                                               low-\pt~calorimeter energy
                                               deposits or tracks,
                                               depending on the soft-term definition, that
                                               are not associated to
                                               physics objects
                                               included in the hard terms.}}\\
& \\
& \\
 \multirow{3}{*}{\parbox{3.2cm}{Pileup-suppressed \met{}}} &
                                                \multirow{3}{*}{\parbox{11.5cm}{All
                                                             \met~reconstruction
                                                             algorithms
                                                             in Section~\ref{sec:soft_term}
                                                             except
                                                             the
                                                             Calorimeter
                                                             Soft Term,
                                                             which
                                                             does not
                                                             apply
                                                             pileup suppression.
                                                              }}  \\
& \\
&  \\
 \multirow{4}{*}{\parbox{3.2cm}{Object-based}} &
                                                \multirow{3}{*}{\parbox{11.5cm}{
                                                    This
                                                    refers to all 
                                                    reconstruction
                                                    algorithms in Section~\ref{sec:soft_term}
                                                    except the Track
                                                    \met{}, namely the
                                                 Calorimeter Soft
                                                 Term, Track Soft Term, Extrapolated Jet Area
                                                with Filter, and Soft-Term Vertex-Fraction algorithms.  
                                                    These consider the physics objects 
                                                    such as electrons, photons, muons, $\tau$-leptons, and jets
                                                    during the \met{} reconstruction. 
}}  \\
& \\
&  \\
&  \\
\hline
		\end{tabular*}
	\end{center}
	\label{tab:metsummarytable}
\end{table}

\section{ATLAS detector}
\label{sec:detector}
\noindent
The ATLAS detector~\cite{ATLAS_PHYS_PAP} is a multi-purpose particle
physics apparatus with a forward-backward symmetric cylindrical
geometry and nearly 4$\pi$ coverage in solid angle. For tracking, the inner detector (ID) covers
the pseudorapidity range of $|\eta|$~$<$~2.5, and consists of a
silicon-based pixel detector, 
a semiconductor tracker (SCT) based on microstrip technology, 
and, for $|\eta|$~$<$~2.0, 
a transition radiation tracker (TRT). 
The ID is surrounded by a thin superconducting solenoid providing a
2 T magnetic field, which allows the measurement of the momenta of charged particles. 
A high-granularity electromagnetic sampling calorimeter based on lead
and liquid argon (LAr) technology covers the region of $|\eta|<3.2$. 
A hadronic calorimeter based on steel absorbers and plastic-scintillator tiles
provides coverage for hadrons, jets, and $\tau$-leptons in the range of $|\eta|$~$<$~1.7. 
LAr technology using a copper absorber is also used for the hadronic calorimeters in the
end-cap region of 1.5~$<$~$|\eta|$~$<$~3.2 and for electromagnetic and 
hadronic measurements with copper and tungsten absorbing materials in
the forward region of 3.1~$<$~$|\eta|$~$<$~4.9. 
The muon spectrometer (MS) surrounds the calorimeters. 
It consists of three 
air-core superconducting toroid magnet systems, 
precision tracking chambers to provide accurate muon tracking
out to $|\eta|$~$=$~2.7, and additional detectors for triggering in the 
region of $|\eta|$~$<$~2.4. A precision measurement of the track coordinates 
is provided by layers of drift tubes at three radial positions within $|\eta|$~$<$~2.0. 
For 2.0~$<$~$|\eta|$~$<$~2.7, cathode-strip 
chambers with high granularity are instead used in the innermost plane.
The muon trigger system consists of resistive-plate chambers in the 
barrel ($|\eta|$~$<$~1.05) and thin-gap chambers in the end-cap 
regions (1.05~$<$~$|\eta|$~$<$~2.4).

\FloatBarrier
\section{Data samples and event selection}
\label{sec:Data}
ATLAS recorded $pp$ collisions at a centre-of-mass energy
of 8 \TeV~with a bunch crossing interval (bunch spacing) of  
$50\,{\textrm ns}$ in 2012. The resulting integrated luminosity is
$20.3$~$\ifb$~\cite{Aaboud2016hhf}.
Multiple inelastic $\pp$ interactions occurred in each bunch crossing, and
the mean number of inelastic collisions per bunch crossing (\muavno)
over the full dataset is 21~\cite{ATLASPU8TeV}, exceptionally reaching as high as about 70.

Data are analysed only if they satisfy the standard 
ATLAS data-quality assessment criteria~\cite{ATLASDQcleaning}.
Jet-cleaning cuts~\cite{ATLASDQcleaning} are applied to 
minimize the impact of instrumental noise and out-of-time 
energy deposits in the calorimeter from cosmic rays or beam-induced
backgrounds.
This ensures that the residual sources of \metmag~mismeasurement due to those
instrumental effects are suppressed. 

\subsection {Track and vertex selection}
\label{sec:evt_reco}

The ATLAS detector measures the momenta of charged particles using the 
ID~\cite{ATLAS-CONF-2014-047}. 
Hits from charged particles are recorded and are used to reconstruct
tracks; these are used to reconstruct
vertices~\cite{ATLAS-CONF-2012-042,Aad2010ac}. 

Each vertex must have at least two tracks with
$\pT$~$>$~0.4~\GeV{}; for the primary hard-scatter vertex (PV),
the requirement on the number of tracks is raised to three. The PV in each event is
selected as the vertex with the largest value of $\sumpTsq$,
where the scalar sum is taken over all the tracks 
matched to the vertex. 
The following track selection criteria\footnote{The
  track reconstruction for electrons and for
  muons does not strictly follow these definitions. For example, a
  Gaussian Sum Filter~\cite{Aad2014fxa} algorithm is used for electrons to improve the
  measurements of its track parameters, which can be degraded due
to Bremsstrahlung losses.}~\cite{ATLAS-CONF-2012-042} are used 
throughout this paper, including the vertex
reconstruction:
 \begin{itemize} 
 \item \pT{}~$>$~0.5 \GeV{} (0.4~\GeV~for vertex reconstruction and
   the calorimeter soft term),
 \item $|\eta|$~$<$~2.5,
 \item Number of hits in the pixel detector $\geq$ 1,
 \item Number of hits in the SCT $\geq$ 6.
 \end{itemize} 
These tracks are then matched to the PV by applying the following
selections: 
\begin{itemize} 
 \item $|d_0|$~$<$~1.5 mm,
 \item $|z_0\sin(\theta$)$|$~$<$~1.5 mm.
 \end{itemize} 
The transverse (longitudinal) impact parameter $d_0$ $(z_0)$ is the
transverse (longitudinal) distance of the track from the PV and is
computed at the point of closest approach to the PV in the plane 
transverse to the beam axis. 
The requirements on the number of hits ensures that
the track has an accurate \pT~measurement. 
The $|\eta|$ requirement keeps only the tracks within the ID acceptance, and the
requirement of \pT~$>$~0.4~\GeV{} 
ensures that the track reaches the outer layers of the ID. 
Tracks with low \pT~have large 
curvature and are more susceptible to multiple scattering. 

The average spread along the beamline
direction for $pp$ collisions
in ATLAS during 2012 data taking is around 50~mm, and the
typical track $z_0$ resolution for those with
$|\eta|~<~0.2$ and $0.5~<~\pt~<~0.6$~\GeV{} is 0.34~mm. The
typical track $d_0$ resolution is around 0.19~mm for the
same $\eta$ and \pt{} ranges, and both the $z_0$ and $d_0$
resolutions improve with higher track \pT{}.

Pileup effects come from two sources: in-time and out-of-time. 
In-time pileup is the result of multiple $pp$ interactions in
the same LHC bunch crossing. 
It is possible to distinguish the in-time pileup interactions
by using their vertex positions, which are spread along the beam
axis. At \muavno~$=$~21, the efficiency to 
reconstruct and select the correct vertex for
\Zmm~simulated events is around 93.5\% and rises to more than 98\% when requiring
two generated muons with \pT~$>$~10 \GeV{} inside the ID
acceptance~\cite{vertex}. When vertices are separated along the beam axis by a distance
smaller than the position resolution, they can be reconstructed as a single vertex. Each
track in the reconstructed vertex is assigned a weight based upon its compatibility
with the fitted vertex, which depends on the $\chi^2$ of the fit. The
fraction of \Zmm~reconstructed vertices with more than 50\% of the sum
of track weights coming from pileup
interactions is around 3\% at \muavno~$=$~21~\cite{ATLAS-CONF-2012-042,vertex}.
Out-of-time pileup 
comes from $pp$ collisions in earlier and later bunch crossings, 
which leave signals in the calorimeters that can take up to 450~ns for
the charge collection time. This is longer than the 
50 ns between subsequent collisions and occurs because
the integration time of the calorimeters is significantly 
larger than the time between the bunch crossings. By contrast the
charge collection time of the silicon tracker is less than 25~ns.

\subsection{Event selection for \Zll{}}
\label{sec:evtselZll}
The ``standard candle'' for evaluation of the \metmag~performance is
\Zll~events ($\ell=e$ or $\mu$). They are produced without neutrinos,
apart from a very small number originating from heavy-flavour  
decays in jets produced in association with the $Z$ boson. 
The intrinsic \metmag~is therefore
expected to be close to zero, and the \metmag~distributions are used to
evaluate the modelling of the effects 
that give rise to fake \met{}.

Candidate \Zll{} events are
required to pass an electron or muon
trigger~\cite{ATLAS-CONF-2012-048,ATLASMuonTrigger}. The lowest \pT~threshold for the
unprescaled single-electron (single-muon) trigger is \pt~$>$~25 (24)~\GeV{}, 
and both triggers apply a track-based isolation as well as quality selection 
criteria for the
particle identification. Triggers with higher \pT~thresholds,
without the isolation requirements, are used to improve
acceptance at high \pT{}. These triggers require
\pt~$>$~60 (36)~\GeV{} for electrons (muons). 
Events are accepted if they pass 
any of the above trigger criteria. 
Each event must contain at least one 
primary vertex with a $z$ displacement from the nominal $pp$
interaction point of less than $200\,{\textrm mm}$ and with at least three
associated tracks.

The offline selection of \Zmm{} events requires the presence of
exactly two identified muons~\cite{ATLASMuonPerf12}. An identified muon is 
reconstructed in the MS and is matched to a track in the
ID. The combined ID$+$MS track must have \pT~$>$~25~\GeV{} and $|\eta|$~$<$~2.5. 
The $z$ displacement of
the muon track from the primary vertex is required to be less than
$10$ mm. An isolation criterion is applied to
the muon track, where the scalar sum of the \pT~of additional tracks within a 
cone of size $\Delta R$~$=$~$\sqrt{(\Delta\eta)^2+(\Delta\phi)^2}$~$=$~0.2 around the 
muon is required to be less than 10\% of the muon \pt{}. 
In addition, the two leptons are
required to have opposite charge, 
and the reconstructed dilepton invariant mass, \mll{}, is required to be
consistent with the $Z$ boson mass: 66~$<$~\mll~$<$~116~\GeV{}.

The \met{} modelling and performance results obtained in \Zmm{} and \Zee{} events
are very similar. For the sake of brevity, 
only the \Zmm{} distributions are shown in all sections
except for Section~\ref{sec:tails}.

\subsection{Event selection for \Wln{}}
\label{sec:evtselWlv}
Leptonically decaying $W$ bosons (\Wln) provide an important event
topology with intrinsic \met; the \met{} distribution for such 
events is presented in Section~\ref{sec:Perf_W}.
Similar to \Zll{} events, a sample dominated by leptonically
decaying $W$ bosons is used to study the
\metmag~scale in Section~\ref{sec:linearity},
the resolution of the \met~direction in Section~\ref{sec:metdir}, and
the impact on a reconstructed kinematic observable in Section~\ref{sec:kinematic_obs}. 

The \met~distributions for $W$ boson events in Section~\ref{sec:Perf_W} use the
electron final state. These electrons are selected with
$|\eta|$~$<$~2.47, are required to meet the ``medium''  
identification criteria~\cite{Aad2014nim} and 
satisfy \pT~$>$~25~\GeV. 
Electron candidates in the region 1.37~$<$~$|\eta|$~$<$~1.52 suffer from degraded 
momentum resolution and particle identification due to the transition
from the barrel to the end-cap detector and are therefore discarded in these studies. 
The electrons are required to be isolated, such that the sum of the
energy in the calorimeter within a cone of size $\Delta R$~$=$~0.3
around the electron is less than 14\%
of the electron \pT{}. The summed \pT{} of other tracks within the
same cone is required to be less than 7\% of the electron \pT{}.
The calorimeter isolation variable~\cite{Aad2014nim} is corrected by subtracting estimated contributions from the
electron itself, the underlying event~\cite{UE_atlas}, and
pileup.  The electron tracks are then matched to
the PV by applying the following selections: 
\begin{itemize} 
 \item $|d_0|$~$<$~5.0 mm,
 \item $|z_0\sin(\theta$)$|$~$<$~0.5 mm.
 \end{itemize} 

The $W$ boson selection is based on the single-lepton
triggers and the same lepton selection criteria as those used in the
\Zll~selection. Events are rejected if they contain more than one reconstructed
lepton. Selections on the \metmag~and transverse mass (\mT{}) are 
applied to reduce the multi-jet background with one jet misidentified
as an isolated lepton. The
transverse mass is calculated from the lepton and the \metvec{}, 
\begin{equation}
  \mT= \sqrt{2p_{\mathrm T} ^{\ell}
    E_{\mathrm{T}}^{\mathrm{miss}}(1-\cos\Delta\phi)},
\label{eq:mT}
\end{equation}
\noindent where $p_{\mathrm T}^{\ell}$ is the transverse momentum of the lepton and $\Delta\phi$ is the azimuthal angle 
between the lepton and \metvec{} directions. Both the \mT{} and \met{}
are required to be greater than 50~\GeV{}. These selections can bias
the event topology and its phase space, so they are only used when 
comparing simulation to data in 
Section~\ref{sec:Perf_W}, as they substantially improve the purity of $W$ bosons in data events.

The \met{} modelling and performance results obtained in \Wen{} and \Wmn{} events
are very similar. For the sake of brevity, only one of the two is considered
in following two sections: \met{} distributions in \Wen{} events are
presented in Section~\ref{sec:Perf_W} and 
the performance studies show \Wmn{} events
in Section~\ref{sec:Performance}. When studying the \met{} tails, 
both final states are considered in Section~\ref{sec:tails}, because
the $\eta$-coverage and reconstruction performance between muons 
and electrons differ. 

\subsection{Monte Carlo simulation samples}
\label{sec:MC}

Table~\ref{tab:gen_samples} summarizes the MC simulation 
samples used in this paper. The \Zll{} and \Wln~samples are generated with \ALPGEN~\cite{alpgen}
interfaced with \PYTHIA~\cite{Sjostrand2006za}
(denoted by \ALPGEN$+$\PYTHIA)
to model the parton shower and hadronization, and underlying event using the PERUGIA2011C set~\cite{perugia}
of tunable parameters. 
One exception is the \Ztautau~sample with leptonically decaying
$\tau$-leptons, which is generated with \ALPGEN~interfaced with \HERWIG{}~\cite{Corcella2000bw} with the 
underlying event modelled using {\JIMMY}~\cite{Jimmy} and the AUET2 tunes~\cite{ATL-PHYS-PUB-2011-008}. 
\ALPGEN~is a multi-leg generator that provides tree-level calculations
for diagrams with up to five additional partons. 
The matrix-element MC calculations are matched to a model of the 
parton shower, underlying event and hadronization.
The main processes that are backgrounds to \Zll~and \Wln~are
events with one or more top quarks (\ttbar~and single-top-quark processes) and diboson 
production ($WW$, $WZ$, $ZZ$). The \ttbar~and $tW$ processes are generated with \POWHEG~\cite{Nason2004rx} interfaced with 
\PYTHIA~\cite{Sjostrand2006za} for hadronization and parton showering, and PERUGIA2011C for the 
underlying event modelling. 
All the diboson processes are generated
with \SHERPA~\cite{Gleisberg2008ta}. 
\POWHEG{} is a leading-order generator with corrections at next-to-leading 
order in $\alphaS$, whereas 
\SHERPA{} is a multi-leg generator at tree level. 

To study event topologies with high jet multiplicities and to investigate 
the tails of the \metmag~distributions, 
\ttbar~events with at least one leptonically decaying $W$ boson are
considered in Section~\ref{sec:tails}. The single top quark ($tW$)
production is considered with at least one leptonically decaying $W$
boson. Both the \ttbar~and $tW$ processes
contribute to the $W$ and $Z$ boson distributions
shown in Section~\ref{sec:Distributions} as well as $Z$ boson distributions
in Sections~\ref{sec:Rec}, \ref{sec:Performance}, and
\ref{sec:systematics} that compare data and simulation. A supersymmetric (SUSY) model comprising pair-produced 500 GeV gluinos each decaying   
to a \ttbar{} pair and a neutralino 
is simulated with \HERWIGPP{}~\cite{Herwigpp}. 
Finally, to study events with forward
jets, the vector-boson fusion (VBF) production of \Htau{}, generated with 
\POWHEG$+$\PYTHIAEight{}~\cite{Pythia8a}, is considered. Both $\tau$-leptons are forced to decay 
leptonically in this sample. 

\begin{table}[h]
\begin{center}
\caption{\label{tab:gen_samples} {Generators, cross-section
    normalizations, PDF sets, and MC tunes used in this analysis. } }
$\newline$ 
\resizebox{1.0\textwidth}{!}{
\begin{tabular}{|l||l|l|l|l|l|} \hline
  Sample & Generator & Use & Cross-section & PDF set & Tune \\ \hline\hline
  \Zmm{} &  \ALPGEN$+$\PYTHIA{}& Signal & NNLO~\cite{Melnikov2006kv} & CTEQ6L1~\cite{Pumplin2002vw} & PERUGIA2011C~\cite{perugia} \\ 
  \Zee{} &  \ALPGEN$+$\PYTHIA{}& Signal & NNLO~\cite{Melnikov2006kv} & CTEQ6L1 & PERUGIA2011C \\ 
  \Ztautau{} &\ALPGEN$+$\HERWIG{} & Signal & NNLO~\cite{Melnikov2006kv} & CTEQ6L1 & AUET2~\cite{ATL-PHYS-PUB-2011-008} \\
  \Wmn{} &  \ALPGEN$+$\PYTHIA{}& Signal & NNLO~\cite{Melnikov2006kv} & CTEQ6L1 & PERUGIA2011C \\ 
  \Wen{} &  \ALPGEN$+$\PYTHIA{}& Signal & NNLO~\cite{Melnikov2006kv} & CTEQ6L1 & PERUGIA2011C \\
  \Wtaunu{} &  \ALPGEN$+$\PYTHIA{}& Signal & NNLO~\cite{Melnikov2006kv} & CTEQ6L1 & PERUGIA2011C \\
  \ttbar{} &  \POWHEG$+$\PYTHIA{}& Signal/Background & NNLO+NNLL~\cite{ttbarxsec1,ttbarxsec2} & CTEQ6L1 & PERUGIA2011C \\ 
  VBF \Htau{} & \POWHEG$+$\PYTHIAEight{} & Signal & -- & NLO CT10~\cite{Lai2010vv} & AU2~\cite{ATL-PHYS-PUB-2012-003} \\ 
  SUSY 500 & \HERWIGPP{} & Signal & -- & CTEQ6L1 & UE EE3~\cite{ueee3} \\\hline 
  $W^{\pm}Z\rightarrow \ell^{\pm}\nu\ell^+\ell^-$ & \SHERPA{} & Background & NLO~\cite{diboson1,diboson2} & NLO CT10 & \SHERPA~default \\
  $ZZ\rightarrow\ell^+\ell^-\nu\bar{\nu}$ &  \SHERPA{} & Background & NLO~\cite{diboson1,diboson2} & NLO CT10 & \SHERPA~default \\
  $W^+W^-\rightarrow\ell^+\nu\ell^-\bar{\nu}$ & \SHERPA{}  & Background & NLO~\cite{diboson1,diboson2} & NLO CT10 & \SHERPA~default \\
  $tW$ &  \POWHEG$+$\PYTHIA{} & Background & NNLO+NNLL~\cite{Kidonakis2010b} & CTEQ6L1 & PERUGIA2011C \\
\hline
  \Zmm{} &  \POWHEG{}$+$\PYTHIAEight{} & Systematic Effects & NNLO~\cite{DYNNLO1,DYNNLO2} & NLO CT10 & AU2 \\ 
  \Zmm{} &  \ALPGEN$+$\HERWIG{} & Systematic Effects & NNLO~\cite{DYNNLO1,DYNNLO2} & CTEQ6L1 & AUET2\\ 
  \Zmm{} &  \SHERPA{}& Systematic Effects & NNLO~\cite{DYNNLO1,DYNNLO2} & NLO CT10 & \SHERPA~default \\ \hline
\end{tabular}
}
\end{center}
\end{table}

To estimate the systematic uncertainties in the data/MC ratio arising from the
 modelling of the soft hadronic recoil, \met{} distributions simulated with different
 MC generators, parton shower and underlying event models are compared. 
The estimation of systematic uncertainties is performed using 
a comparison of data and MC simulation,
as shown in Section~\ref{sec:track_syst}. The following
combinations of generators and parton shower models are considered: \SHERPA,
\ALPGEN{}$+$\HERWIG{}, \ALPGEN{}$+$\PYTHIA{},
and \POWHEG{}$+$\PYTHIAEight{}. The corresponding underlying event tunes 
are mentioned in Table~\ref{tab:gen_samples}.
Parton distribution functions 
are taken from \CTTen~\cite{Lai2010vv} for \POWHEG~and \SHERPA~samples and 
\CTEQSix~\cite{cteq6} for \ALPGEN~samples.

Generated events are propagated through a \GEANTFour~simulation~\cite{GEANT4,Geant4ATLAS} 
of the ATLAS detector. Pileup collisions are generated with \PYTHIAEight{} for all samples,
and are overlaid on top of simulated hard-scatter events
before event reconstruction. Each simulation sample is weighted by its corresponding cross-section and normalized to the integrated luminosity of the data.


\section {Reconstruction and calibration of the \met{}}
\label{sec:Rec}
Several algorithms have been developed to reconstruct the \met~in ATLAS. They differ in the information used to reconstruct the \pt{}
of the particles, using either energy deposits in the calorimeters,
tracks reconstructed in the ID, or both.
This section describes these various reconstruction algorithms, 
and the remaining sections discuss the agreement between data and MC
simulation as well as performance studies. 

\subsection {Reconstruction of the \met{}}
\label{sec:MetTerms}

The \met~reconstruction uses calibrated physics
objects to estimate the amount of missing transverse momentum in the
detector. The \met{} is calculated using the components along the $x$ and $y$ axes:
\begin{eqnarray}
       E_{{x(y)}}^{\mathrm{miss}} =
            E_{{x(y)}}^{\mathrm{miss},e}         +
              E_{{x(y)}}^{\mathrm{miss},\gamma}    +
              E_{{x(y)}}^{\mathrm{miss},\tau}      +
              E_{{x(y)}}^{\mathrm{miss,jets}}  +
               E_{{x(y)}}^{\mathrm{miss},\mu} +
               E_{{x(y)}}^{\mathrm{miss,soft}} ,
\label{eq7} 
\end{eqnarray}
where each term is calculated as the negative vectorial sum of
transverse momenta of energy deposits and/or tracks.
To avoid double counting, energy deposits in the calorimeters and
tracks are matched to reconstructed physics objects in the following order:
electrons ($e$), photons ($\gamma$), the visible parts of 
hadronically decaying $\tau$-leptons ($\tau_{\mathrm{had-vis}}$; labelled as $\tau$),
jets and muons ($\mu$). Each type of physics object is represented by a
separate term in Eq.~(\ref{eq7}). 
The signals not associated with physics objects form the ``soft term'', whereas those associated with the physics objects are collectively referred to as the ``hard term''.

The magnitude and azimuthal angle\footnote{The $\textrm{arctan}$
  function returns values from $[-\pi,+\pi]$ and uses the sign of both
  coordinates to determine the quadrant.} (\phimiss) of \metvec{} are calculated as:
\begin{eqnarray}
 E_{\mathrm{T}}^{\mathrm{miss}}=\sqrt{(E_{{x}}
 ^{\mathrm{miss}})^{2} +(E_{{y}}^{\mathrm{miss}})^{2}}, \nonumber \\ 
\phi^{\textrm miss}=\textrm{arctan}(E_{{y}}^{\mathrm{miss}} /
  E_{{x}}^{\mathrm{miss}}).
\label{eq12} 
\end{eqnarray}

The total transverse energy in the detector, 
labelled as \sumet, quantifies the total event activity and is
an important observable for understanding the resolution of the
\met{}, especially with increasing pileup contributions.
It is defined as: 
\begin{eqnarray}
       \sum E_{\mathrm{T}} =
              \sum p_{\mathrm{T}}^{e}         +
              \sum p_{\mathrm{T}}^{\gamma}    +
              \sum p_{\mathrm{T}}^{\tau}      +
               \sum p_{\mathrm{T}}^{\mathrm{jets}}     +
               \sum p_{\mathrm{T}}^{\mu} +
               \sum p_{\mathrm{T}}^{\mathrm{soft}},
 \label{eqsumet}
\end{eqnarray}
\noindent{}which is the scalar sum of the transverse momenta of
reconstructed physics objects and soft-term signals that contribute to
the \met{} reconstruction. The physics objects included in 
$\sum p_{\mathrm{T}}^{\mathrm{soft}}$ depend on the \met~definition, 
so both calorimeter objects and track-based
objects may be included in the sum, despite differences in \pT~resolution.

\subsubsection {Reconstruction and calibration of the \metmag~hard terms}
\label{sec:hard_term}

The hard term of the \met{}, which is computed from the reconstructed electrons, photons, 
muons, $\tau$-leptons, and jets, is described in more detail in this section.

Electrons are reconstructed from clusters in the
electromagnetic (EM) calorimeter which are associated with an ID
track~\cite{Aad2014nim}. Electron identification is restricted to
the range of $|\eta|$~$<$~$2.47$, excluding the transition region between the
barrel and end-cap EM calorimeters, $1.37$~$<$~$|\eta|$~$<$~$1.52$. They are calibrated at the EM scale\footnote{The EM scale is the basic signal scale for the ATLAS calorimeters. It accounts correctly for the energy deposited by EM showers in the calorimeter, but it does not consider energy losses in the un-instrumented material.} with the default electron
calibration, and those satisfying the
``medium'' selection criteria~\cite{Aad2014nim} with $\pt>10$~\GeV{} are
included in the \met{} reconstruction. 

The photon reconstruction is also seeded from clusters of
energy deposited in the EM calorimeter and is designed to separate electrons from photons.
 Photons are calibrated at the EM scale and are required to satisfy the ``tight'' photon selection criteria 
 with \pt~$>$~10~\GeV{}~\cite{Aad2014nim}.  

Muon candidates are identified by matching an ID track
with an MS track or segment~\cite{ATLASMuonPerf12}. 
MS tracks
are used for $2.5$~$<$~$|\eta|$~$<$~$2.7$ to
extend the $\eta$ coverage.
Muons are required to satisfy $\pT$~$>$~$5$~\GeV~to be included in the \met{} reconstruction. 
The contribution of muon energy
deposited in the calorimeter is taken into account using either
parameterized estimates or direct
measurements, to avoid double counting a small fraction of their momenta.

Jets are reconstructed from three-dimensional topological 
 clusters (topoclusters)~\cite{TopoClusters} of energy deposits in the
 calorimeter using the anti-$k_t$
 algorithm~\cite{CacciariAntiKt} with a distance parameter
 $R$~$=$~0.4. The topological clustering algorithm suppresses noise by
 forming contiguous clusters of calorimeter cells with significant 
 energy deposits. The local cluster weighting (LCW)~\cite{Aad2011he,ATLAS-CONF-2015-017} calibration is used to account for different
 calorimeter responses to electrons, photons and hadrons. Each
 cluster is classified as coming
 from an EM or hadronic shower, using information from its
 shape and energy density, and calibrated accordingly. 
 The jets are reconstructed from
 calibrated topoclusters and then corrected for in-time and out-of-time
  pileup as well as the position of
  the PV~\cite{ATLASPU8TeV}. Finally, the jet energy scale
 (JES) corrects for jet-level effects by restoring, on average, the energy of 
 reconstructed jets to that of the MC generator-level jets. The complete procedure is referred to as the LCW$+$JES
 scheme~\cite{Aad2011he,ATLAS-CONF-2015-017}. Without changing the
 average calibration, additional corrections are made based upon the
 internal properties of the jet (global sequential calibration) to 
 reduce the flavour dependence and 
 energy leakage effects~\cite{ATLAS-CONF-2015-017}. 
 Only jets with calibrated
 \pT~greater than $20$~\GeV~are used to calculate 
  the jet term $E_{{x(y)}}^{\mathrm{miss,jets}}$ in
  Eq.~(\ref{eq7}), and the optimization of the 20~\GeV{} threshold is
  discussed in Section~\ref{sec:jet_opt}. 

To suppress contributions from jets originating from pileup interactions, 
a requirement on the jet vertex-fraction (JVF)~\cite{ATLASPU8TeV} may
be applied to 
selected jet candidates. Tracks matched to jets are extrapolated 
back to the beamline to ascertain whether they originate from the hard 
scatter or from a pileup collision. The JVF is then computed as the ratio shown below:
\begin{eqnarray}
{\mathrm {JVF}} = \sum_{\mathrm {track,PV,jet}} p_{\mathrm{T}} \mathlarger{\mathlarger{\mathlarger{/}}}  \sum_{\mathrm {track,jet}}  p_{\mathrm{T}}.
\label{eqn:jvf} 
\end{eqnarray}
This is the ratio of the scalar sum of transverse momentum of all tracks
matched to the jet and the primary vertex to the
\pT~sum of all tracks matched to the jet, where the sum is performed over all tracks with \pt~$>$~0.5 \GeV~and
$|\eta|$~$<$~2.5 and the matching is performed using the ``ghost-association'' procedure~\cite{CacciariJetArea,Cacciari2008gn}.

The JVF distribution is peaked toward 1 for hard-scatter jets and 
toward 0 for pileup jets. 
No JVF selection requirement is applied to jets that have no associated tracks. 
Requirements on the JVF are made in 
the STVF, EJAF, and TST \met~algorithms
as described in Table~\ref{tab:softtermsummarytable} and Section~\ref{sec:jet_selec}.

Hadronically decaying $\tau$-leptons are seeded by calorimeter jets
with $|\eta|$~$<$~2.5 and \pt~$>$~10 \GeV{}. As described for jets, the
LCW calibration is applied, corrections are made to subtract the energy due to pileup
interactions, and the energy of the hadronically
decaying $\tau$ candidates is calibrated at the $\tau$-lepton energy
scale (TES)~\cite{Aad2014rga}. The TES is independent of the JES and is determined using an MC-based procedure. Hadronically decaying $\tau$-leptons 
passing the ``medium'' requirements~\cite{Aad2014rga} and 
having \pt~$>$ 20 \GeV{} after TES corrections are considered
for the \met{} reconstruction.

\subsubsection {Reconstruction and calibration of the \metmag~soft term}
\label{sec:soft_term}

The soft term is a necessary but challenging ingredient of the \met{}
reconstruction. It comprises all the detector signals not
matched to the physics objects defined above and can
contain contributions from the hard scatter as well as the underlying event and pileup interactions. Several algorithms designed to 
reconstruct and calibrate the soft term 
have been developed, as well as methods to suppress the pileup
contributions. 
A summary of the \met{} and soft-term reconstruction algorithms is given 
in Table~\ref{tab:softtermsummarytable}. 

\begin{table}
	\caption{Summary of \met{} and soft-term reconstruction algorithms
          used in this paper.}
	\footnotesize
	\begin{center}
\begin{tabular*}{\linewidth}{@{\extracolsep{\fill}}|c|c|l|}
\hline
		Term 	& Brief Description & Section list  \\ \hline
 \multirow{5}{*}{\parbox{1.7cm}{CST \met{}}} &
                                               \multirow{5}{*}{\parbox{9cm}{The Calorimeter Soft Term (CST) \met{} takes its soft term from energy deposits in the calorimeter which are not matched to high-\pT{} physics objects. Although noise suppression is applied to reduce fake signals, no additional pileup suppression techniques are used. }}&  \\
& & Sect.~\ref{sec:cst} (Definition) \\
& & Sect.~\ref{sec:Perf_Z} (\Zmumu~modelling)\\
& & Sect.~\ref{sec:Perf_W} (\Wen~modelling)\\
& & Sect.~\ref{sec:Performance} (Perf. studies) \\
 \multirow{5}{*}{\parbox{1.7cm}{TST \met{}}} &
                                               \multirow{5}{*}{\parbox{9cm}{The
                                               Track Soft Term (TST) \met{}
                                               algorithm uses a soft
                                               term that is calculated
                                               using tracks within the
                                               inner detector that are not associated with 
                                               high-\pT{} physics objects. The JVF selection 
                                               requirement is applied to jets.}}&  \\
& &Sect.~\ref{sec:tst} (Definition) \\
& & Sect.~\ref{sec:Perf_Z} (\Zmumu~modelling)\\
& & Sect.~\ref{sec:Perf_W} (\Wen~modelling)\\
& & Sect.~\ref{sec:Performance} (Perf. studies) \\
 \multirow{5}{*}{\parbox{1.7cm}{EJAF \met{}}} &
                                                \multirow{5}{*}{\parbox{9cm}{The
                                                Extrapolated Jet Area
                                                with Filter \met{}
                                                algorithm applies
                                                pileup subtraction to
                                                the CST based on the idea of jet-area corrections. 
                                                The JVF selection 
                                                requirement is applied to jets.}}&  \\
& &Sect.~\ref{sec:PU_JetArea0} (Definition) \\
& & Sect.~\ref{sec:Perf_Z} (\Zmumu~modelling)\\
& & Sect.~\ref{sec:Performance} (Perf. studies) \\
& & \\
 \multirow{5}{*}{\parbox{1.7cm}{STVF \met{}}} &
                                                \multirow{5}{*}{\parbox{9cm}{The
                                                Soft-Term Vertex-Fraction (STVF) \met{}
                                                algorithm suppresses
                                                pileup effects in the
                                                CST by scaling the
                                                soft term by a
                                                multiplicative factor
                                                calculated based on
                                                the fraction of scalar-summed 
                                                track \pt~not
                                                associated with
                                                high-\pT{} physics
                                                objects that can be
                                                matched to the primary vertex.
                                                The JVF selection 
                                               requirement is applied to jets.}}&  \\
& & Sect.~\ref{sec:stvf} (Definition) \\
& & Sect.~\ref{sec:Perf_Z} (\Zmumu~modelling)\\
& & Sect.~\ref{sec:Performance} (Perf. studies) \\
& & \\
 \multirow{5}{*}{\parbox{1.7cm}{Track \met{}}} & \multirow{5}{*}{\parbox{9cm}{The Track \met{} is reconstructed entirely from tracks to avoid pileup contamination that affects the other algorithms.}}&  \\
& & Sect.~\ref{sec:trkmetdef} (Definition) \\
& & Sect.~\ref{sec:Perf_Z} (\Zmumu~modelling)\\
& & Sect.~\ref{sec:Performance} (Perf. studies) \\
& & \\
\hline
		\end{tabular*}
	\end{center}
	\label{tab:softtermsummarytable}
\end{table}

Four soft-term reconstruction algorithms are considered in this
paper. Below the first two are defined, and then some motivation is given
for the remaining two prior to their definition.

\begin{itemize}
\item Calorimeter Soft Term (CST)$\newline$
\label{sec:cst}
\noindent{}This reconstruction algorithm~\cite{ATLASMETPaper2011}
uses information mainly from the calorimeter and 
is widely used by ATLAS. The algorithm also includes corrections based
on tracks but does not attempt to resolve the various $pp$
interactions based on the track $z_0$ measurement. The soft term is referred to as the
CST, whereas the entire \met{} is written as CST \met{}. Corresponding
naming schemes are used for the other reconstruction algorithms. The CST is reconstructed using energy deposits in the
calorimeter which are not matched to the high-\pT{} physics objects used in the
\met{}. To avoid fake signals in the calorimeter, noise suppression is
important. This is achieved by calculating the soft term using only cells belonging to topoclusters, which
are calibrated at the LCW scale~\cite{Aad2011he,ATLAS-CONF-2015-017}. 
The tracker and calorimeter provide redundant \pT{} measurements for
charged particles, so
an energy-flow algorithm is used to determine which measurement to use. Tracks with \pT~$>$~0.4~\GeV{} that
are not matched to a high-\pT~physics objects are used instead of the
calorimeter \pT~measurement, if their \pT~resolution is better than the expected calorimeter \pt~resolution. 
The calorimeter resolution is estimated as $0.4\cdot\sqrt{\pt}~\GeV{}$, in which the
\pt~is the transverse momentum of the reconstructed track. 
Geometrical matching between tracks and topoclusters (or 
high-\pT~physics objects) is performed using the $\Delta R$ significance
defined as $\Delta R / \sigma_{\Delta R}$, where $\sigma_{\Delta R}$ is the $\Delta R$ 
resolution, parameterized as a function of the track \pT{}. 
A track is considered to be associated to a topocluster in the soft term 
when its minimum $\Delta R / \sigma_{\Delta R}$ is less than 4. 
To veto tracks matched to high-\pT~physics objects,
tracks are required to have $\Delta R / \sigma_{\Delta R}$~$>$~8. 
The \metmag~calculated using the CST algorithm is
documented in previous publications such as Ref.~\cite{ATLASMETPaper2011} 
and is the standard algorithm in most ATLAS 8~\TeV~analyses.

\item Track Soft Term (TST)$\newline$
\label{sec:tst}
\noindent{}The TST is 
reconstructed purely from tracks that pass the selections 
outlined in Section~\ref{sec:evt_reco} and are not associated with the high-\pT{} physics
objects defined in Section~\ref{sec:hard_term}. The detector coverage of the TST is the
ID tracking volume ($|\eta|$~$<$~2.5), and no calorimeter topoclusters
inside or beyond this region are included.
This algorithm allows excellent vertex matching for the soft term,
which almost completely removes the in-time pileup dependence, but
misses contributions from soft neutral particles. The track-based
reconstruction also entirely removes the out-of-time pileup contributions
that affect the CST.

To avoid double counting the \pT{} of particles, the tracks matched to the 
high-\pT{} physics objects need to be removed from the soft
term. All of the following classes of tracks are excluded from the soft term:
\begin{itemize}
\item[-] tracks within a cone of size $\Delta R$~$=$~0.05 around
  electrons and photons
\item[-] tracks within a cone of size $\Delta R$~$=$~0.2 around \tauvis{}
\item[-] ID tracks associated with identified muons
\item[-] tracks matched to jets using the ghost-association technique described in Section~\ref{sec:hard_term}
\item[-] isolated tracks with $\pT~\geq~120$~\GeV~($\geq~200$~\GeV~for
  $|\eta|$~$<$~1.5) having transverse momentum uncertainties larger than
  40\% or having no associated calorimeter energy deposit with \pt~larger than 
  65\% of the track \pT{}. The \pT~thresholds are chosen to ensure
  that muons not in the coverage of the MS are still included in the
  soft term. This is a cleaning cut to remove mismeasured tracks.
\end{itemize}

\end{itemize}

\noindent{}
A deterioration of the CST \met~resolution 
is observed as the average
number of pileup interactions increases~\cite{ATLASMETPaper2011}. All
\met~terms in Eq.~(\ref{eq7}) are affected by
pileup, but the terms which are most affected are the jet term and CST, 
because their constituents are spread over larger regions in the calorimeters
than those of the \met~hard terms. 
Methods to suppress pileup are therefore needed, which can restore
the \metmag~resolution to values similar to those observed in the
absence of pileup. 

The TST algorithm is very stable with respect to pileup but does not include neutral
particles. 
Two other pileup-suppressing algorithms were developed, which consider contributions from neutral particles. 
One uses an $\eta$-dependent event-by-event estimator for the transverse 
momentum density from pileup, using calorimeter information, while
the other applies an event-by-event global correction based on the
amount of charged-particle \pT{} from the hard-scatter vertex, relative
to all other $pp$ collisions. 
The definitions of these two soft-term algorithms are described in the following:

\begin{itemize}

\item Extrapolated Jet Area with Filter (EJAF) $\newline$
\label{sec:PU_JetArea0}
\noindent{}The jet-area method for the pileup subtraction uses a soft
term based on the idea of jet-area corrections~\cite{CacciariJetArea}.
This technique uses direct event-by-event measurements of
the energy flow throughout the entire ATLAS detector to estimate the
\pT~density of pileup energy deposits and was developed from the strategy applied to
jets as described in Ref.~\cite{ATLASPU8TeV}.

The topoclusters belonging to the soft term are used for jet finding with 
the $k_{t}$ algorithm~\cite{CATANI1993187,PhysRevD.48.3160} with distance parameter $R$~$=$~0.6 and jet \pt~$>$~0. 
The catchment areas~\cite{CacciariJetArea,Cacciari2008gn} for these reconstructed
jets are labelled $A_{\mathrm{jet}}$; this provides a measure of the
jet's susceptibility to contamination from pileup. Jets with
\pT~$<$~20~\GeV{} are referred to as soft-term jets, and
the \pT-density of each soft-term jet $i$ is then measured by computing:
\begin{equation}
\rho_{\mathrm{jet}, i} = \frac{p_{\mathrm{T}, i}^{\mathrm{jet}}}{A_{\mathrm{jet}, i}} \label{eq:rho_jet}.
\end{equation}
In a given event, the median \pT-density \rhoevtmed{} for all soft-term $k_{t}$ jets in the event
($N_{\mathrm{jets}}$) found within a given range $-\etamax < \etajet < \etamax$ can be calculated as
\begin{equation}
\rhoevtmed = \mathrm{median}\left\{\rhojeti\right\}\mathrm{\ for\ } i
= 1\ldots N_{\mathrm{jets}}\mathrm{\ in\ } |\etajet| < \eta_{\mathrm{max}}\,.
\label{eq:rhomedian}
\end{equation}
This median \pT-density \rhoevtmed{} gives a good estimate of the
in-time pileup activity in each detector region. If determined with
\etamax~$=$~2, it is found to also be an appropriate indicator of out-of-time 
pileup contributions~\cite{CacciariJetArea}. A lower value for \rhoevtmed{} is
computed by using jets with $|\etajet|$ larger than 2, which is mostly due to the particular geometry of the ATLAS
calorimeters and their cluster reconstruction algorithms.\footnote{The
  forward ATLAS calorimeters are less granular than those in the
  central region, which leads to fewer clusters being reconstructed.}

In order to extrapolate \rhoevtmed~into the forward regions of the
detector, the average topocluster \pt~in slices of $\eta$, \Npv{}, and \muavno{} is
converted to an average \pt~density \rhoavg{} for the
soft term. As described for
the \rhoevtmed{}, \rhoavg{} is found to be uniform in the central
region of the detector with $|\eta|$~$<$~\etaplateau{}~$=$~1.8. The transverse
momentum density profile is then computed as 
\begin{equation}
P^\rho(\eta,N_{\text{PV}},\muavno)  = \frac{\rhoavg}{\rhocenavg}
\label{eq:avg_rho}
\end{equation}
\noindent{}where \rhocenavg~is the average \rhoavg{} for
$|\eta|$~$<$~\etaplateau{}. The $P^\rho(\eta,N_{\text{PV}},$\muavno{}$)$ is 
therefore 1, by definition, for $|\eta|$~$<$~\etaplateau{} and decreases for
larger $|\eta|$.

A functional form of $P^\rho(\eta,N_{\text{PV}},$\muavno$)$ is used to
parameterize its dependence on $\eta$, \Npv{}, and \muavno{} and is defined
as
\begin{equation}
P_{\textrm{fct}}^\rho(\eta,N_{\text{PV}},\muavno) = 
\left\{
\begin{array}{ll} 
1 & \left(|\eta|~<~\etaplateau \right) \\
\left(1 - G_{\textrm{base}}(\etaplateau) \right) \cdot G_{\textrm{core}}(|\eta|-\etaplateau) + G_{\textrm{base}}(\eta) & \left(|\eta|~\geq~\etaplateau \right)
\end{array} \right.
\label{eq:ejaf_pu_density}
\end{equation}
\noindent{}where the central region $|\eta|$~$<$~\etaplateau{}~$=$~1.8
is plateaued at 1, and then a pair of
Gaussian functions $G_{\textrm{core}}(|\eta|-\etaplateau)$ and $ G_{\textrm{base}}(\eta)$ are 
added for the fit in the forward regions of the calorimeter. The value
of $G_{\textrm{core}}(0)~=~1$ so that
Eq.~(\ref{eq:ejaf_pu_density}) is continuous at $\eta~=~\etaplateau$.
Two example
fits are shown in Figure~\ref{fig:rho_jets} for \Npv{}~$=$~3 and 8
with \muavno~$=$~7.5--9.5 interactions per bunch crossing. For both
distributions the value is defined to be unity in the central region ($|\eta|$~$<$~\etaplateau), and the sum of two
Gaussian functions provides a good description of the change in the
amount of in-time pileup beyond \etaplateau{}.
The baseline Gaussian function $G_{\textrm{base}}(\eta)$ has a larger width and is
used to describe the larger amount of in-time 
pileup in the forward region as seen in
Figure~\ref{fig:rho_jets}. Fitting with Eq.~(\ref{eq:ejaf_pu_density}) provides a parameterized function for in-time and out-of-time pileup which is valid for
the whole 2012 dataset.
\begin{figure}
\centering
  \subfigure[]{\includegraphics[height=50mm]{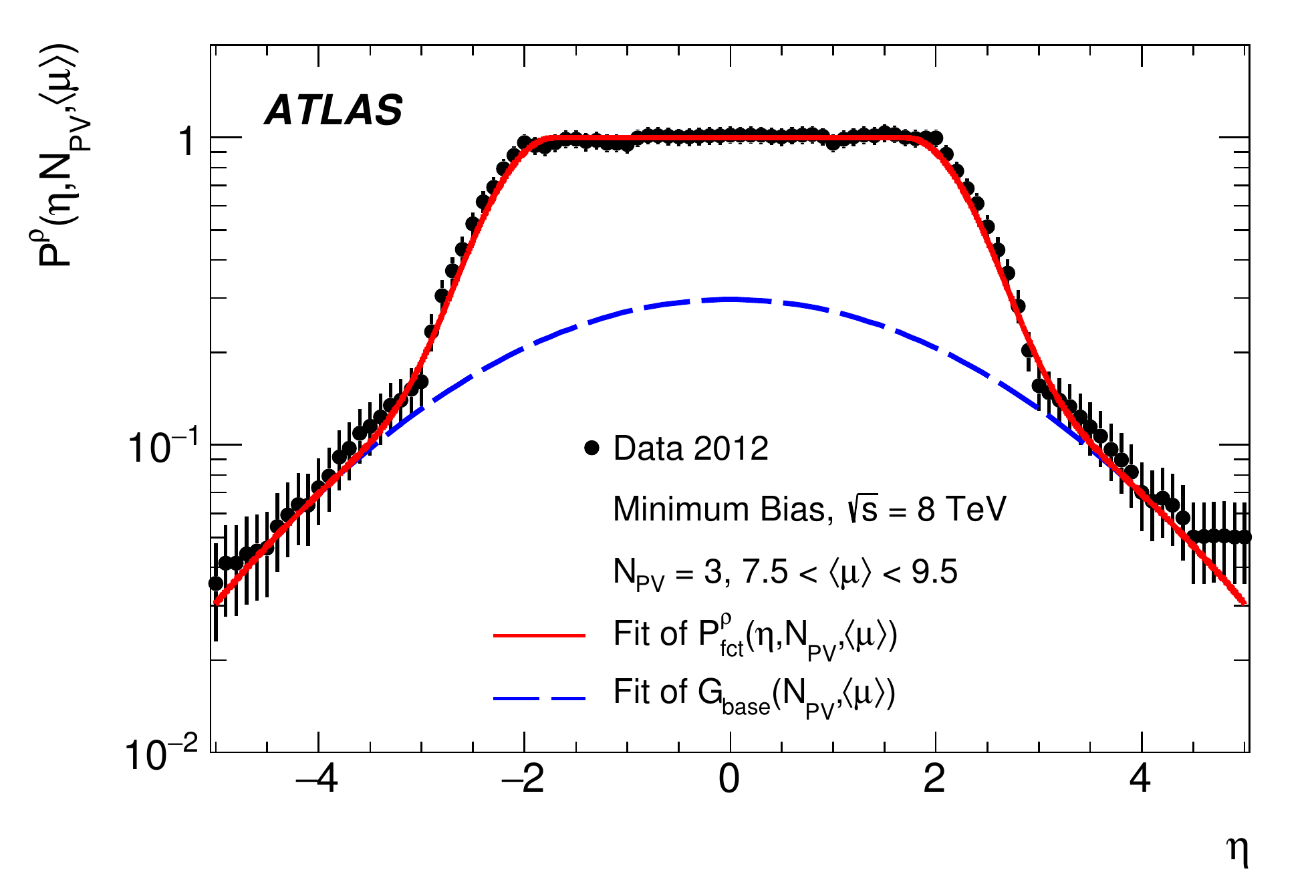}}
  \subfigure[]{\includegraphics[height=50mm]{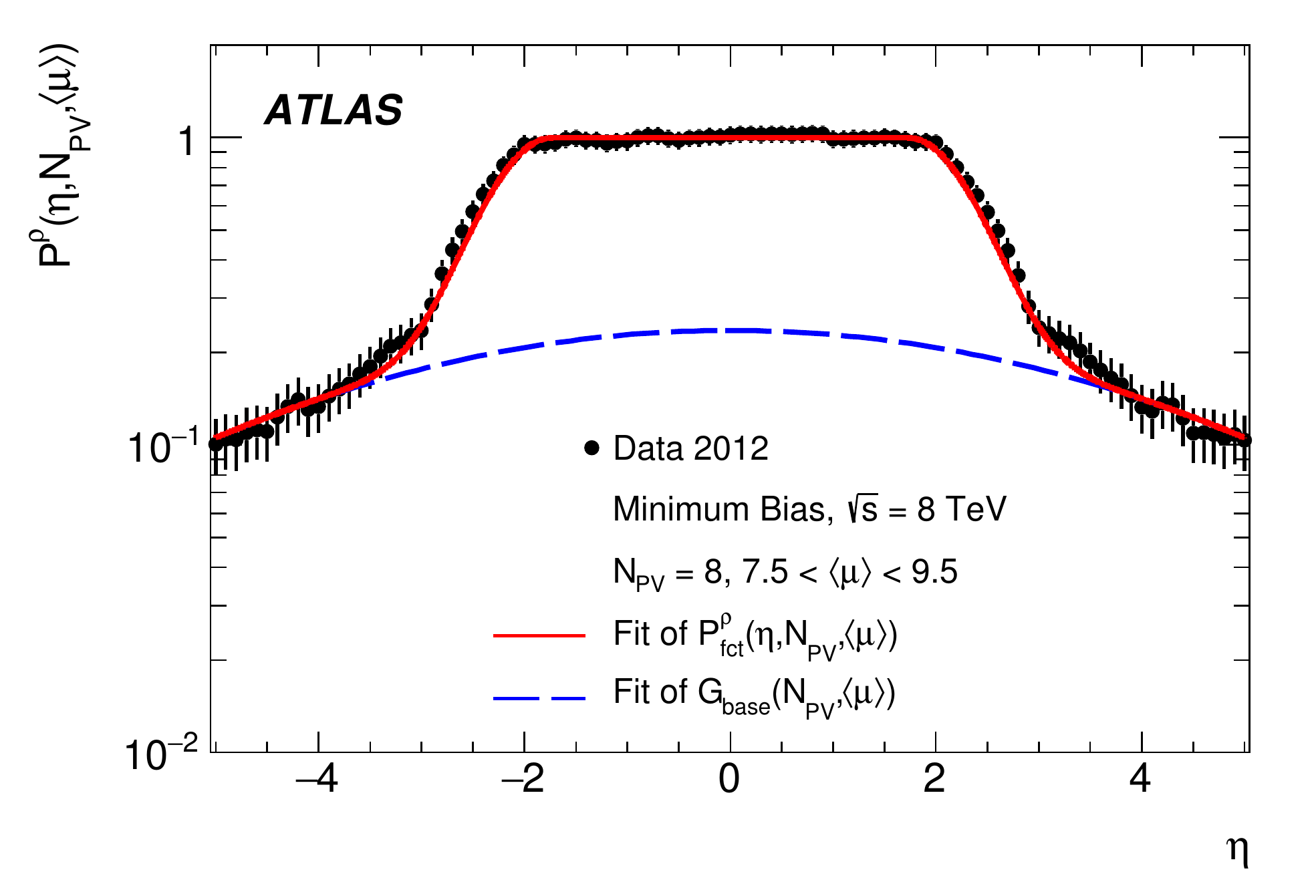}}
\caption{The average transverse momentum density shape
  $P^\rho(\eta,N_{\text{PV}},$\muavno$)$ for jets in data is compared to the
  model in Eq.~(\ref{eq:ejaf_pu_density}) with \muavno~$=$~7.5--9.5 and with (a) three
  reconstructed vertices and (b) eight reconstructed vertices. The
  increase of jet activity in the forward regions coming from more
  in-time pileup with \Npv~$=$~8 in (b) can be seen by the flatter shape of the
  Gaussian fit of the forward activity $G_{\mathrm{base}}($\Npv$,$\muavno$)$ (blue dashed line).
}
\label{fig:rho_jets}
\end{figure}

The soft term for the EJAF \met~algorithm is calculated as
\begin{equation}
\metsoftxy = - \sum_{i=0}^{\Nfilterjet} \pxyjetcorri, \label{eq:ejaf_met}
\end{equation}
which sums the transverse momenta, labelled \pxyjetcorri{}, of the
corrected soft-term jets matched to the
primary vertex. The number of these filtered jets, which are selected after the
pileup correction based on their JVF and \pT{}, is labelled
\Nfilterjet{}. More details of the jet selection and the
application of the pileup correction to the jets are given in Appendix~\ref{sec:ejaf}.

\item Soft-Term Vertex-Fraction (STVF)$\newline$
\label{sec:stvf}
\noindent The algorithm, called the soft-term vertex-fraction, utilizes an event-level
parameter computed from the ID track
information, which can be reliably matched to the hard-scatter collision, to 
suppress pileup effects in the CST. 
This correction is applied as a multiplicative factor (\stvf{}) to
the CST, event by event, and the resulting STVF-corrected CST is simply
referred to as STVF. The \stvf{} is calculated as
\begin{equation}
\stvf ={\sum_{\mathrm {tracks,PV}} p_{\mathrm{T}}
}\mathlarger{\mathlarger{\mathlarger{/}}} \sum_{\mathrm {tracks}} p_{\mathrm{T}},
\label{eqpileup} 
\end{equation}
\noindent{}which is the scalar
sum of \pT~of tracks matched to the PV divided by the total
scalar sum of track \pT~in the event, including pileup. The sums are
taken over the tracks that do not match
high-\pt{} physics objects belonging to the hard term. 
The mean \stvf~value is shown versus the number of reconstructed
vertices (\Npv{}) in Figure~\ref{fig:stvf_weight}. Data and simulation
(including $Z$, diboson, \ttbar{}, and $tW$ samples) are shown with
only statistical uncertainties and agree
within 4--7\% across the full range of
\Npv~in the 8~\TeV~dataset. The differences mostly arise from
the modelling of the amount of the underlying event and \ptZ{}. The 0-jet and inclusive samples
have similar values of \stvf{}, with that for the inclusive sample being around
2\% larger. 

\begin{figure*}
\begin{center}
  \includegraphics[height=55mm]{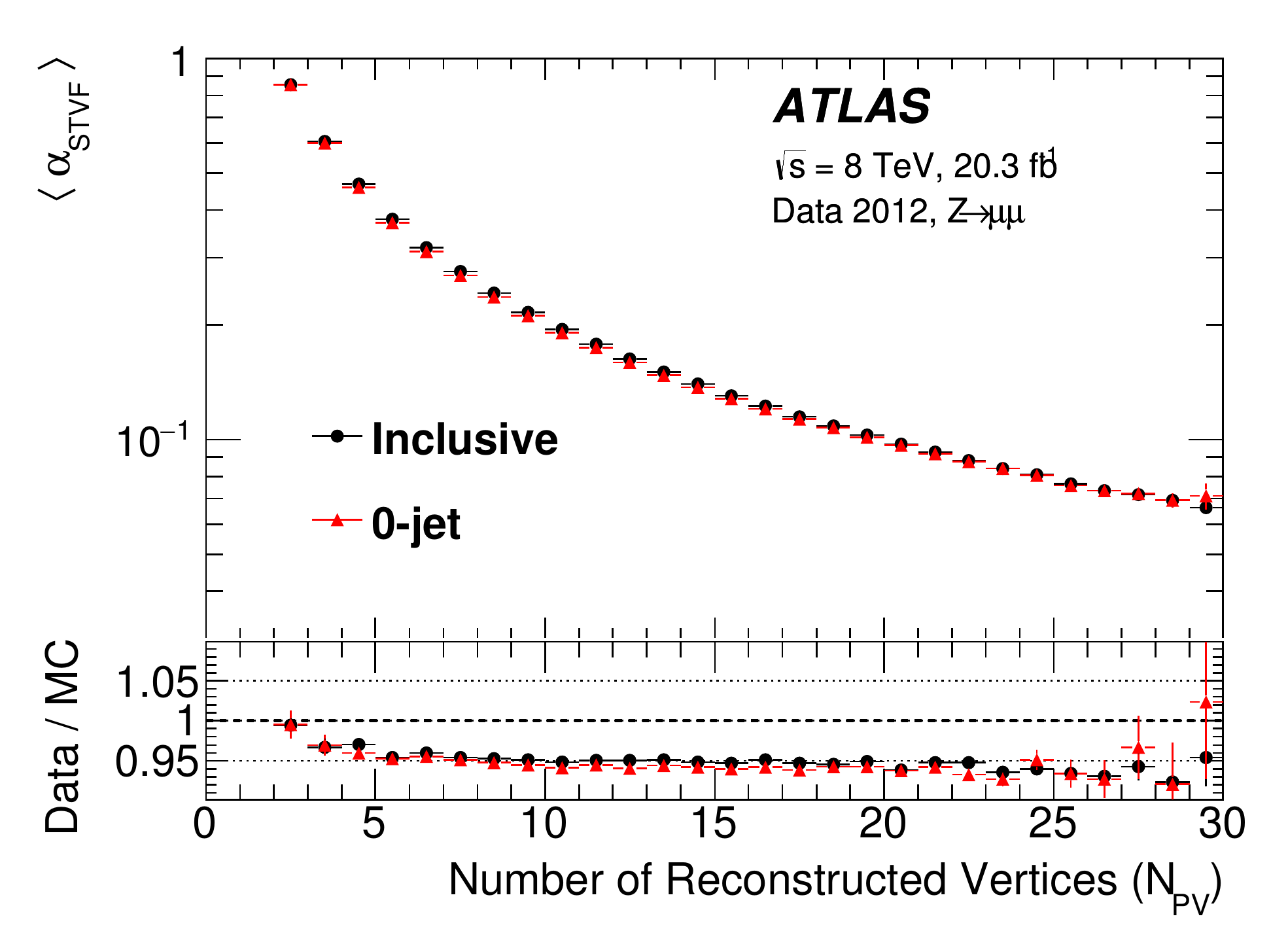}
\end{center}
\caption{The mean \stvf~weight is shown versus the number of
  reconstructed vertices (\Npv{}) for 0-jet and inclusive
  events in \Zmm~data. The inset at the bottom of the figure shows the ratio of the data
  to the MC predictions with only the statistical
  uncertainties on the data and MC simulation. The bin boundary always
  includes the lower edge and not the upper edge. 
}  
\label{fig:stvf_weight}
\end{figure*}

\end{itemize}

\subsubsection {Jet \pT~threshold and JVF selection}
\label{sec:jet_selec}
The TST, STVF, and EJAF \met~algorithms complement the pileup
reduction in the soft term with additional requirements on the jets
entering the \met~hard term, which are also aimed at reducing pileup dependence. These \met~reconstruction algorithms apply a requirement of $\text{JVF}$~$>$~0.25 
to jets with $\pT$~$<$~50~\GeV{} and $|\eta|$~$<$~2.4
in order to suppress those originating from pileup interactions. 
The maximum $|\eta|$ value is lowered to 2.4 to ensure that the core of 
each jet 
is within the tracking volume ($|\eta|$~$<$~2.5)~\cite{ATLASJetPileup2012}.
Charged particles from jets below the \pt~threshold are considered in 
the soft terms for the STVF, TST, and EJAF (see
Section~\ref{sec:soft_term} for details).

The same $\text{JVF}$ requirements are not applied to the CST
\met~because its soft term includes the soft recoil from all
interactions, so removing jets not associated with the hard-scatter
interaction could create an imbalance. The procedure for
choosing the jet \pT{} and $\text{JVF}$ criteria is summarized in
Section~\ref{sec:jet_opt}.

Throughout most of this paper 
the number of jets is computed without a $\text{JVF}$ requirement
so that the \met~algorithms are compared
on the same subset of events. 
However, the $\text{JVF}$~$>$~0.25 requirement is
applied in jet counting when 1-jet and $\geq$~2-jet samples are
studied using the TST \met{} reconstruction, which includes
Figures~\ref{fig:TST_njet_dependence} and
\ref{fig:ptsoft_fits_example}. The $\text{JVF}$ removes pileup jets
that obscure trends in samples with different jet multiplicities.

\par\noindent
\subsection {Track \met{}}
\label{sec:trkmetdef}

Extending the philosophy of the TST definition to the full
event, the \met~is reconstructed from tracks alone, reducing the pileup
contamination that afflicts the other object-based algorithms. 
While a
purely track-based \met{}, designated Track \met{}, has almost no
pileup dependence, it is insensitive to neutral particles, which do not form tracks in the ID. 
This can degrade the \met{} calibration, especially in
event topologies with numerous or highly energetic jets. The
$\eta$ coverage of the Track \met~is also limited to the ID acceptance
of $|\eta|$~$<$~2.5, which is substantially smaller than the calorimeter
coverage, which extends to $|\eta|$~$=$~$4.9$.

Track \met{} is calculated by taking the negative vectorial sum of \ptvec{} of tracks satisfying the same quality criteria as the TST tracks. Similar to the TST, tracks with poor momentum resolution or without
corresponding calorimeter deposits are removed. Because of
Bremsstrahlung within the ID, the electron \pT{} is determined more precisely
by the calorimeter than by the ID. 
Therefore, the Track \met~algorithm uses the electron \pT{}
measurement in the calorimeter and removes tracks overlapping its
shower. Calorimeter deposits from photons are not added because they cannot be reliably associated to particular $pp$ interactions.
For muons, the ID track
\pT{} is used and not the fits combining the ID and MS \pT{}. 
For events without any reconstructed jets, the Track and 
TST \met~would have similar values, but differences could still originate
from muon track measurements as well as reconstructed photons or
calorimeter deposits from $\tauvis$, which are only included in the TST.

The soft term for the Track \met~is defined to be identical to the TST
by excluding tracks associated with the high-\pT{} physics objects used
in Eq.~(\ref{eq7}).

\section{Comparison of \met~distributions in data and MC simulation}
\label{sec:Distributions}

In this section, basic \met~distributions before and after pileup
suppression in \Zll~and \Wln~data
events are compared to the distributions from the MC signal
plus relevant background samples. 
All distributions in this section include the dominant 
systematic uncertainties on the high-\pt{} objects, 
the \metsoftvec{} (described in Section~\ref{sec:systematics})
and pileup modelling~\cite{ATLAS-CONF-2012-042}. 
The systematics listed above are the largest systematic uncertainties in the \metmag~for $Z$ and $W$ samples. 

\subsection{Modelling of \Zll~events}
\label{sec:Perf_Z}

The CST, EJAF, TST, STVF,
and Track \metmag{}~distributions for \Zmm{} data and simulation 
are shown in Figure~\ref{fig:METZ_mag}. The $Z$ boson signal
region, which is defined in Section~\ref{sec:evtselZll}, 
has better than 99\% signal purity. 
The MC simulation agrees with data for all
\metmag~reconstruction algorithms within the assigned systematic
uncertainties. The mean and the
standard deviation of the \met~distribution is shown for all of 
the \met~algorithms in \Zmm~inclusive
simulation in Table~\ref{tab:met_mean}. The CST \met~has the
highest mean \met{} and thus the 
broadest \metmag~distribution.
All of the \metmag~algorithms with pileup suppression  
have narrower \metmag~distributions as shown by their smaller
mean \met{} values.
However, those algorithms also have non-Gaussian tails in the 
\metx{} and \mety{} distributions, which contribute to the region with
\metmag~$\gtrsim$~50 \GeV{}. The Track
\metmag~has the largest tail because it does not include contributions 
from the neutral particles, and this results in it having the 
largest standard deviation. 

\begin{table}[h]
\begin{center}
\caption{\label{tab:met_mean} The mean and standard deviation of the \met~distributions
  in \Zmm~inclusive simulation.}
\begin{tabular}{l||l} 
\met~Alg. & Mean $\pm$ Std. Dev. [GeV]\\ \hline \hline
CST \met{} & 20.4~$\pm$~12.5 \\
EJAF \met{} & 16.8~$\pm$~11.5 \\
TST \met{} & 13.2~$\pm$~10.3 \\
STVF \met{} & 13.8~$\pm$~10.8 \\
Track \met{} & 13.9~$\pm$~14.4 \\
\end{tabular}
\end{center}
\end{table}

The tails of the \metmag~distributions in Figure~\ref{fig:METZ_mag} for \Zmm~data are observed to be
compatible with the sum of expected signal and background contributions, namely
\ttbar~and the summed diboson ($VV$) processes including $WW$, $WZ$, and $ZZ$, which all have high-\pT~neutrinos in
their final states. 
Instrumental effects can show up in the tails of the
\metmag{}, but such effects are small.

The \met{} $\phi$ distribution is not shown in this paper but is very uniform,
having less than 4 parts in a thousand difference from positive and
negative $\phi$. Thus the $\phi$-asymmetry is greatly reduced from that observed in Ref.~\cite{ATLASMETPaper2011}.

The increase in systematic
uncertainties in the range 50--120~\GeV{} in Figure~\ref{fig:METZ_mag}
comes from the tail of the \met~distribution for the simulated
\Zmm~events. The increased width in the
uncertainty band is asymmetric because many systematic uncertainties
increase the \met~tail in \Zmm~events by creating an imbalance in the transverse
momentum. The largest of these
systematic uncertainties are those associated with the jet energy resolution, 
the jet energy scale, and pileup.
The pileup systematic
uncertainties affect mostly the CST and EJAF \met, while the jet energy scale uncertainty causes the
larger systematic uncertainty for the TST and STVF \met{}. The Track \met~does
not have the same increase in systematic uncertainties 
because it does not make use of reconstructed jets. Above 120~\GeV{}, most
events have a large intrinsic \met{}, and the systematic uncertainties
on the \met{}, especially the soft term, are smaller.

\begin{figure*}[htbp]
\begin{center}
\subfigure[]{\includegraphics[width=0.39\textwidth]{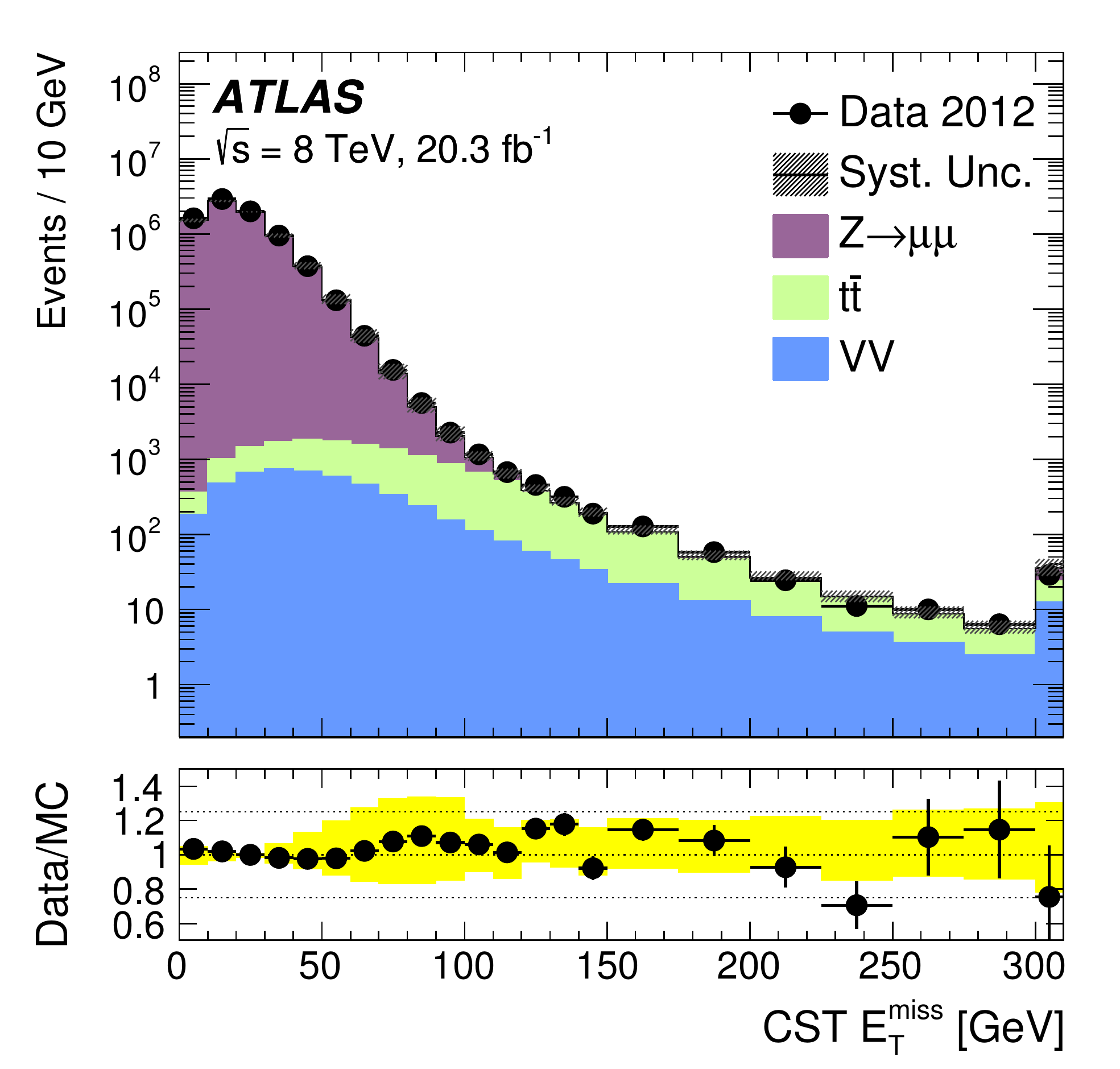}}
\subfigure[]{\includegraphics[width=0.39\textwidth]{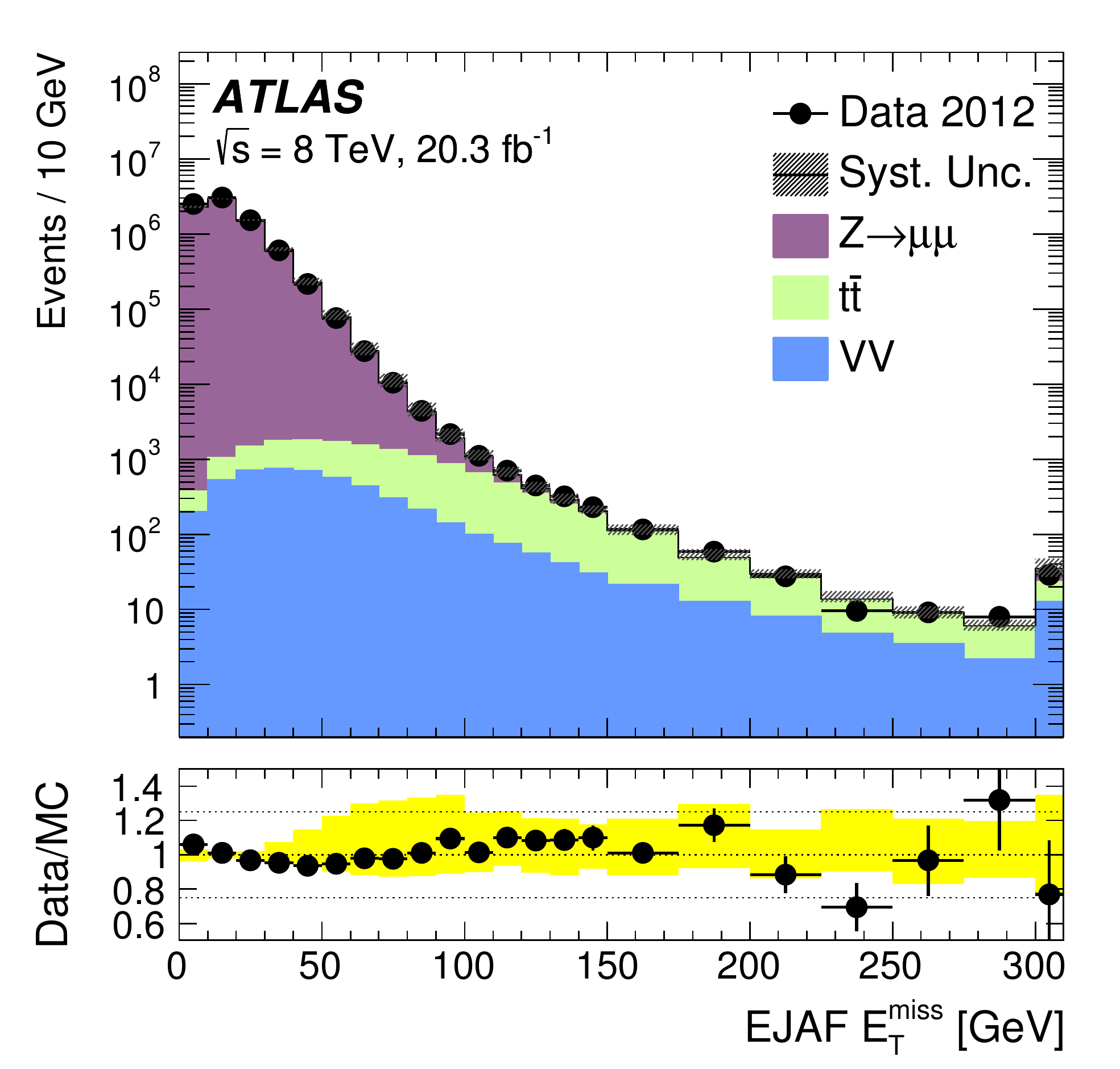}}\\
\subfigure[]{\includegraphics[width=0.39\textwidth]{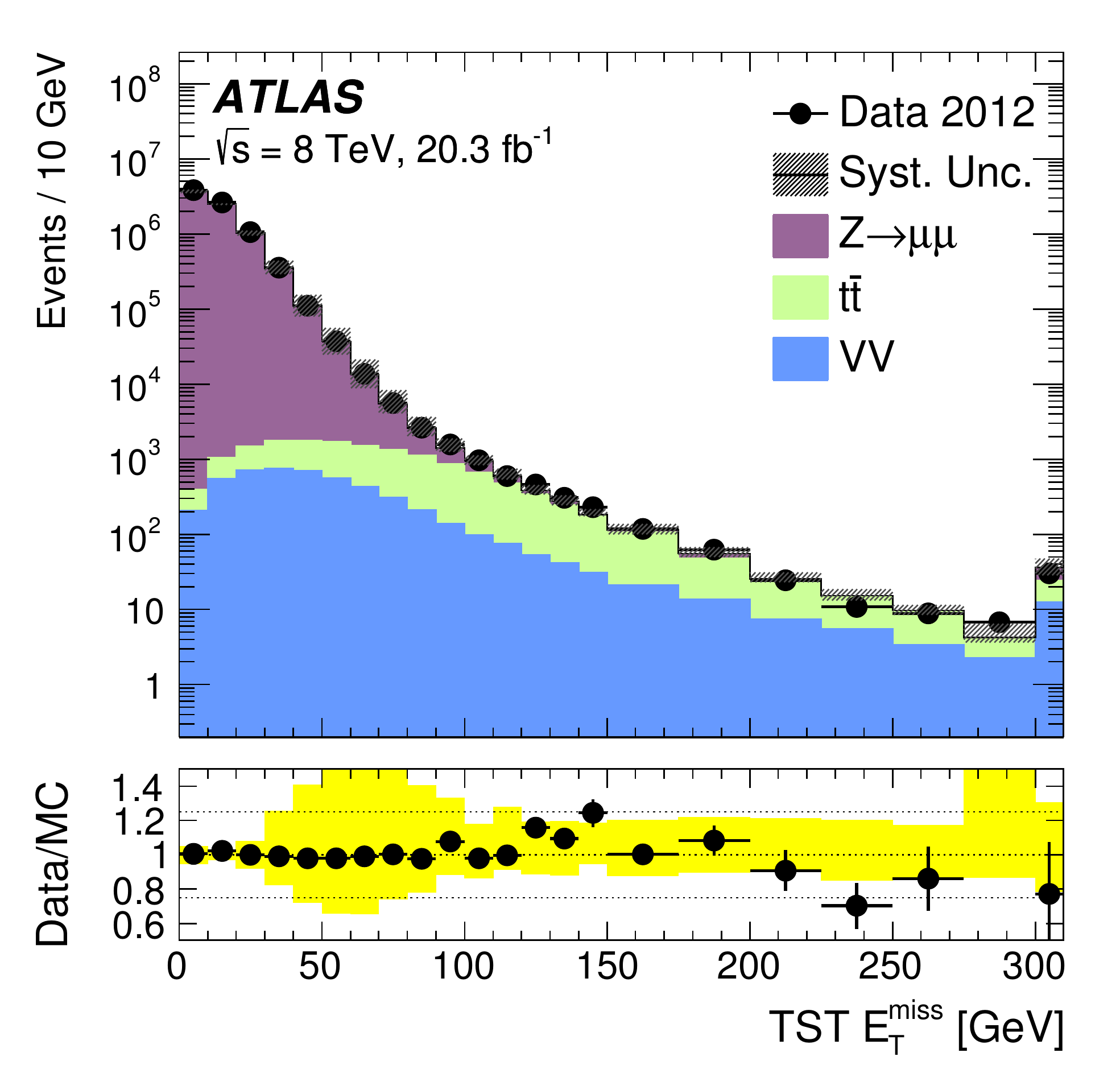}}
\subfigure[]{\includegraphics[width=0.39\textwidth]{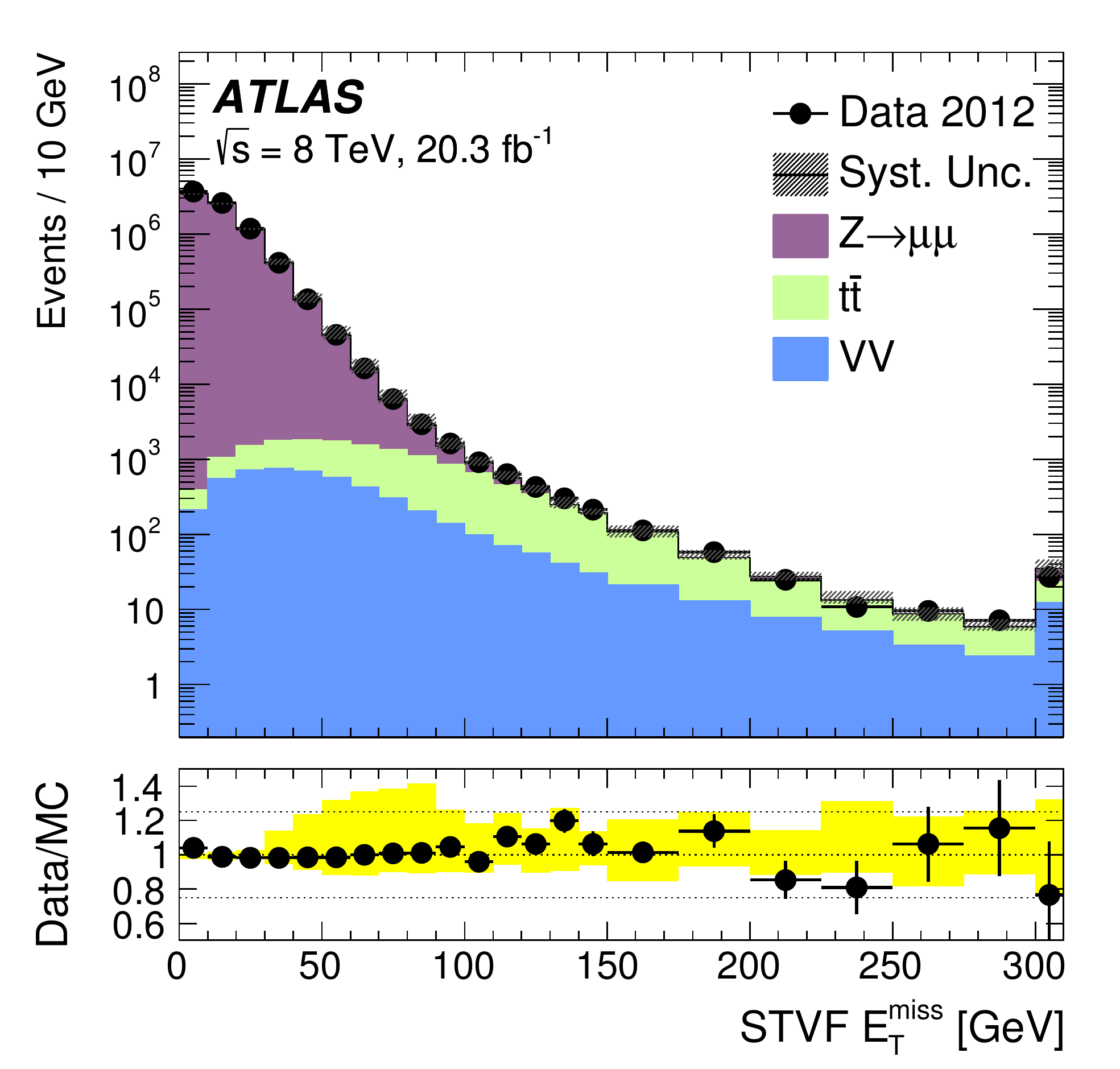}}\\
\subfigure[]{\includegraphics[width=0.39\textwidth]{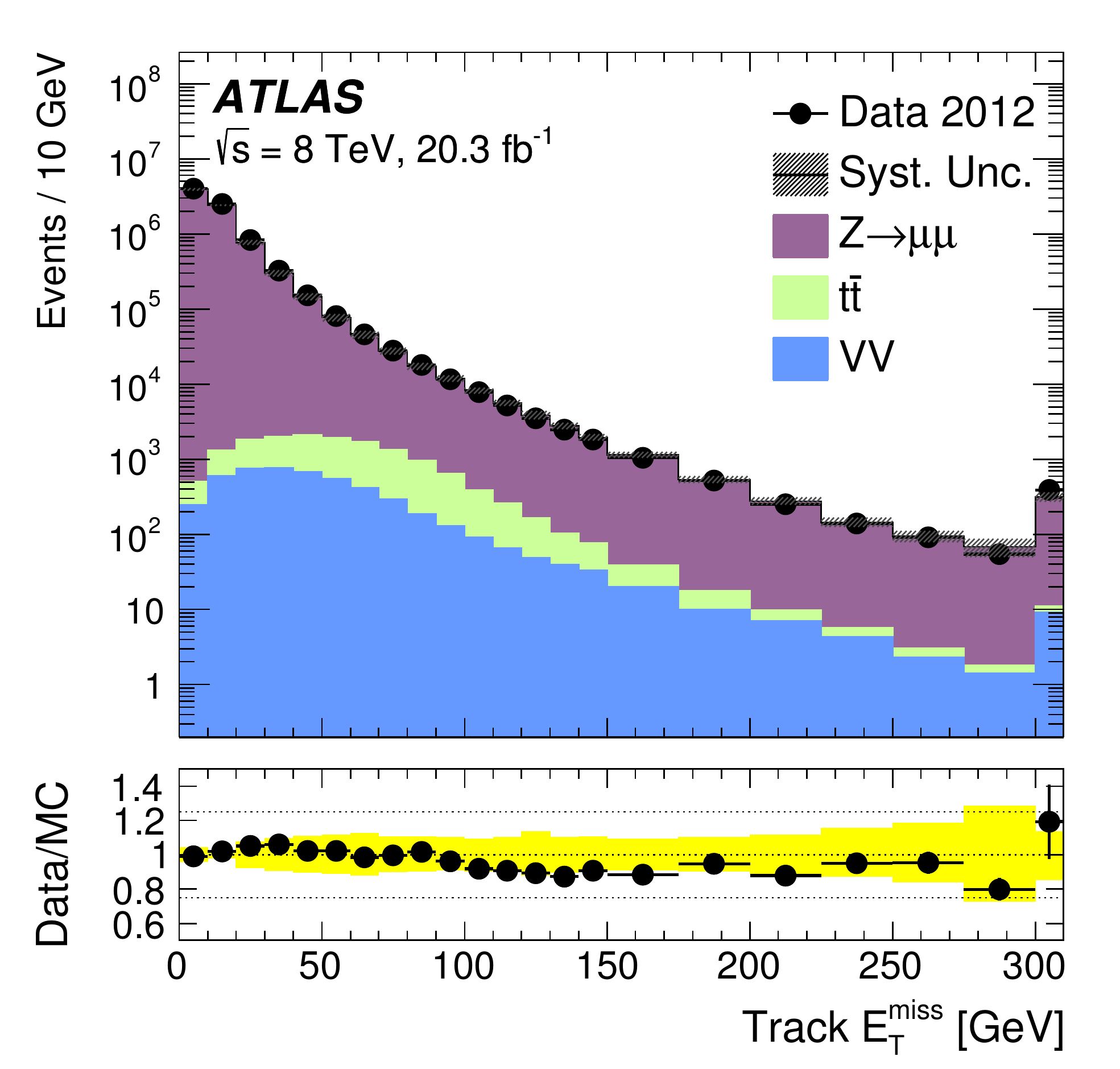}} 
\end{center}
\caption{Distributions of the \metmag~with the (a) CST,
  (b) EJAF, (c) TST, (d) STVF, and (e) Track \met{} are shown in data and MC simulation events satisfying the \Zmm{} selection.
The lower panel of the figures shows the ratio of data to MC
simulation, and the bands correspond to the combined systematic and MC
statistical uncertainties. The far right bin includes the integral of
all events with \met{} above 300~\GeV{}.
}
\label{fig:METZ_mag}
\end{figure*}

Figure~\ref{fig:METZ_soft_mag} shows the soft-term distributions. The pileup-suppressed \met~algorithms
generally have a smaller mean soft term as well as 
a sharper peak near zero compared to the CST. 
Among the \met~algorithms, the soft term from the EJAF algorithm shows the smallest
change relative to the CST.
The TST has a sharp peak near zero similar to the STVF but with a
longer tail, which mostly comes from individual
tracks. These tracks are possibly mismeasured and further studies are
planned. The simulation under-predicts the TST relative to the
observed data between 60--85 \GeV{}, and the differences exceed the
assigned systematic uncertainties. This region corresponds to the
transition from the narrow core to the tail coming from
high-\pt{} tracks. The differences between data and simulation could be due to mismodelling of the rate of mismeasured tracks, for which no systematic uncertainty is applied. The mismeasured-track cleaning, as discussed in Section~\ref{sec:tst}, reduces the TST tail starting at 120 \GeV{}, and this region is modelled within the assigned uncertainties. The mismeasured-track cleaning for tracks below 120 \GeV{} and entering the TST is not optimal, and future studies aim to improve this. 

\begin{figure*}[!htbp]
\begin{center}
\subfigure[]{\includegraphics[width=0.39\textwidth]{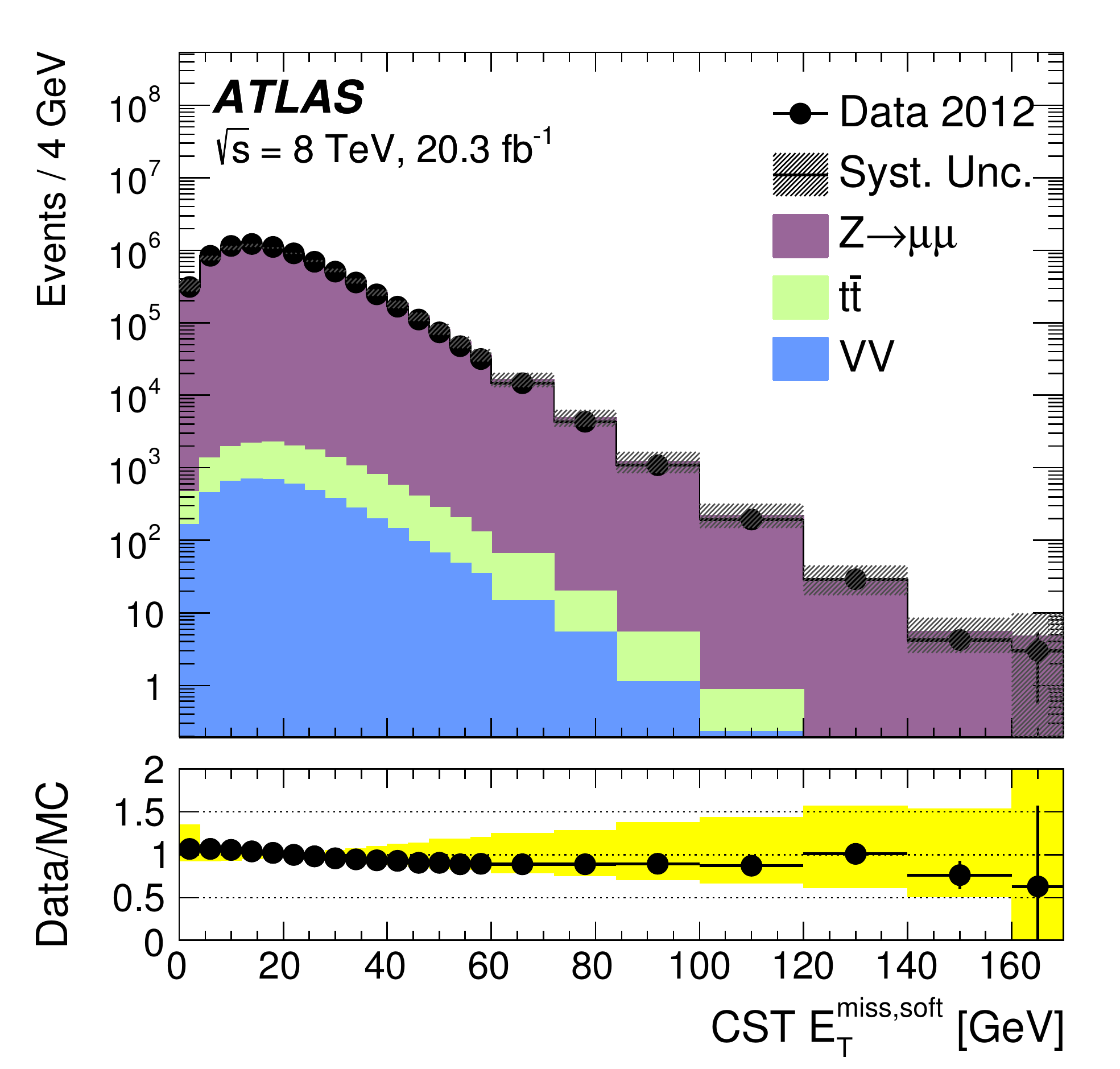}}
\subfigure[]{\includegraphics[width=0.39\textwidth]{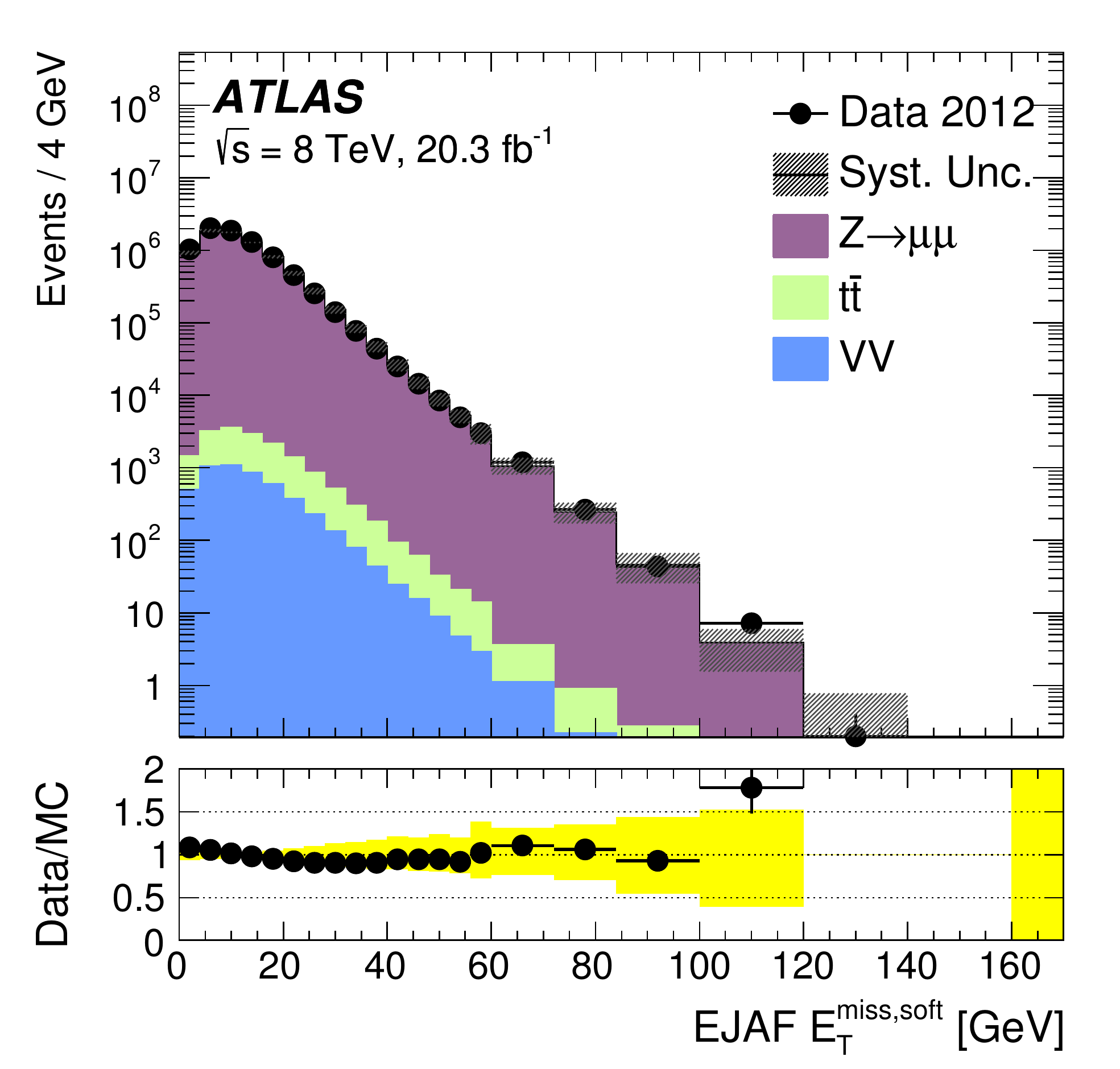}}\\
\subfigure[]{\includegraphics[width=0.39\textwidth]{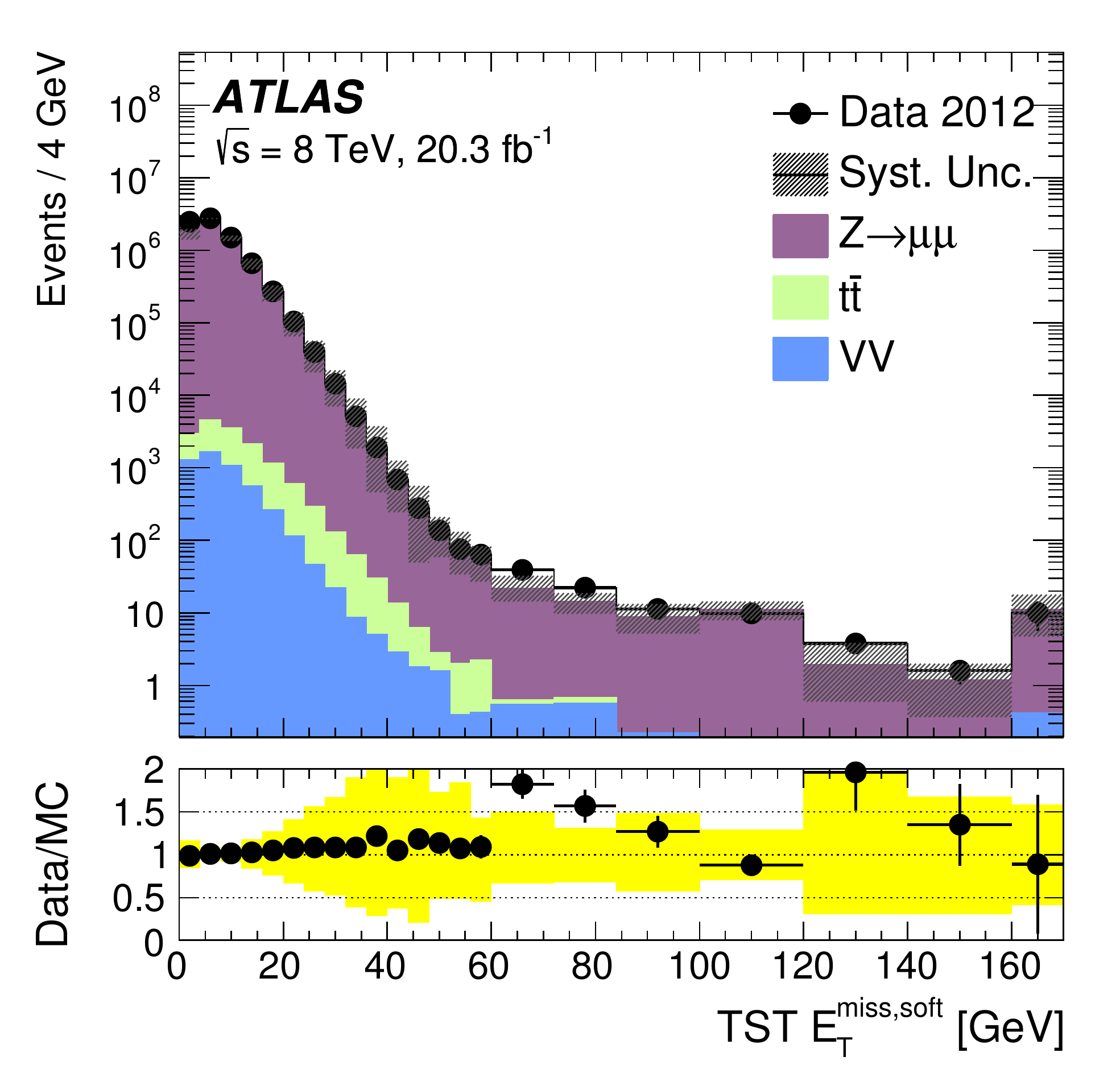}}
\subfigure[]{\includegraphics[width=0.39\textwidth]{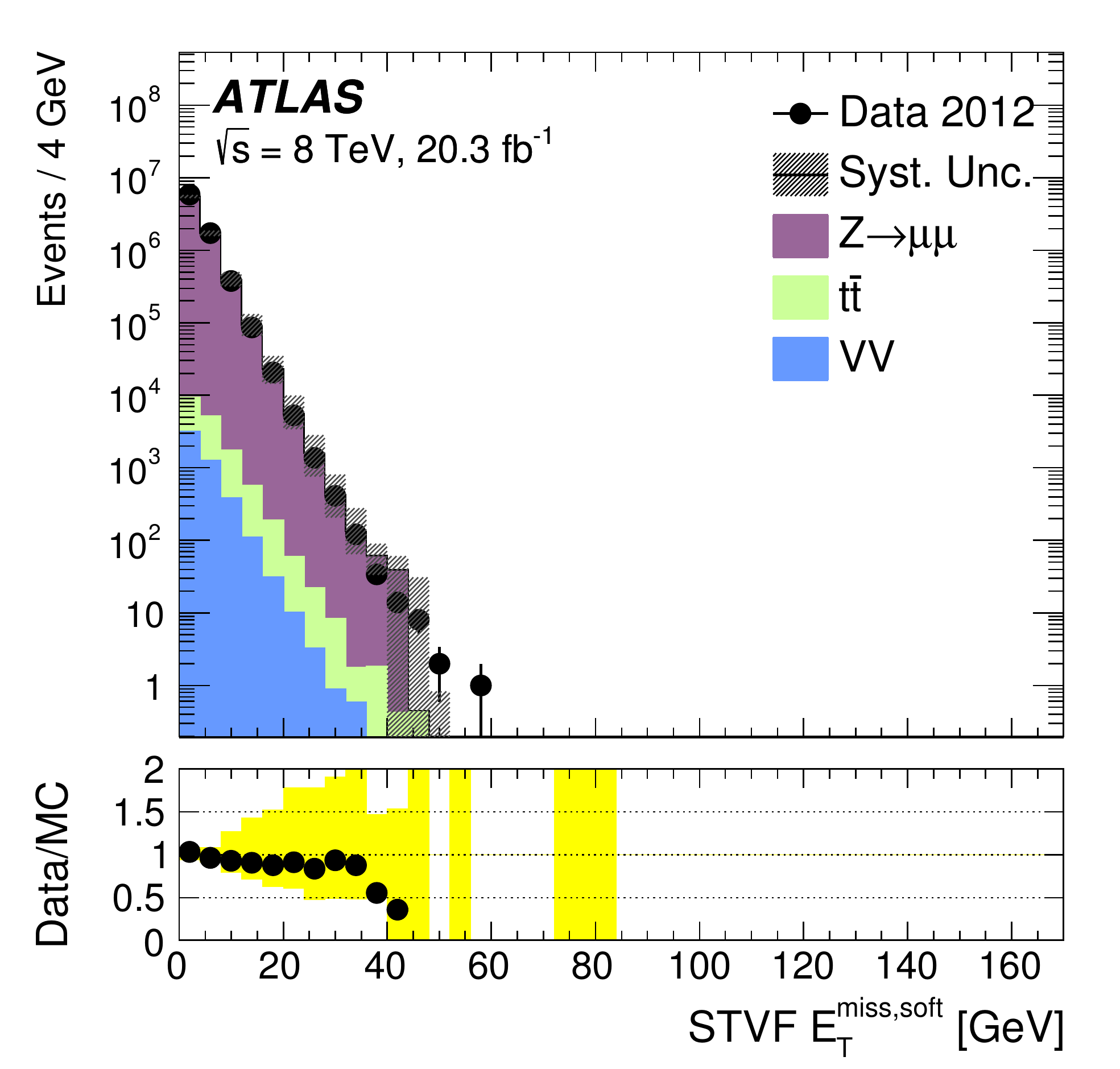}}\\
\end{center}
\caption{Distributions of the soft term for the (a) CST, (b) EJAF,
  (c) TST, and (d) STVF are shown in data and MC simulation events satisfying the \Zmm{} selection.
The lower panel of the figures show the ratio of data to MC
simulation, and the bands correspond to the combined systematic and MC
statistical uncertainties. The far right bin includes the integral of
all events with \metsoft{} above 160~\GeV{}.
}
\label{fig:METZ_soft_mag}
\end{figure*}

The \met{} resolution is expected to be proportional to $\sqrt{\sumetevt}$
when both quantities are measured with the calorimeter alone~\cite{ATLASMETPaper2011}. While this proportionality does not hold for tracks, it is
nevertheless interesting to understand the modelling of \sumetevt{} 
and the dependence of \met{} resolution on it. 
Figure~\ref{fig:METZ_sumet} shows the \sumetevt{}
distribution for \Zmm{} data and
MC simulation both for the TST and the CST algorithms. The \sumetevt~is typically larger for the CST algorithm
than for the TST because the former includes energy deposits from
pileup as well as neutral particles 
and forward contributions beyond the ID volume.
The reduction of pileup contributions in the soft and jet terms leads
to the \sumetevt(TST)~having a sharper peak at around 100~\GeV{} followed by
a large tail, due to high-\pT~muons and 
large $\sum p_{\mathrm{T}}^{\mathrm{jets}}$. 
The data and simulation agree within the uncertainties for the 
\sumetevt(CST) and \sumetevt(TST) distributions.

\begin{figure*}[!htbp]
\begin{center}
\subfigure[]{\includegraphics[width=0.39\textwidth]{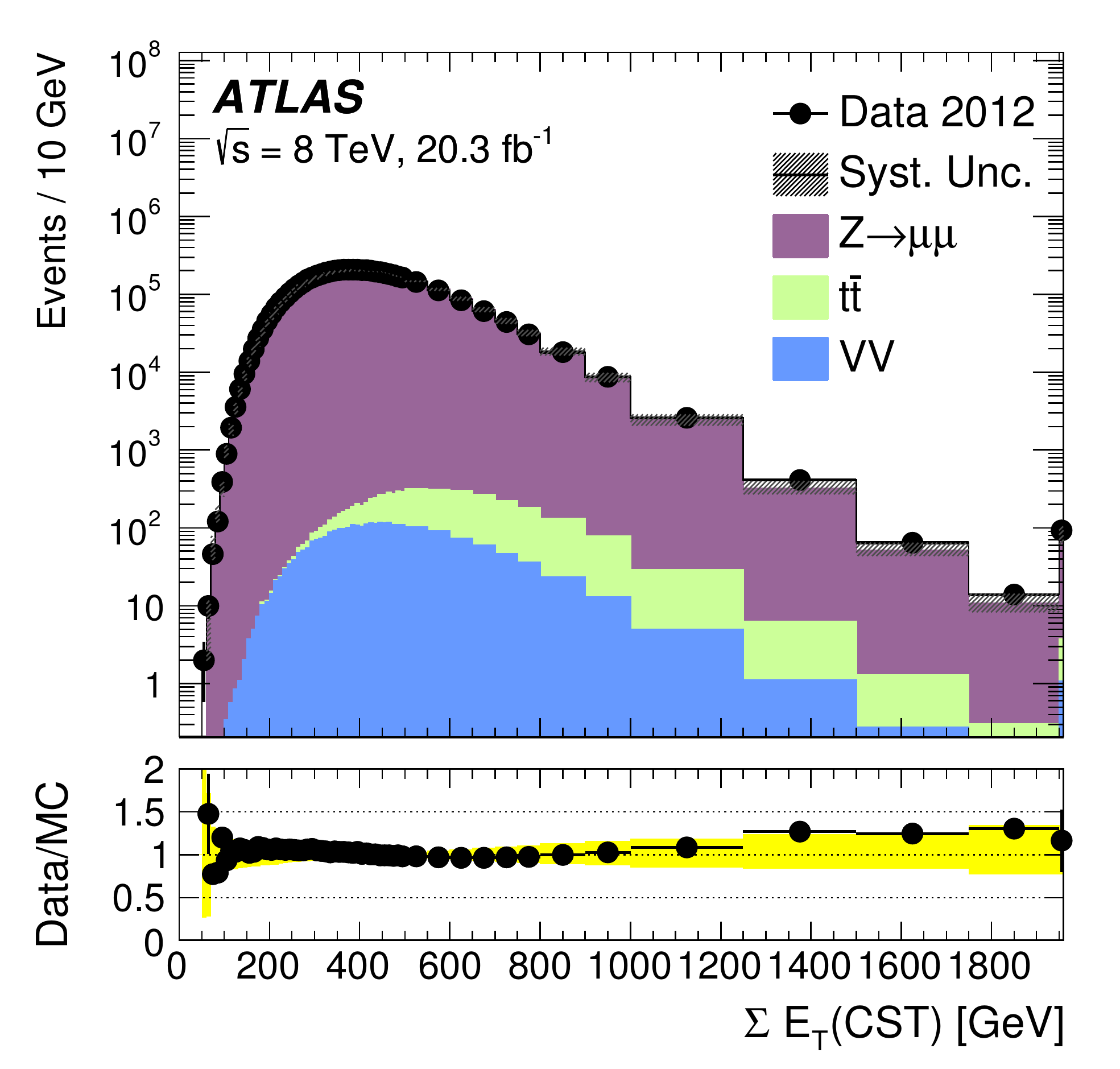}}
\subfigure[]{\includegraphics[width=0.39\textwidth]{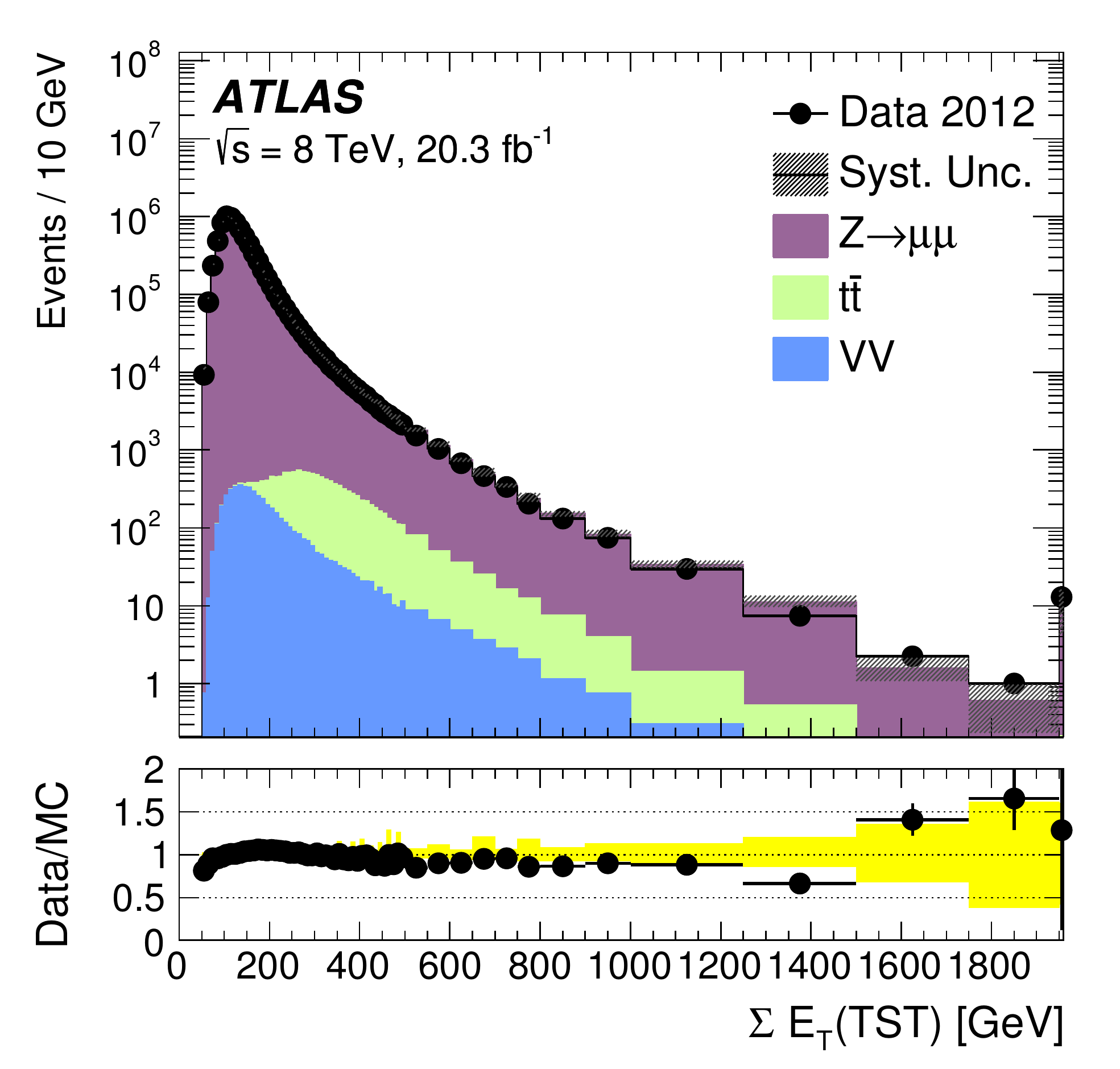}}
\end{center}
\caption{Distributions of (a) \sumetevt(CST) and
  (b) \sumetevt(TST) are shown in data and MC simulation events satisfying the \Zmm{} selection.
The lower panel of the figures show the ratio of data to MC simulation,
and the bands correspond to the combined systematic and MC
statistical uncertainties. The far right bin includes the integral of
all events with \sumetevt{} above 2000~\GeV{}.
}
\label{fig:METZ_sumet}
\end{figure*}

\subsection{Modelling of \Wln~events}
\label{sec:Perf_W}

In this section, the selection
requirements for the \mT~and \met~distributions are defined using the
same \met~algorithm as that labelling the distribution (e.g. selection
criteria are applied to the CST
\met~for distributions showing the CST \met{}). The intrinsic \met~in
\Wln~events allows a comparison of the \metmag{} scale  between data and simulation. 
The level of agreement between data and MC simulation for the
\met~reconstruction algorithms is studied using \Wen~events with the selection defined in
Section~\ref{sec:evtselWlv}. 

The CST and TST \met~distributions in \Wen~events are shown in
Figure~\ref{fig:METW_basic}. The \Wtaunu~contributions are combined
with \Wen~events in the figure. The data and MC simulation
agree within the assigned systematic uncertainties for both the CST
and TST \met~algorithms. The other \met{} algorithms show similar
levels of agreement between data and MC simulation.

\begin{figure*}[htbp]
\begin{center}
\subfigure[]{\includegraphics[width=0.39\textwidth]{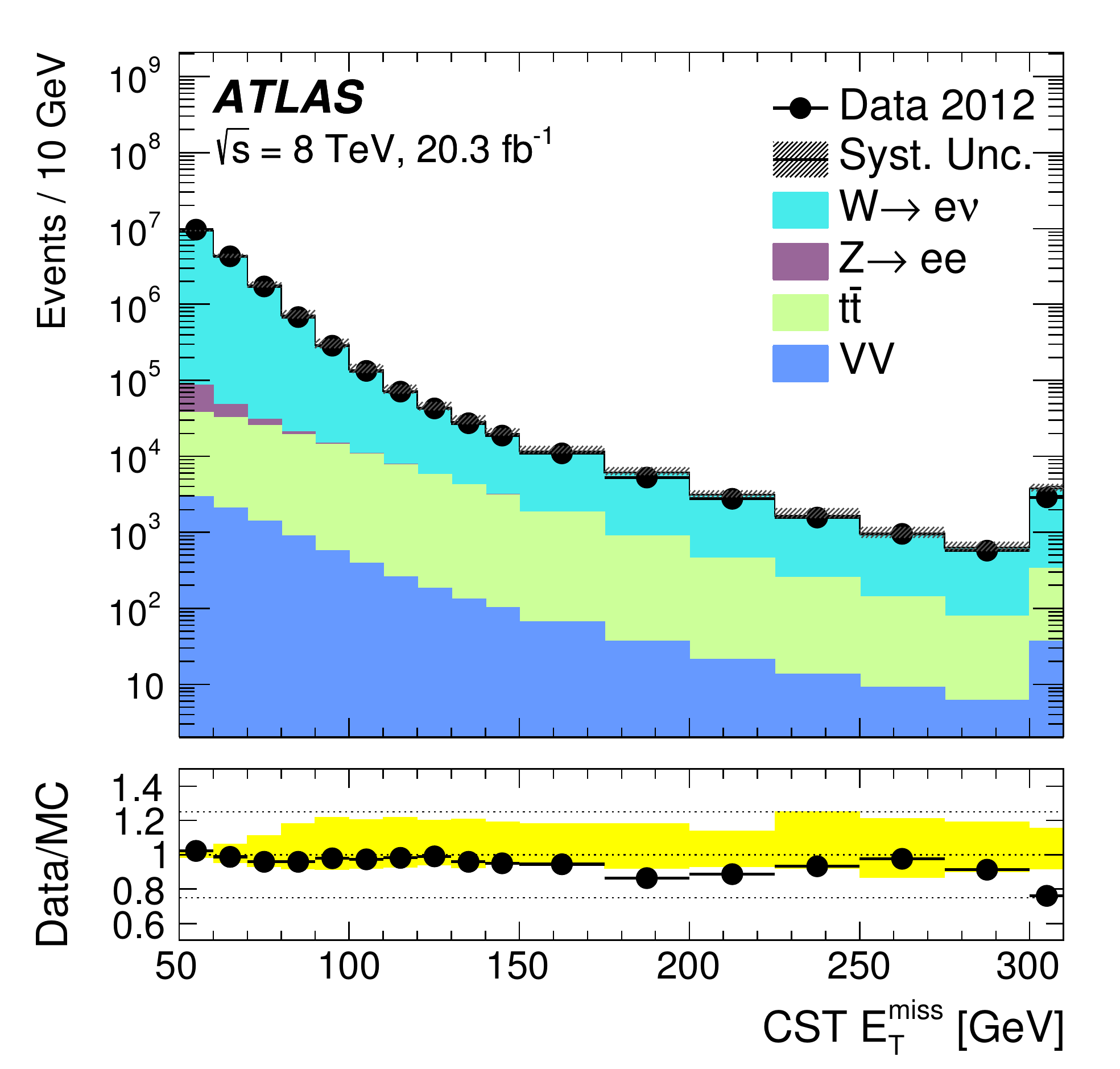}}
\subfigure[]{\includegraphics[width=0.39\textwidth]{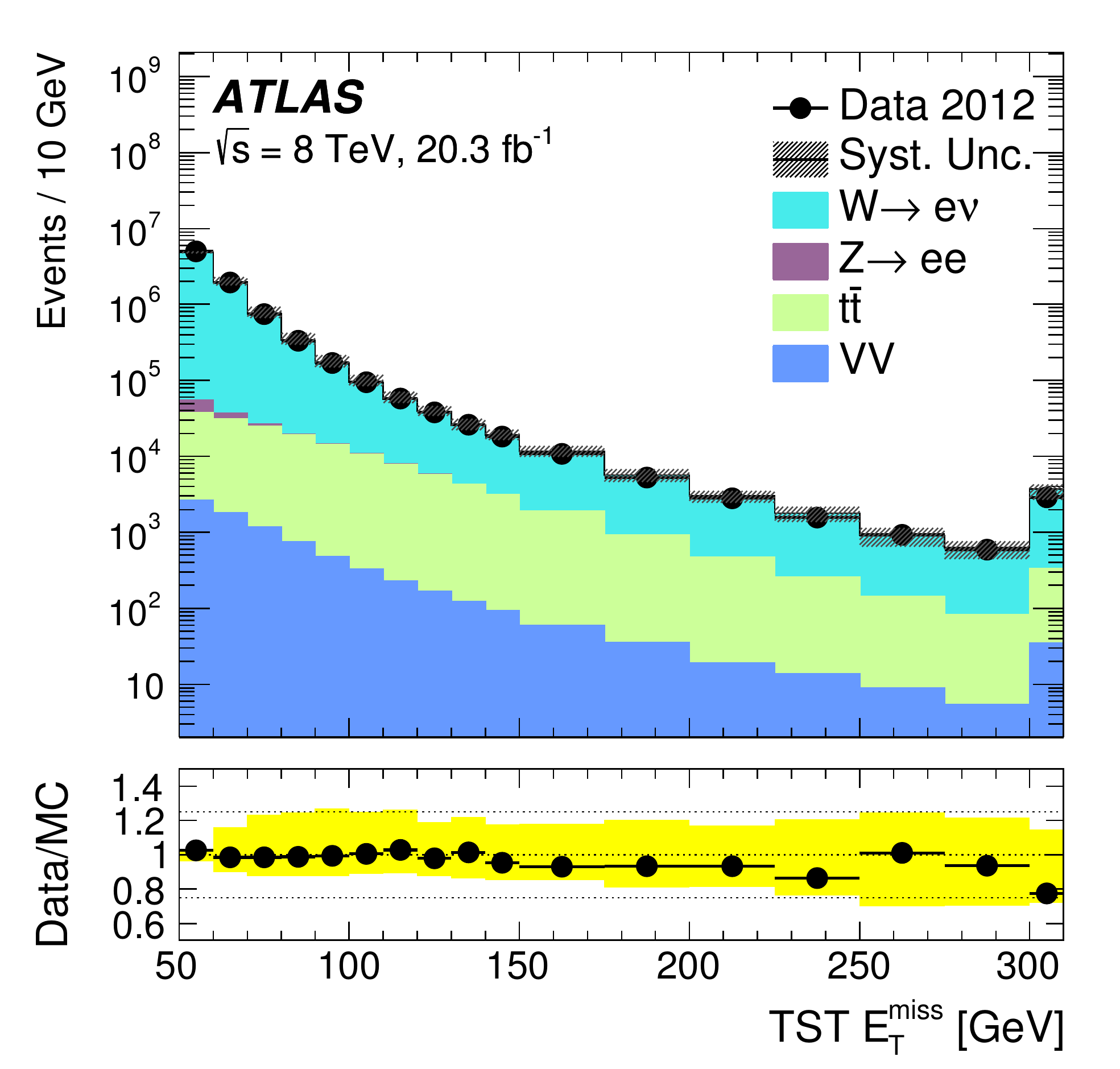}}
\end{center}
\caption{Distributions of the (a) CST and (b) TST \met~as measured in a data sample of \Wen~events.
The lower panel of the figures show the ratio of data to MC simulation,
and the bands correspond to the combined systematic and MC
statistical uncertainties. The far right bin includes the integral of
all events with \met{} above 300~\GeV{}.
}
\label{fig:METW_basic}
\end{figure*}

\FloatBarrier
\section{Performance of the \met~in data and MC simulation}
\label{sec:Performance}

\subsection{Resolution of \met}
\label{resol}

The \metx{} and \mety{} are expected to be approximately 
Gaussian distributed for \Zll~events as discussed in
Ref.~\cite{ATLASMETPaper2011}.
However, because of the non-Gaussian tails in these distributions, especially for the pileup-suppressing \met~algorithms, the root-mean-square (RMS) is used to estimate the resolution.
This includes important information about the tails, which would be
lost if the result of a Gaussian fit over only the core of the
distribution were used instead. The resolution of
the \met~distribution is extracted using the RMS from
the combined distribution of \metx~and \mety{}, which are determined to be independent from correlation studies. The previous ATLAS \met~performance
paper~\cite{ATLASMETPaper2011} studied the resolution defined by the
width of Gaussian fits in a narrow range of $\pm2$RMS around the mean
and used a separate study to investigate the tails. Therefore, the results of this
paper are not directly comparable to those of the previous study. The resolutions presented in this paper are expected to be 
larger than the width of the Gaussian fitted in this manner because the RMS 
takes into account the tails.

In this section, 
the resolution for the \met{} is presented for 
\Zmm{}~events using both data and MC simulation. Unless it is a
simulation-only figure (labelled with ``Simulation'' under the ATLAS label), the MC distribution includes the signal sample
(e.g. \Zmm{}) as well as diboson, \ttbar{}, and $tW$ samples. 

\subsubsection{Resolution of the \met~as a function of the number of
  reconstructed vertices}
\label{sec:resol_npv}

The stability of the \met~performance as a function of the amount of pileup is
estimated by studying the \metmag~resolution as a function of the number of
reconstructed vertices (\Npv{}) for \Zmm~events as shown in
Figure~\ref{fig:Z_resol_npv}. The bin edge is always including the
lower edge and not the upper. For example, the events with \Npv{} in the inclusive range 30--39 are combined because of small sample size. In
addition, very few events were collected below \Npv{} of 2 during 2012
data taking.
Events in which there are no reconstructed
jets with \pt~$>$~20 \GeV{} are referred to collectively as the 0-jet sample. 
Distributions are shown here for both the 0-jet and inclusive samples. 
For both samples, the data and MC simulation agree
within 2\% up to around \Npv~$=$~15 but the deviation grows to around
5--10\% for \Npv{}~$>$~25, which might be attributed to the decreasing 
sample size.
All of the \met~distributions show a similar level of agreement between
data and simulation across the
full range of \Npv{}.

For the 0-jet sample in Figure~\ref{fig:Z_resol_npv}(a), the STVF,
TST, and Track \met~resolutions all have a small slope with respect to \Npv{}, which implies stability
of the resolution against pileup. In addition, their resolutions agree within 1~\GeV{} throughout the \Npv{} range. In the
0-jet sample, the TST and Track \metmag{} are both primarily
reconstructed from tracks; however, 
small differences arise mostly
from accounting for photons in the TST \met~reconstruction algorithm. The CST
\met~is directly affected by the pileup as its reconstruction does not 
apply any pileup suppression techniques. Therefore, the CST
\met~has the largest dependence on \Npv, with a resolution ranging from
7 \GeV~at \Npv~$=$~2 to around 23 \GeV~at \Npv~$=$~25. The 
\met~resolution of the EJAF distribution, while better than that of the CST
\met{}, is not as good as that of the other pileup-suppressing algorithms.

For the inclusive sample in Figure~\ref{fig:Z_resol_npv}(b), the Track
\metmag~is the most stable with respect to pileup with almost no
dependence on \Npv{}. 
For \Npv~$>$~20, the Track \met~has the best resolution showing that pileup creates a larger degradation in the resolution of the other \met~distributions than
excluding neutral particles, as the Track \met~algorithm does. The EJAF \met~algorithm does not 
reduce the pileup dependence
as much as the TST and STVF \met~algorithms, and the CST \met~again
has the largest dependence on \Npv{}.

\begin{figure}[htbp]
\subfigure[]{\includegraphics[height=60mm]{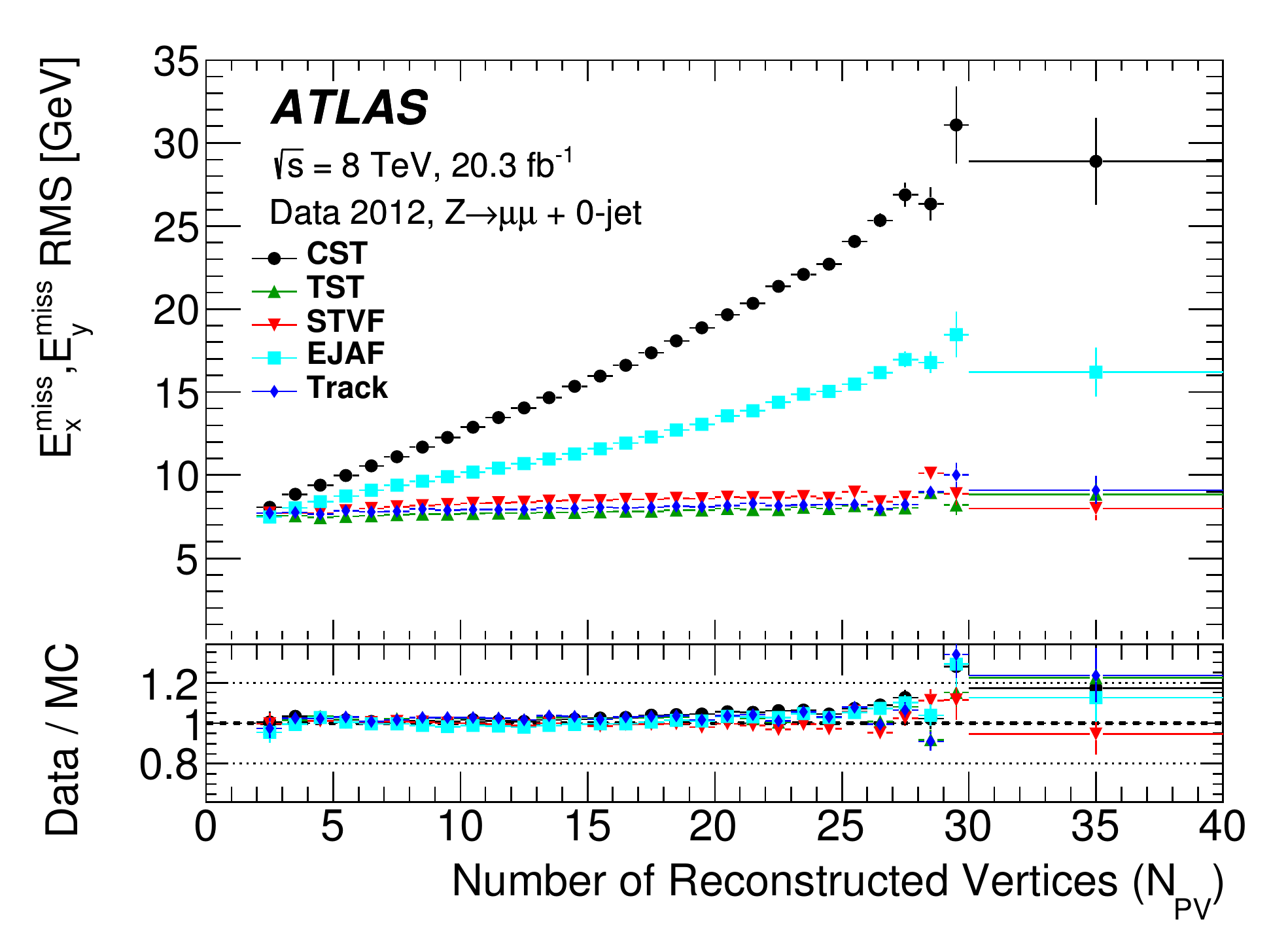}}
\subfigure[]{\includegraphics[height=60mm]{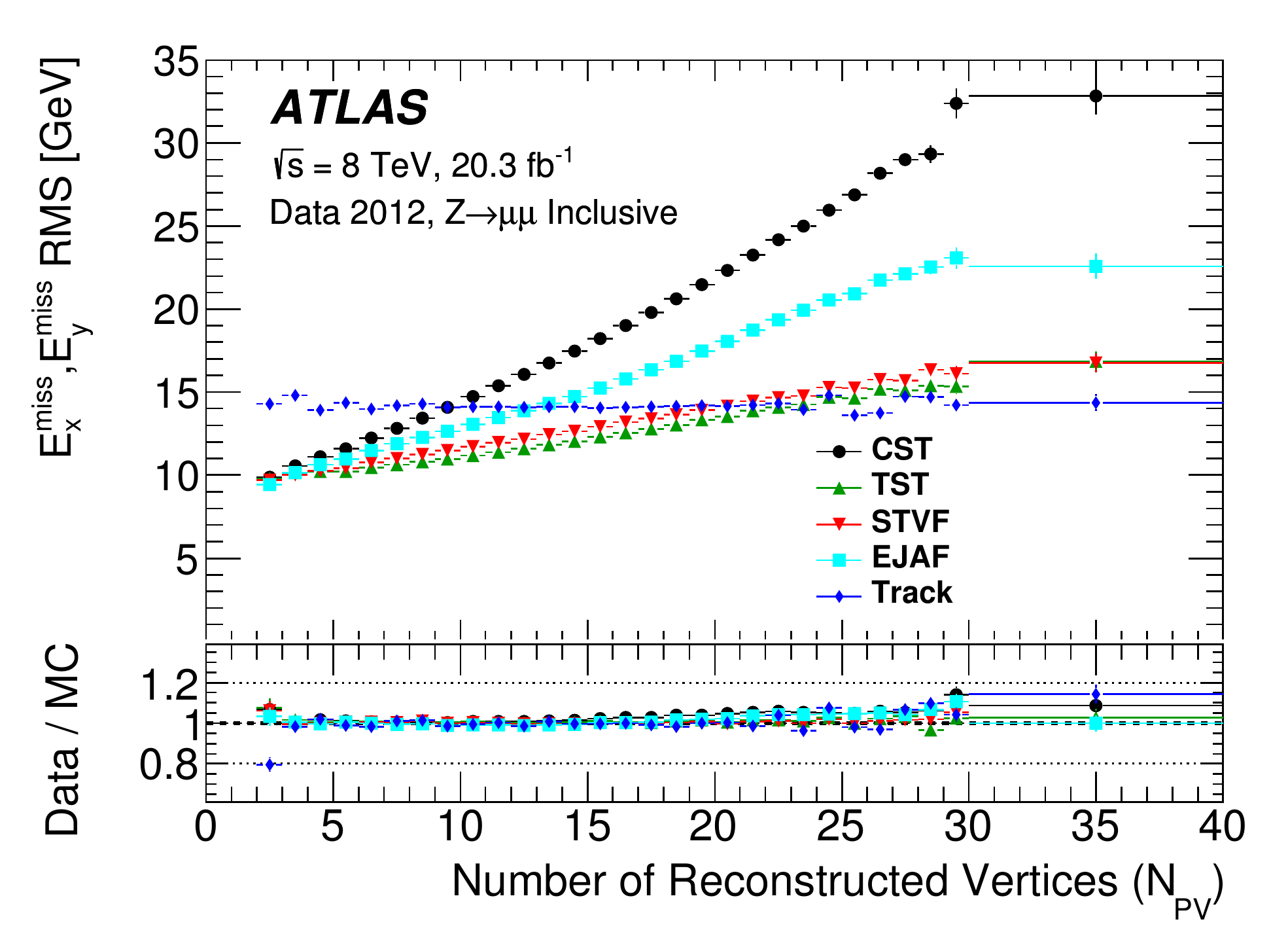}}
\caption{The resolution obtained from the combined distribution of 
  \metx~and \mety~for the
  CST, STVF, EJAF, TST, and Track \metmag{} algorithms as a
  function of \Npv~in (a) 0-jet and (b) inclusive \Zmm~events in data. 
  The insets at the bottom of the figures show the ratios of the data 
  to the MC predictions.} 
\label{fig:Z_resol_npv}
\end{figure}

Figure~\ref{fig:Z_resol_npv} also shows that 
the pileup dependence of the TST, CST, EJAF and STVF \met~is
smaller in the 0-jet sample than in the inclusive
sample. Hence, the evolution of the \met~resolution is shown for
different numbers of jets in Figure~\ref{fig:TST_njet_dependence} with
the TST \met~algorithm as a representative example. The jet counting
for this figure includes only the jets used by the TST \met~algorithm, so the
$\text{JVF}$ criterion discussed in Section~\ref{sec:jet_selec} is applied. Comparing 
the 0-jet, 1-jet and $\geq$2-jet distributions, the resolution is
degraded by 4--5 \GeV{} with each additional jet, which is much larger than any
dependence on \Npv{}. 
The inclusive distribution has a larger slope with respect to 
\Npv~than the
individual jet categories, which indicates that the behaviour seen in the inclusive sample is driven by
an increased number of pileup jets included in the \met~calculation at
larger \Npv{}.

\begin{figure}[htbp]
  \begin{centering}
    \includegraphics[height=60mm]{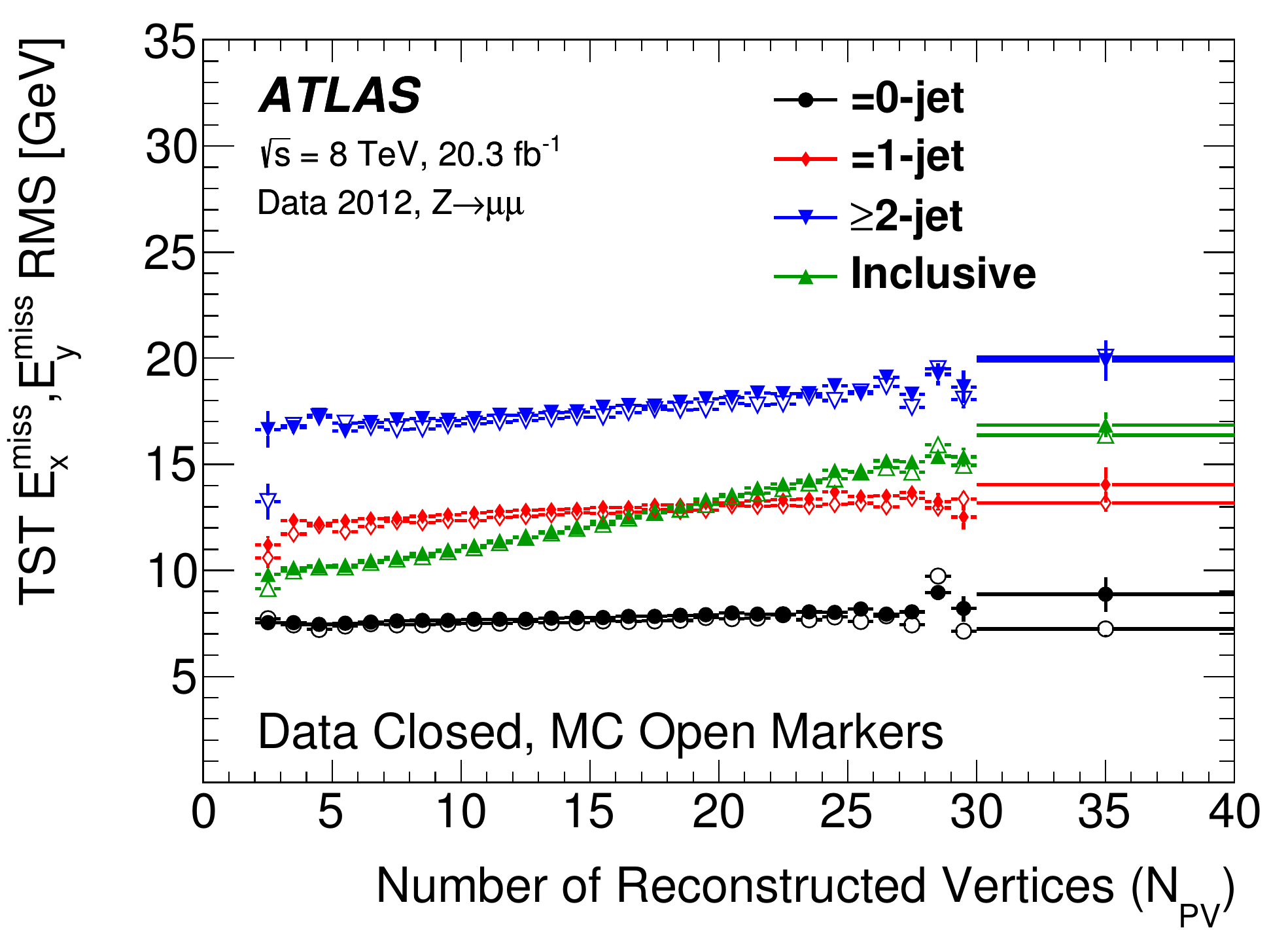}
    \caption{\label{fig:TST_njet_dependence} The resolution of the combined distribution of \metx~and 
      \mety~for the TST \metmag~as a
      function of \Npv~for the 0-jet, 1-jet, $\geq$ 2-jet, and inclusive 
      \Zmm~samples. The data (closed markers) and MC simulation (open
      markers) are overlaid. The jet counting uses the same
      $\text{JVF}$ criterion as the TST \met~reconstruction algorithm.} 
\end{centering}
\end{figure}

\subsubsection{Resolution of the \met~as a function of \sumetevt{}}
\label{sec:resol_Z}

The resolutions of \met{}, resulting from the different reconstruction
algorithms, are compared
as a function of the scalar sum of
transverse momentum in the event, as calculated using
Eq.~(\ref{eqsumet}). The CST \met~resolution is observed to depend linearly
on the square root of the \sumet~computed with the CST \met~components in
Ref.~\cite{ATLASMETPaper2011}. However, the \sumetevt~used in this subsection
is calculated with the TST \met~algorithm. This allows studies of the
resolution as a function of the momenta of particles from the selected
PV without including the amount of pileup activity in the
event. Figure~\ref{fig:resol_all_sumet_TST} shows 
the resolution as a function of \sumetevt(TST)
for \Zmumu~data and MC
simulation in the 0-jet and inclusive samples. 

In the 0-jet sample shown in Figure~\ref{fig:resol_all_sumet_TST}(a), 
the use of tracking information in the soft term, especially for the STVF, TST, and
Track \metmag{}, greatly improves the 
resolution relative to the
CST \met{}. The EJAF \met{} has a better resolution than that of 
the CST \met{}
but does not perform as well as the other reconstruction algorithms.
All of the resolution curves have an approximately linear 
increase with \sumetevt(TST); however, the Track \met{} resolution
increases sharply starting at \sumetevt(TST)~$=$~200 \GeV{} due to
missed neutral contributions like photons. The resolution predicted by the simulation
is about 5\% larger than in data for all \met~algorithms at
\sumetevt(TST)~$=$~50 \GeV{}, but agreement improves as \sumetevt(TST) increases
until around \sumetevt(TST)~$=$~200~\GeV{}. Events with jets can end
up in the 0-jet event selection, for example, if a jet is misidentified as a
hadronically decaying $\tau$-lepton. The $\sum
p_{\mathrm{T}}^{\tau}$ increases with \sumetevt(TST), and the rate of
jets misreconstructed as hadronically decaying $\tau$-leptons is not well modelled
by the simulation, which leads to larger \met{}
resolution at high \sumetevt(TST) than that observed in the data. The Track \met{} can be more
strongly affected by misidentified jets because neutral particles from the high-\pT{}
jets are not included. 

For the inclusive sample in Figure~\ref{fig:resol_all_sumet_TST}(b), 
the pileup-suppressed \metmag~distributions have better resolution than 
the CST \met~for \sumetevt(TST)~$<$~200~\GeV{}, but these
events are mostly those with no associated jets. For higher
\sumetevt(TST), the impact from the 
\sumjetet~term starts to dominate the resolution as well as the \sumetevt(TST).
Since the vector sum
of jet momenta is mostly common\footnote{As defined in
  Section~\ref{sec:jet_selec}, the CST \met~does not apply a JVF
  requirement on the jets like the TST, EJAF, and STVF
  \met{}. However, large \sumjetet{} tends to come from hard-scatter
  jets and not from pileup.} to all \metmag~algorithms except 
for the Track \metmag{}, those algorithms show similar 
performance in terms of the resolution. 
At larger \sumetevt(TST), the Track \metmag~resolution begins to degrade
relative to the other algorithms because it does not include the
high-\pT{} neutral particles coming from jets.
The ratio of data to MC simulation for the Track
\met~distribution is close to one, while for other algorithms the
MC simulation is below the data by about 5\% at large \sumetevt(TST). While
the Track \met~appears well modelled for the \ALPGEN$+$\PYTHIA{}
simulation used in this figure, the modelling depends strongly on the parton
shower model. 

\begin{figure}
\subfigure[]{\includegraphics[height=60mm]{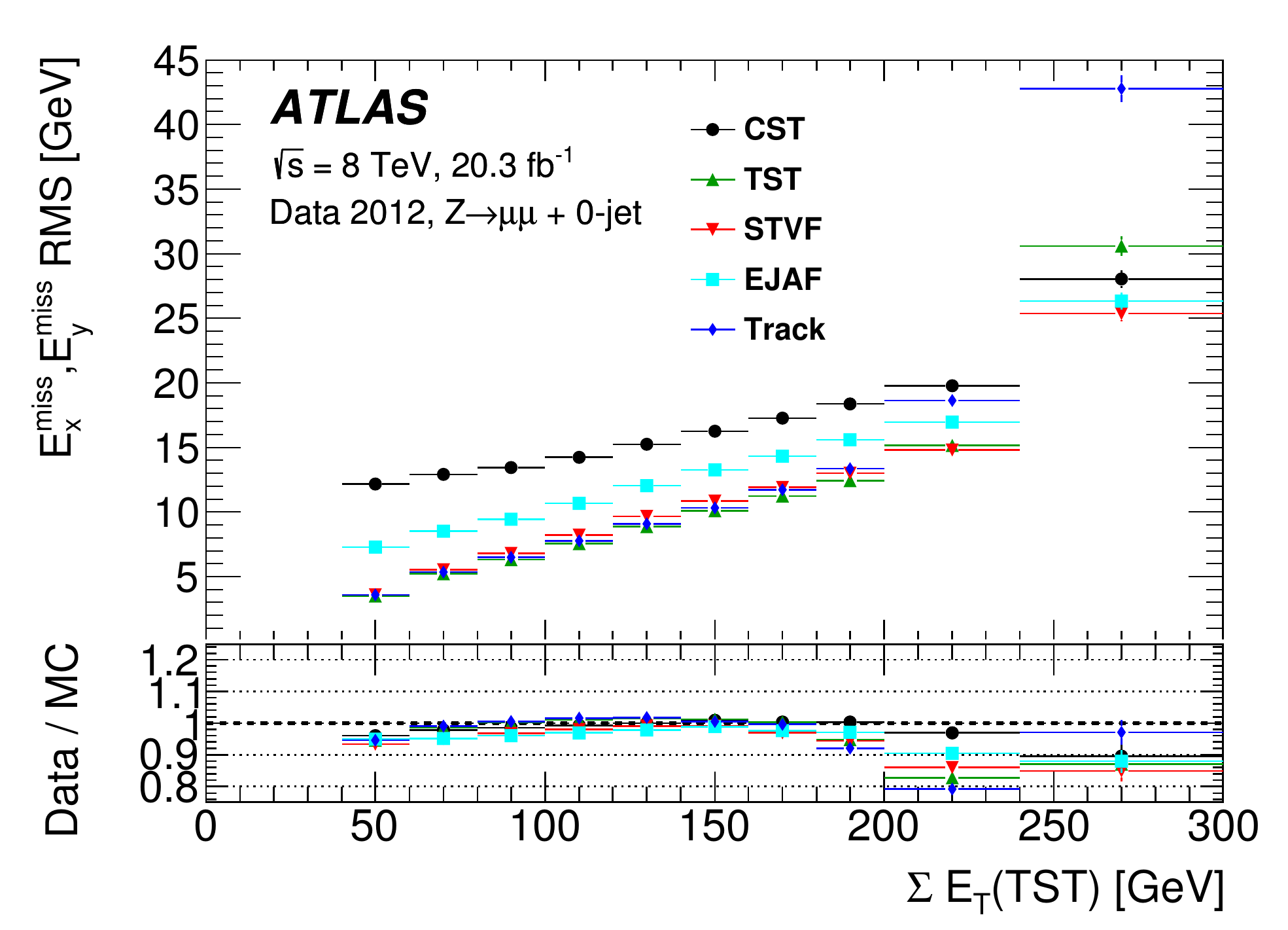}}
\subfigure[]{\includegraphics[height=60mm]{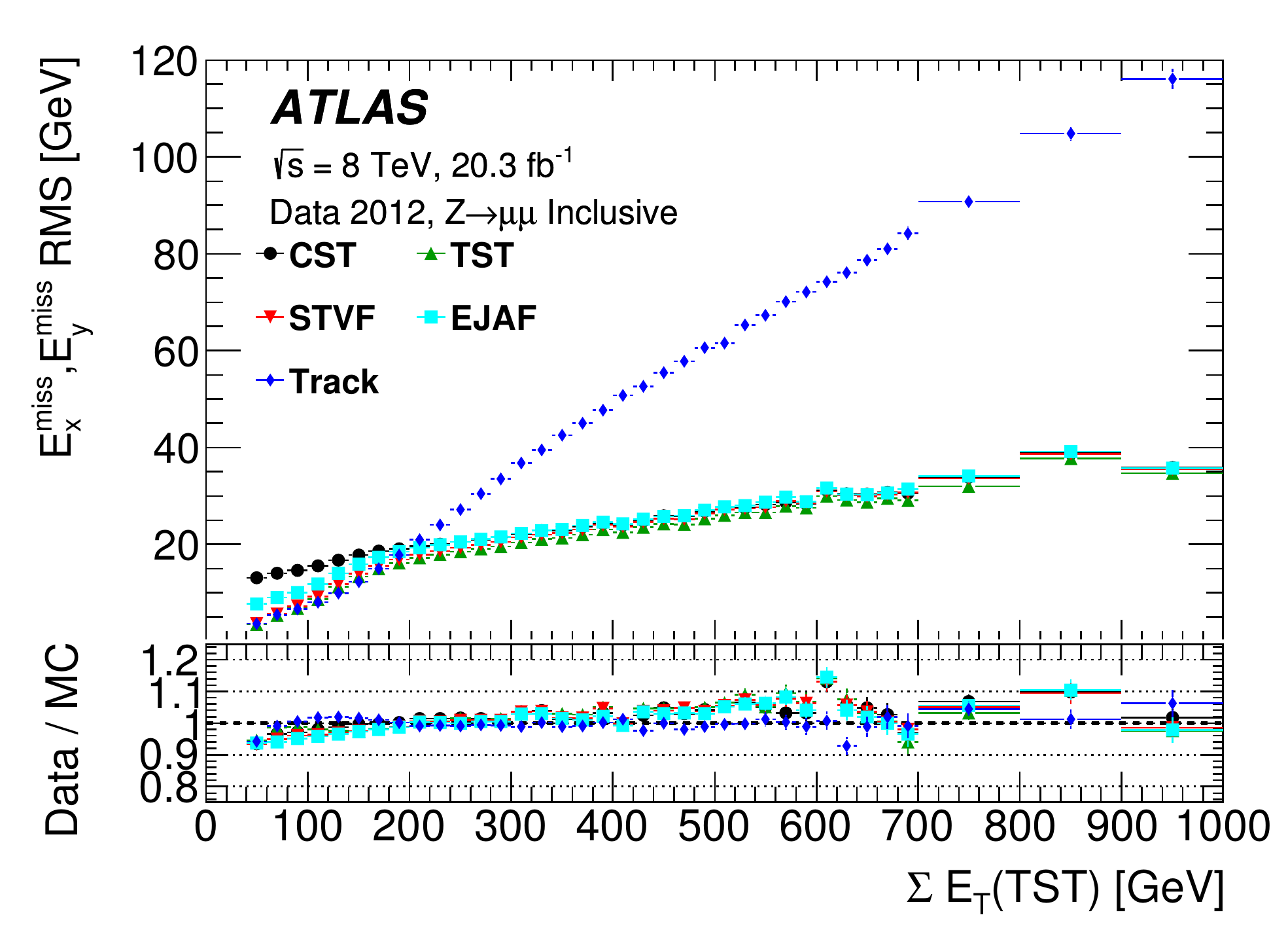}}
\caption{The resolution of the combined distribution of \metx~and \mety~for the
  CST, STVF, EJAF, TST, and Track \metmag{} as a
  function of \sumet(TST) in \Zmm~events in data for the 
  (a) 0-jet and (b) inclusive
  samples. The insets at the bottom of the figures show the ratios of the data
  to the MC predictions.}
 \label{fig:resol_all_sumet_TST}
 \end{figure}

\subsection{The \met{} response}
\label{sec:met_response}

The balance of \metvec~against the vector boson \ptvec~in
$W/Z+$jets events is used to evaluate the \met~response. 
A lack of balance is a global indicator of biases in \met~reconstruction and implies a
systematic misestimation of at least one of the \met{} terms,
possibly coming from an imperfect selection or calibration of the
reconstructed physics objects. 
The procedure to evaluate the response differs between \Zjets~events
(Section~\ref{sec:recoil}) and \Wjets~events (Section~\ref{sec:linearity}) because of the high-\pt~neutrino in the leptonic
decay of the $W$ boson.

\subsubsection{Measuring \met~recoil versus \ptZ{}}
\label{sec:recoil}

In events with \Zmm~decays, the \ptvec~of the $Z$ boson 
defines an axis in the transverse plane of the ATLAS detector, and for
events with 0-jets, the \metvec~should balance the \ptvec~of the $Z$ boson
(\ptZvec{}) along this axis. Comparing the response in events with and
without jets allows distinction between the jet and soft-term responses.
The component of the \metvec~along the \ptZvec{} axis is sensitive to
biases in detector responses~\cite{ATLASCSC}. 
The unit vector of \ptZvec{}
is labelled as \Az{} and is defined as:
\begin{equation}
\Az =\frac{\vec{p_{\mathrm{T}}}^{\ell^+}+\vec{p_{\mathrm{T}}}^{\ell^-}}{|\vec{p_{\mathrm{T}}}^{\ell^+}+\vec{p_{\mathrm{T}}}^{\ell^-}|},
\end{equation}
where $\vec{p_{\mathrm{T}}}^{\ell^+}$ and $\vec{p_{\mathrm{T}}}^{\ell^-}$ are the transverse
momentum vectors of the leptons from the $Z$ boson decay.

The recoil of the
$Z$ boson is measured by removing the $Z$ boson decay products from
the \metvec{} and is computed as
\begin{equation}
  \recoil{}= \metvec{}+\ptZvec{}.
\end{equation}
Since the \metvec~includes a negative vector sum over the lepton momenta, the
addition of \ptZvec{} removes its contribution. With an ideal
detector and \met~reconstruction algorithm, \Zll~events have no \met{}, and the \recoil{} balances with
\ptZvec{} exactly. For the real detector and \met~reconstruction algorithm, the degree of balance is measured by projecting the recoil onto
\Az{}, and the relative recoil is defined as the projection $\recoil{}\cdot\Az{}$
divided by \ptZ{}, which gives a dimensionless estimate that is unity if
the \met~is ideally reconstructed and calibrated. 
Figure~\ref{fig:diagnzscale} shows the mean relative recoil 
versus \ptZ~for \Zmm{} events where the average value is indicated by angle brackets. The data and MC
simulation agree within around 10\% for all \met{} algorithms for all
\ptZ{}; however, the
agreement is a few percent worse for \ptZ~$>$~50 \GeV{} in the 0-jet sample.

The \Zmm~events in the 0-jet sample in Figure~\ref{fig:diagnzscale}(a) 
have a relative recoil significantly lower than unity 
($\langle\recoil{}\cdot\Az{}/\ptZ{}\rangle$~$<$~1)  
throughout the \ptZ~range. In the 0-jet sample, the relative recoil
estimates how well the soft term balances the \ptvec{} of muons from
the $Z$ decay, which are better measured than the soft term. The
relative recoil below one indicates that the soft term is underestimated. 
The CST \met{} has a relative recoil measurement of 
$\langle\recoil{}\cdot\Az{}/\ptZ{}\rangle$~$\sim$~0.5 throughout the 
\ptZ~range, giving it the best recoil performance among the \met~algorithms. 
The TST and Track \met~have slightly larger biases than the 
CST \met{} because neutral particles are 
not considered in the soft term. The TST \met~recoil improves
relative to that of the Track \met~for \ptZ~$>$~40 \GeV{} because of the
inclusion of photons in its reconstruction.
The relative recoil distribution for the STVF \met~shows the largest 
bias for \ptZ{}~$<$~60 \GeV{}. 
The STVF algorithm scales the recoil down globally 
by the factor \stvf~as defined in Eq.~(\ref{eqpileup}), and this
correction decreases the already underestimated soft term. The
\stvf~does increase with \ptZ{} going from 0.06 at \ptZ~$=$ 0
\GeV{} to around 0.15 at \ptZ~$=$~50~\GeV{}, and this results in a rise in
the recoil, which approaches the TST \met~near \ptZ~$\sim$~70 \GeV{}.

In Figure~\ref{fig:diagnzscale}(b), the inclusive \Zmm~events have
a significantly underestimated relative recoil for \ptZ~$<$ 40 \GeV{}. The balance between the \recoil~and
\ptZvec~improves with \ptZ{} because of an increase in events having
high-\pT~calibrated jets recoiling against the $Z$ boson. The presence
of jets included in the hard term also reduces the sensitivity to the soft term, which is
difficult to measure accurately. The difficulty in isolating effects
from soft-term contributions from high-\pT~physics
objects is one reason why the soft term is not corrected. As with the 0-jet sample, the
CST \met{} has a significantly under-calibrated relative
recoil in the low-\ptZ~region, and all of the other \met~algorithms have
a lower relative recoil than the CST \met{}. Of the pileup-suppressing \met~algorithms, the TST
\met~is closest to the relative recoil of the CST \met{}.
The relative recoil of the Track \met~is significantly lower than unity because 
the neutral particles recoiling from the $Z$ boson are not included in its
reconstruction. Finally, the STVF \met{} shows the lowest relative recoil among the object-based 
\met~algorithms as discussed above for
Figure~\ref{fig:diagnzscale}(a), even lower than the Track \met~for
\ptZ~$<$~16 \GeV{}. 

\begin{figure}[htbp]
  \subfigure[]{\includegraphics[height=60mm]{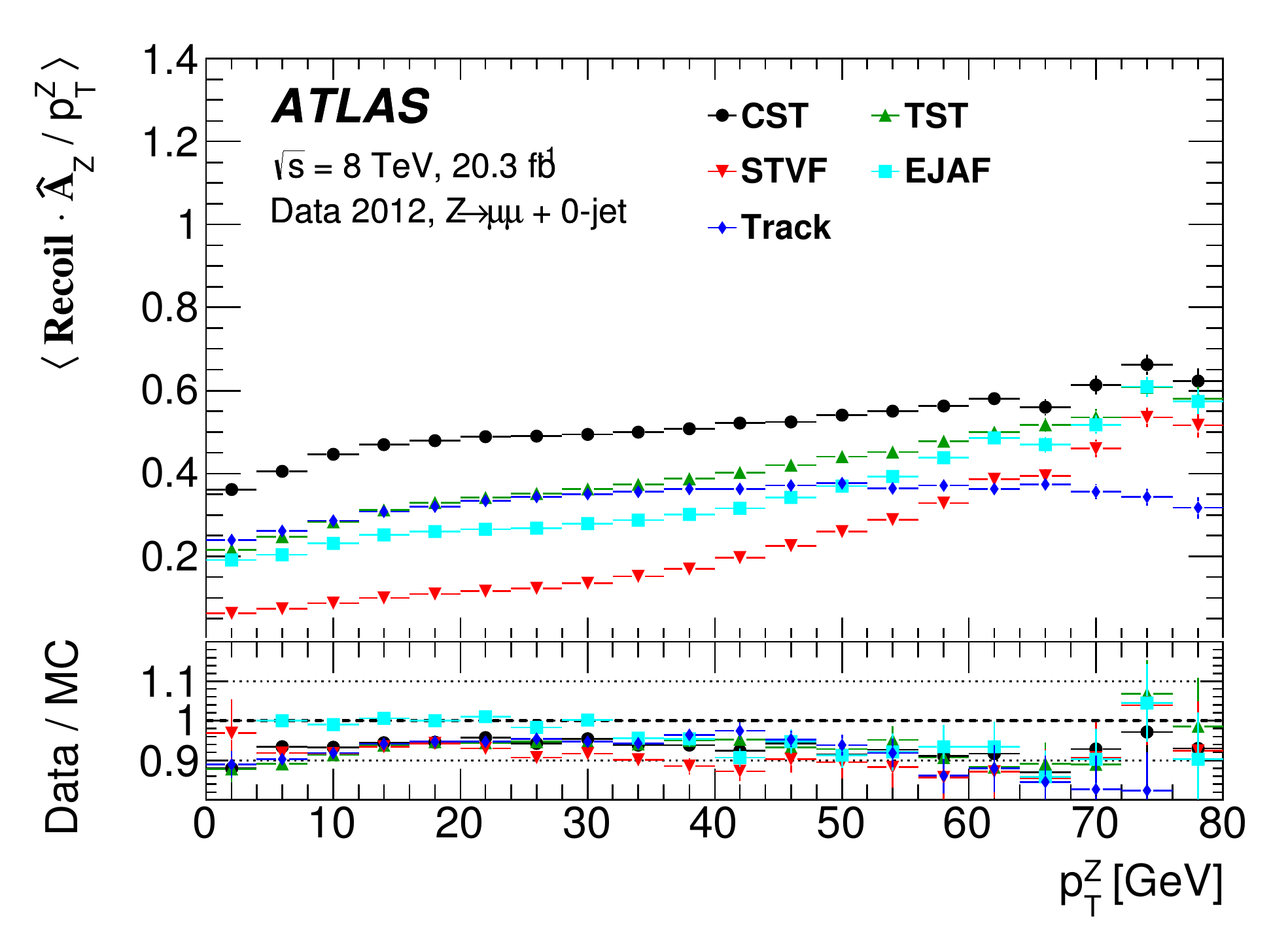}} 
  \subfigure[]{\includegraphics[height=60mm]{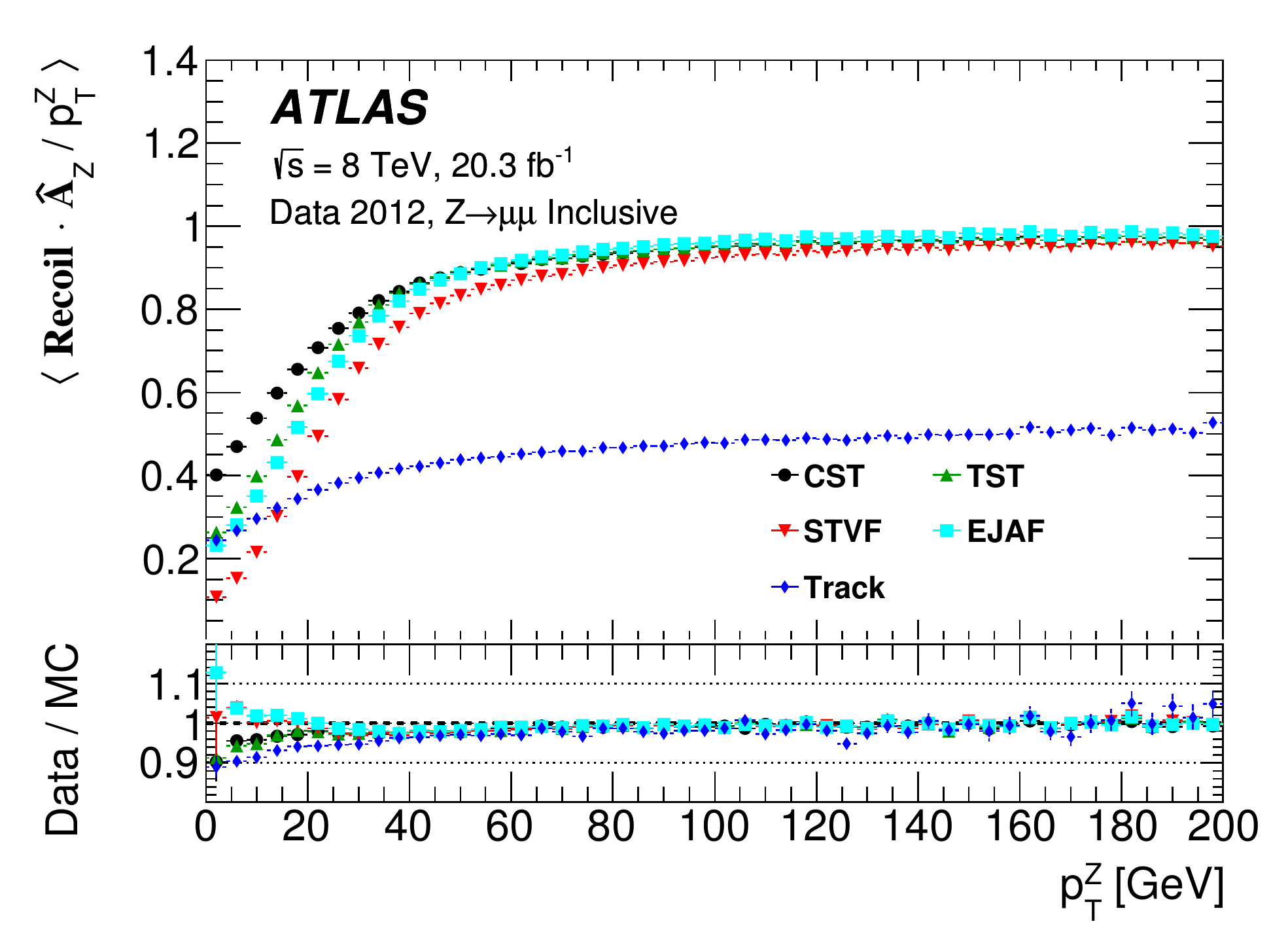}} 
  \caption{$\langle\recoil{}\cdot\Az{}/\ptZ{}\rangle$
    as a function \ptZ~for the (a) 0-jet 
    and (b) inclusive events in \Zmm~data. The insets at the bottom of the figures show the ratios of the data to the MC predictions.}
  \label{fig:diagnzscale}
\end{figure}

\subsubsection{Measuring \metmag~response in simulated \Wln~events}
\label{sec:linearity}
For simulated events with intrinsic \met{}, the response is studied by
looking at the relative mismeasurement of the reconstructed
\met{}. This is referred to here as the ``linearity'', and 
is a measure of how consistent
the reconstructed \met{} is with the \mettruemag{}.
The linearity is defined as the mean value of the ratio, $(E_{\mathrm{T}}^{\textrm miss}-E_{\mathrm{T}}^{\textrm miss,True})/E_{\mathrm{T}}^{\textrm miss,True}$
and is expected to be zero if the \metmag~is reconstructed at the correct 
scale.

For the linearity studies, no selection on the \metmag~or \mT~is 
applied, in order to avoid biases as these are purely simulation-based studies. In Figure~\ref{fig:MET_Lin_wmunu}, the linearity for
\Wmn~simulated events is presented as a function of the
\mettruemag{}. Despite the relaxed selection, a positive linearity is evident
 for \mettruemag $<$ 40 \GeV, due to the finite resolution of the
 \metmag~reconstruction and the fact that the
 reconstructed \metmag~is positive by definition. 
 The CST \met{} has the largest deviation from zero at low \mettruemag{} 
 because it has the largest 
\metmag~resolution. 

For the events in the 0-jet sample in Figure~\ref{fig:MET_Lin_wmunu}(a), 
all \met~algorithms have a negative linearity
for \mettruemag~$>$ 40 \GeV{}, which diminishes for
\mettruemag$\gtrsim 60$~\GeV{}. 
The region of \mettruemag{} between 40 and 60~\GeV{} mostly includes
events lying in the Jacobian peak of the $W$ transverse mass, and
these events include
mostly on-shell $W$ bosons. For \met~$\gtrsim$~40
\GeV{}, the on-shell $W$ boson must have non-zero \pT{}, which typically comes
from its recoil against jets. However, no reconstructed or generator-level jets are found in this
0-jet sample. Therefore, most of the events with 40~$<$~\mettruemag{}~$<$~60
\GeV{} have jets below the 20 \GeV~threshold contributing to the
soft term, and the soft term is not calibrated. The under-estimation
of the soft term, described in
Section~\ref{sec:recoil}, causes the linearity to deviate further from zero in this region. 
Events with \mettrue~$>$~60 \GeV{} are mostly
off-shell $W$ bosons that are produced with very low \pT{}. For these events, the \ptvec~contributions to the
\met~reconstruction come mostly from the well-measured muon \ptvec{}, and the soft term
plays a much smaller role. Hence, the linearity improves as the impact
of the soft term decreases with larger \mettrue{}.

For inclusive events in Figure~\ref{fig:MET_Lin_wmunu}(b) with \mettruemag~$>40$ \GeV{}, the deviation of the linearity from zero 
is smaller than 5\% for the CST \met. The linearity of the TST \met~is
within 10\% of unity in the range of 40--60 \GeV{} and improves for
higher \mettruemag~values. The STVF \met~has the most negative
bias in the linearity among the object-based 
\met~algorithms for \mettruemag~$>$~40~\GeV. The TST, CST, STVF, and EJAF
\met~algorithms perform similarly for all \mettruemag~values. As expected, the linearity of the Track \met~settles below zero due to not accounting for neutral particles in jets.

\begin{figure}[htbp]
 \subfigure[]{\includegraphics[height=60mm]{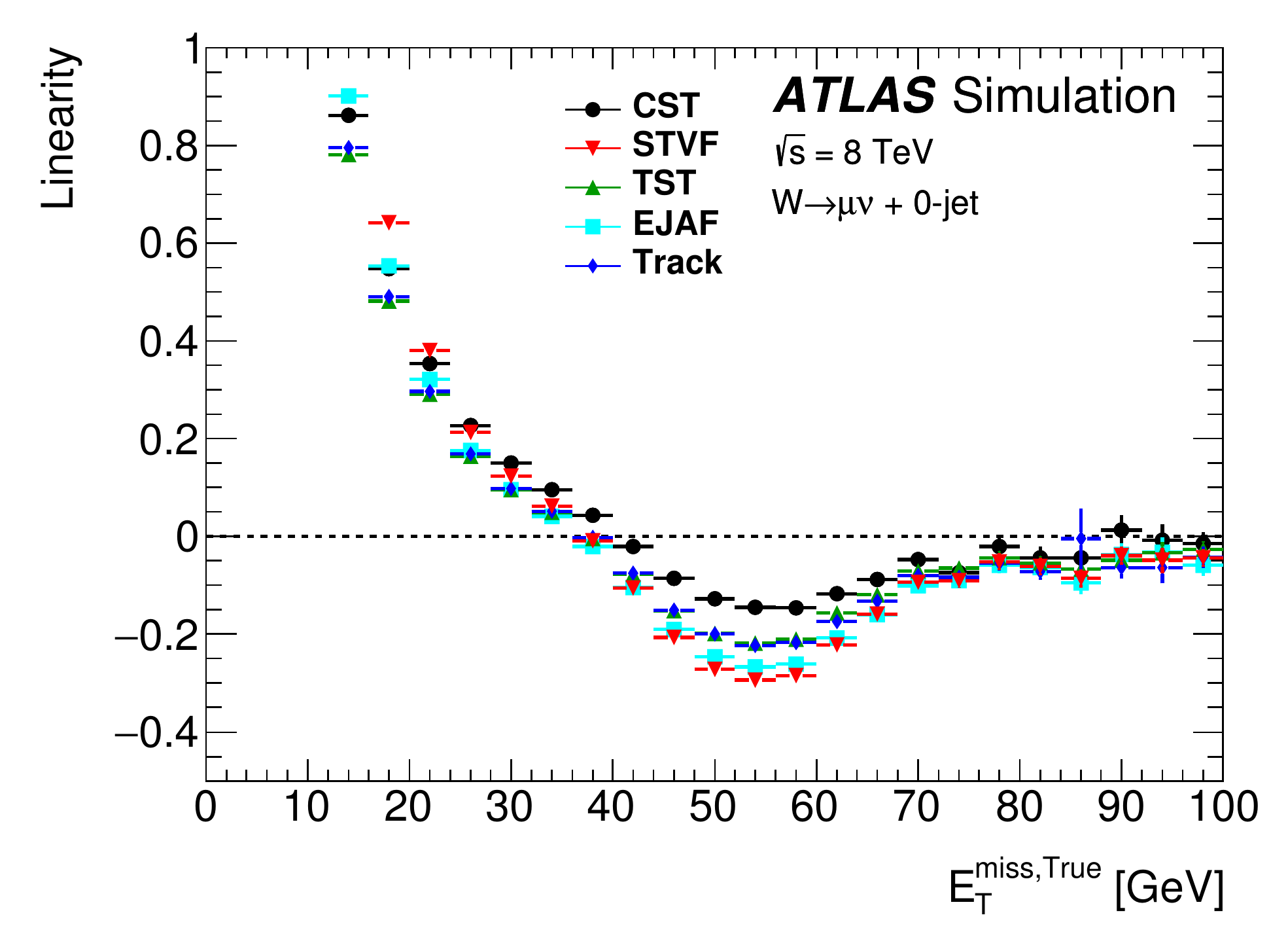}}
 \subfigure[]{\includegraphics[height=60mm]{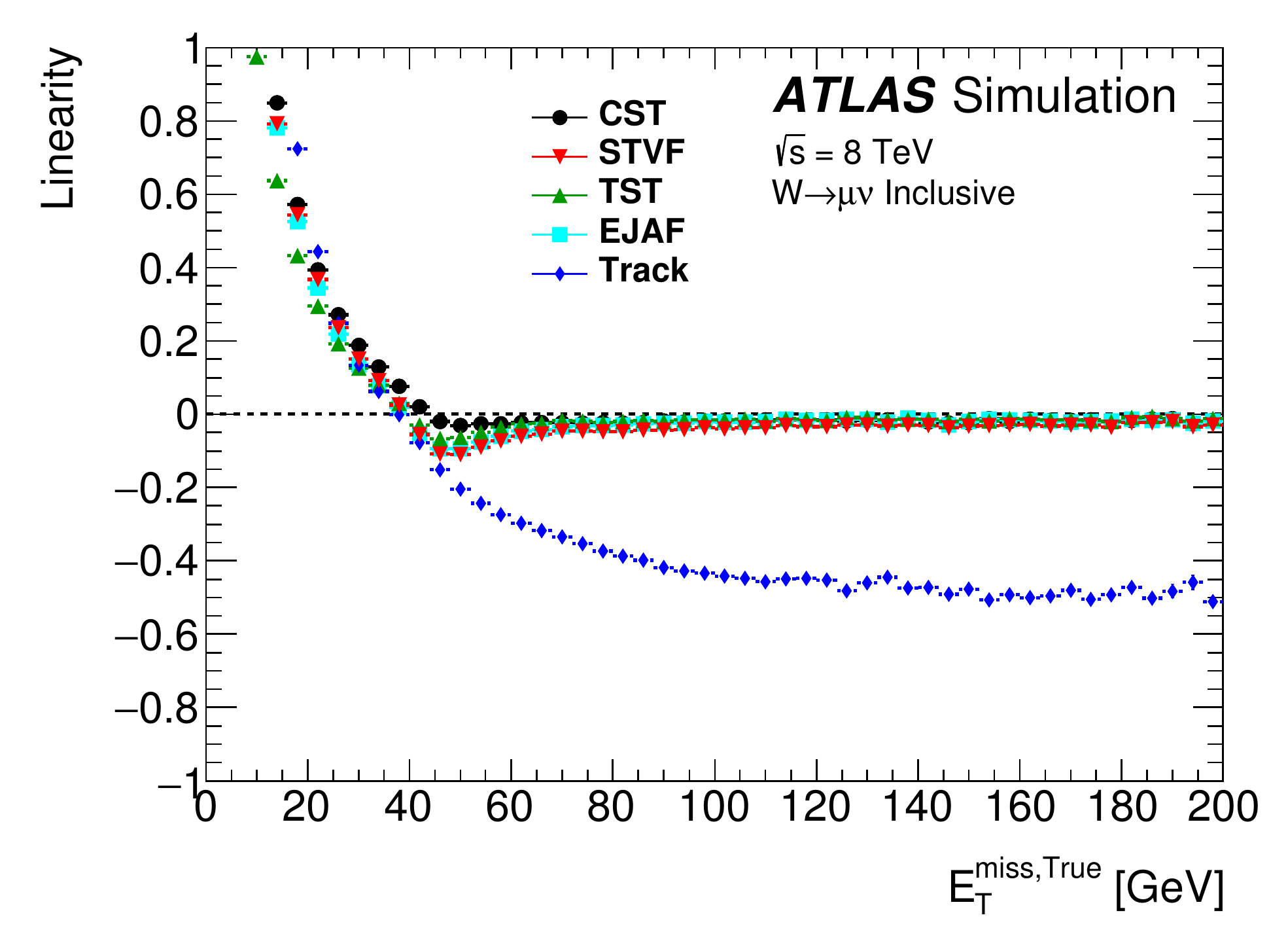}}
\caption{\metmag~linearity in \Wmn~MC simulation is shown versus
  \mettruemag~in the (a) 0-jet and (b) inclusive events.}
\label{fig:MET_Lin_wmunu}
\end{figure}

\subsection{The \metvec~angular resolution}
\label{sec:metdir}

The angular resolution is important
for the reconstruction of kinematic observables such as the transverse mass of the 
$W$ boson and the invariant mass in \Htau~events \cite{htautaupub}. For simulated \Wln{} events, the direction of the reconstructed \metvec~is compared to the \mettruevec~for 
each \met~reconstruction algorithm using the difference
in the azimuthal angles, \dphimettruth{}, which has a mean value of
zero. The RMS of the distribution is taken as the resolution, which is labelled \rmsdphi{}. 

No selection on the \metmag~or \mT~is 
applied in order to avoid biases. The \rmsdphi~is shown as a function of \mettruemag{} in
Figure~\ref{fig:dphimet_truemet_w}(a) for the 0-jet sample in
\Wmn~simulation; the angular resolution generally improves 
as the \mettruemag~increases, for all algorithms.
For \mettruemag~$\lesssim$ 120 \GeV{}, the pileup-suppressing
algorithms improve the
resolution over the CST \met~algorithm, but all of
the algorithms produce distributions with similar resolutions in the higher
\mettruemag~region. 
The increase in 
\rmsdphi~at around 40--60 \GeV~in the 0-jet sample is due to the larger
contribution of jets below 20 \GeV~entering the soft term as mentioned in Section~\ref{sec:linearity}.
The distribution from the inclusive sample shown in Figure~\ref{fig:dphimet_truemet_w}(b) has the same 
pattern as the one from the 0-jet sample, except that the performance of the 
Track \met~algorithm is again significantly worse. 
In addition, the transition region near 40~$<$~\mettrue~$<$~60 \GeV{} is
smoother as 
the under-estimation of the soft term becomes less significant 
due to the presence of events with high-\pT~calibrated jets.
The TST \met~algorithm has the best angular resolution for both the
0-jet and inclusive topologies throughout the entire range of
\mettruemag{}. 

\begin{figure}[htbp]
 \subfigure[]{\includegraphics[height=60mm]{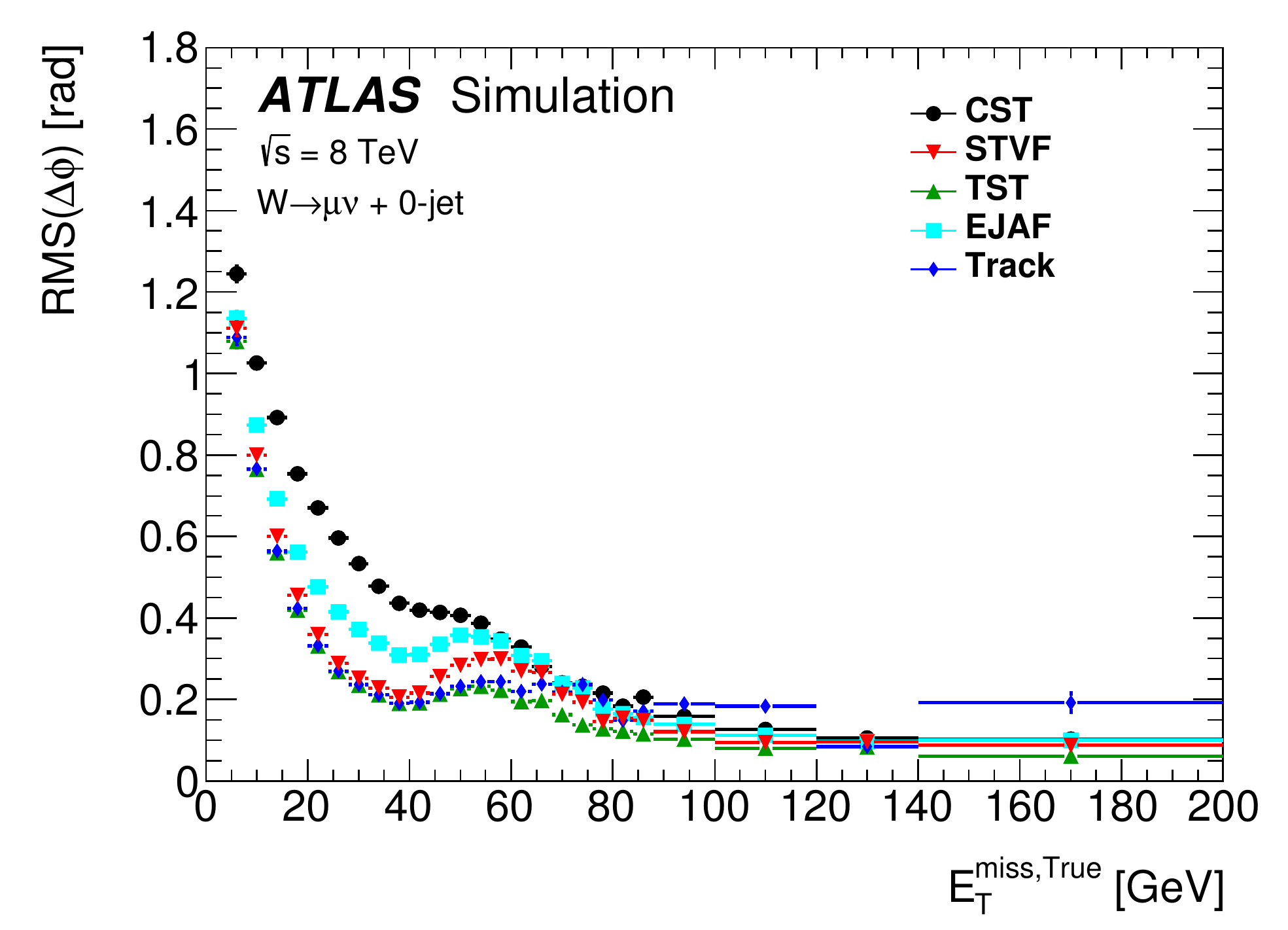}}
 \subfigure[]{\includegraphics[height=60mm]{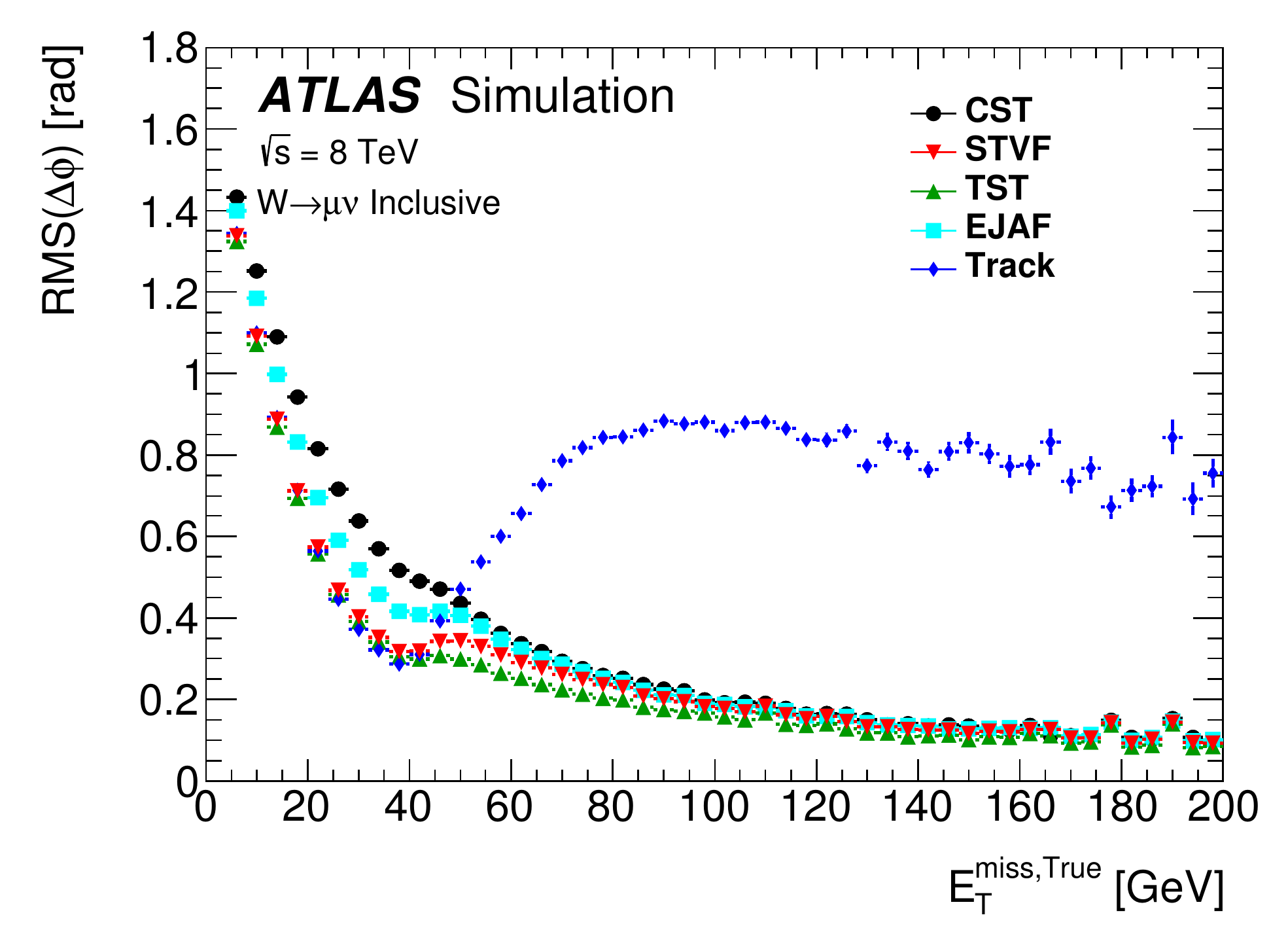}} 
\caption{The resolution of \dphimettruth{}, labelled as \rmsdphi{}, is shown for
  \Wmn~MC simulation for the (a) 0-jet and (b) inclusive samples.}
\label{fig:dphimet_truemet_w}
\end{figure}

\subsection{Transverse mass in \Wln~events}
\label{sec:kinematic_obs}

The $W$ boson events are selected using kinematic
observables that are computed from the \metvec~and lepton transverse
momentum. This section evaluates the scale of the \mT, as defined in
Eq.~(\ref{eq:mT}), reconstructed with each \met~definition. The
\mT~computed using the reconstructed \metvec~is compared to the
\mTtrue{}, which is calculated using the
\mettruevec{} in \Wmn~MC simulation. The mean of the difference
between the reconstructed and generator-level \mT{}, ($\langle\mT - \mTtrue \rangle$), 
is shown as a function of \mTtrue{} in
Figure~\ref{fig:wmt_reso_vs_mttruth} for the 0-jet and 
inclusive samples. No \metmag~or \mT~selection is made in these figures, to avoid biases. 
All distributions for the \met~algorithms have a positive bias at low values of
\mTtrue~coming from the positive-definite nature of the \mT{} and the
finite \met~resolution. For the 0-jet sample, the CST algorithm
has the smallest bias for \mT~$\lesssim$~60 \GeV{} because it includes
the neutral particles with no corrections for pileup. However, for the inclusive sample the TST
\met{} has the smallest bias as the \met~resolution plays a larger
role. The STVF and Track \met~have the largest bias for $\mTtrue$ $<$ 50 \GeV~in the 0-jet 
and inclusive samples, respectively. This is due to the over-correction in the soft term by \stvf{} for the former and from the missing neutral particles in the latter case. 
For events with \mT~$\gtrsim$~60 \GeV{}, all of the \met~algorithms have
$\langle\mT - \mTtrue \rangle$ close to zero, with a spread of less than 3~\GeV{}.

\begin{figure}[htbp]
 \subfigure[]{\includegraphics[height=60mm]{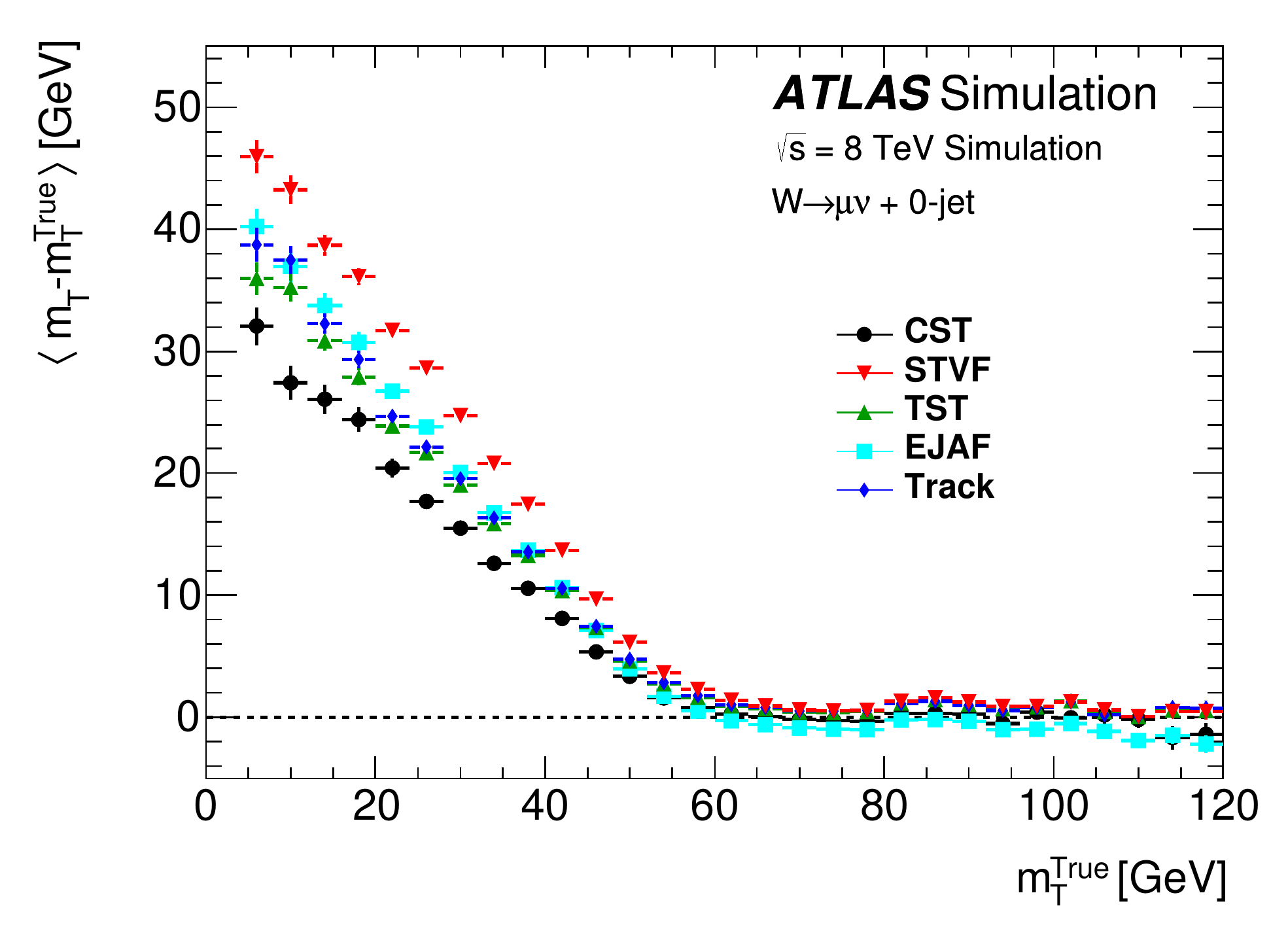}}
 \subfigure[]{\includegraphics[height=60mm]{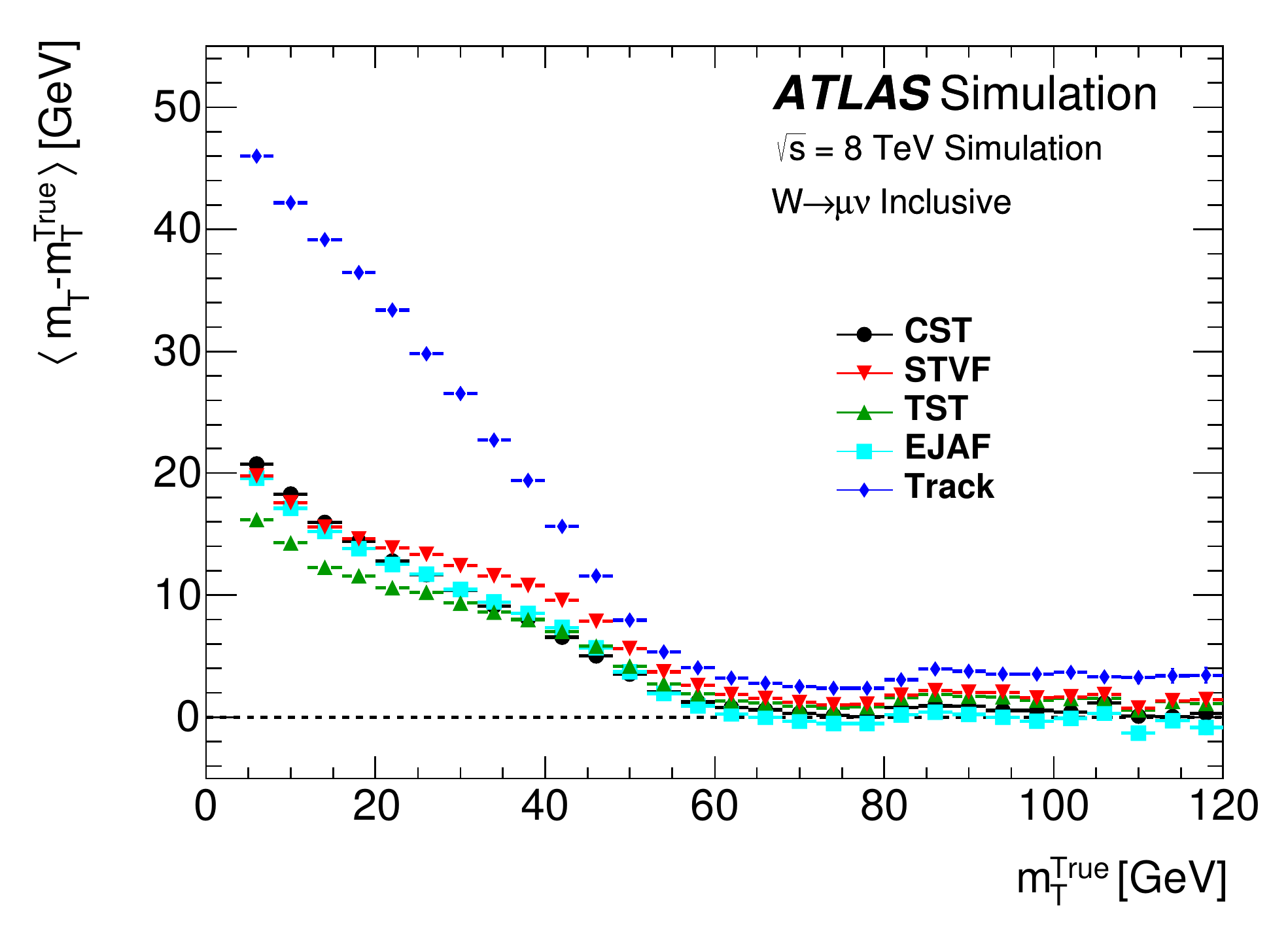}} 
\caption{The $\langle\mT - \mTtrue \rangle$ is shown versus \mTtrue~for
  \Wmn~MC simulation in the (a) 0-jet and (b) inclusive samples.
}
\label{fig:wmt_reso_vs_mttruth}
\end{figure}

\subsection{Proxy for \met{} Significance}
\label{sec:significance}

The \met~significance is a metric defined to quantify how likely it is
that a given event contains intrinsic \met{} and is
computed by dividing the measured \metmag~by an estimate of its
uncertainty. Using 7~\TeV{} data, it was shown that
the CST \metmag~resolution follows an approximately stochastic
behaviour as a function of \sumet{}, computed with the CST components, and is described by 
\begin{equation}
  \sigma(\met) = a \cdot \sqrt{\sumet},
  \label{eq:stoch}
\end{equation}
where $\sigma(\met)$ is the CST \met~resolution \cite{ATLASMETPaper2011}. The typical value of
$a$ in the 8~\TeV~dataset is around 0.97~$\GeV{}^{1/2}$ for the CST
\met{}. 
The proxy of the \met~significance presented in this section is
defined as the $\frac{1}{a}\cdot$\met{}/$\sqrt{\sumet}$. This choice is motivated by
the linear relationship for the CST \met{} between its $\sqrt{\sumet}$
and its \met~resolution. The same procedure does not work for the TST
\met{} resolution, so a value of
2.27~$\GeV{}^{1/2}$ is used to tune the $x$-axis so that integral of \Zmm{} simulation fits the
multiples of the standard deviation of a normal distribution at the
value of 2.
Ideally, only events with large intrinsic 
\met~have large values of $\frac{1}{a}\cdot$\met{}/$\sqrt{\sumet}$, while events with no
intrinsic \met~such as \Zmm~have low values. It is important to point
out that in general \Zmm{} is not a process with large
\met~uncertainties or large $\sqrt{\sumet}$. However, when there
are many additional jets (large \sumet{}), there is a significant probability that one
of them is mismeasured, which generates fake \met{}. 

The distribution of $\frac{1}{a}\cdot$\met{}/$\sqrt{\sumet}$ is shown for the CST and TST \met~algorithms in
Figure~\ref{fig:metSig} in \Zmm~data and MC simulation. 
The data and MC simulation agree within the assigned uncertainties for both 
algorithms. The CST \met{} distribution in Figure~\ref{fig:metSig}(a) has a very narrow core for the
\Zmm~process, having 97\% of data events with
1.03$\cdot$\met{}/$\sqrt{\sumet}$~$<$~2.
The proxy of the \met~significance, therefore, provides discrimination power
between events with intrinsic \met{} (e.g. \ttbar~and dibosons) and those with fake \met{} (e.g. poorly
measured \Zmm~events with a large number of jets).

\begin{figure}[htbp]
  \centering
  \subfigure[]{\includegraphics[height=70mm]{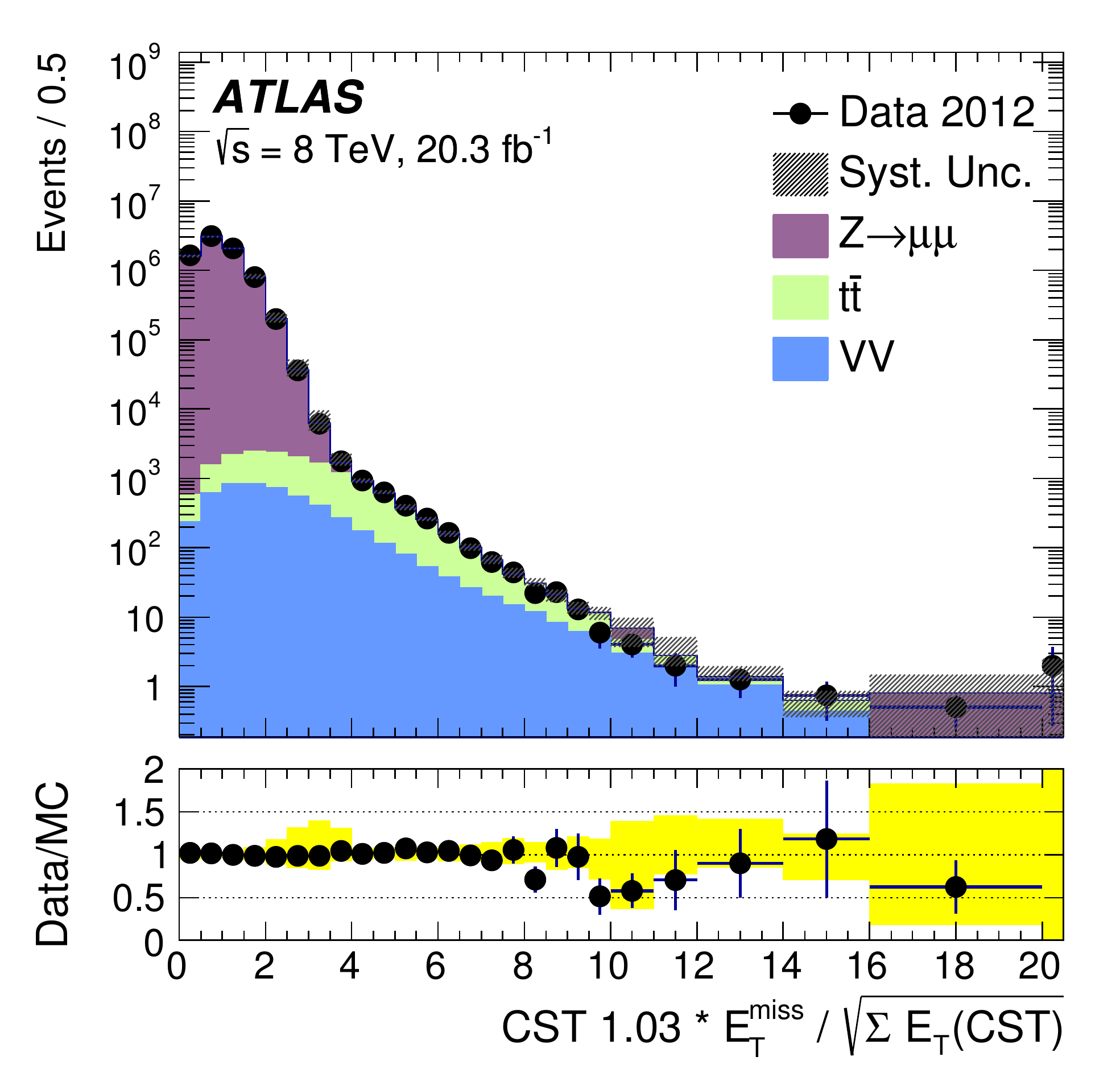}}
  \subfigure[]{\includegraphics[height=70mm]{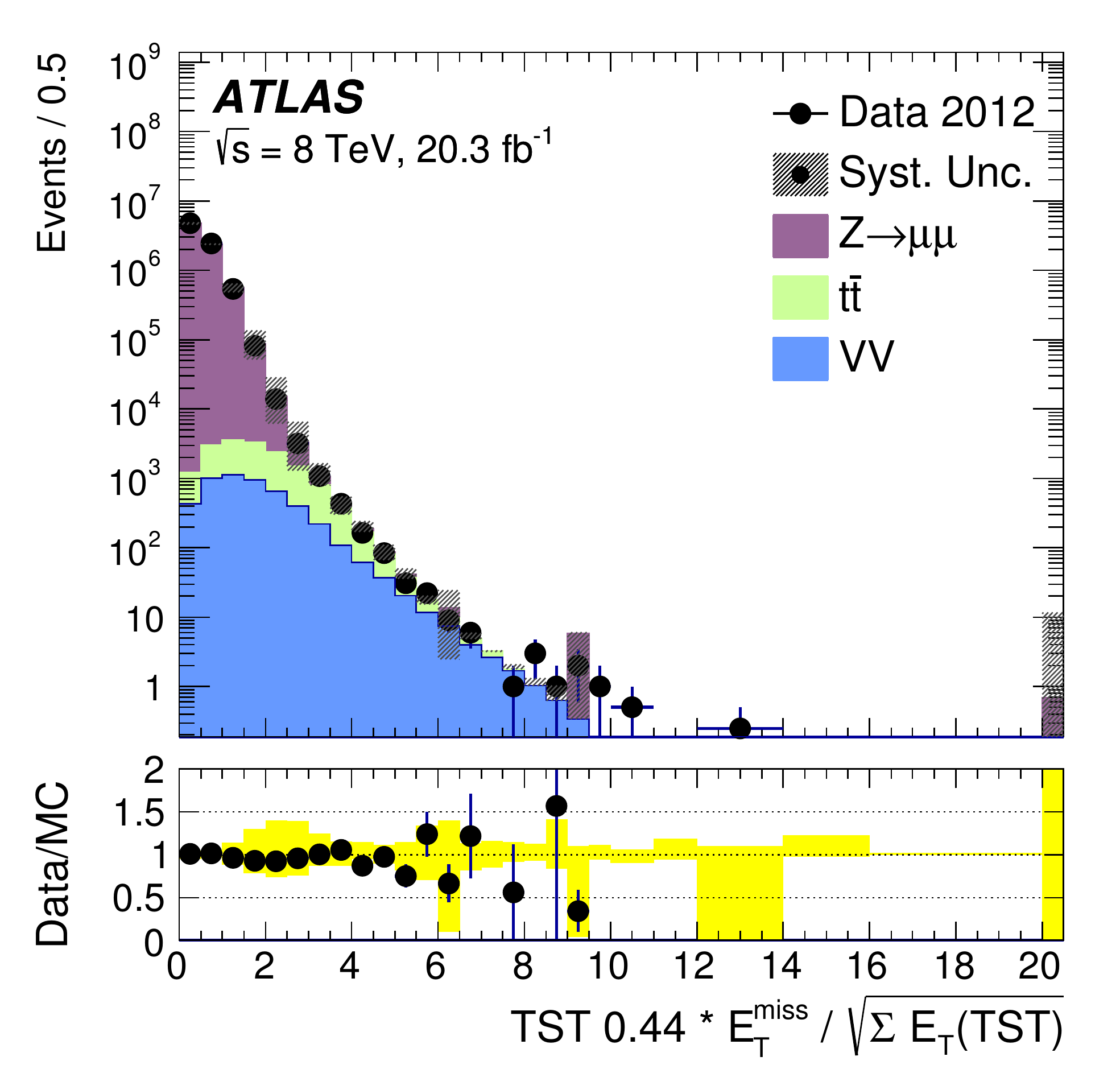}}\\
  \caption{The proxy for \met~significance is shown in data and MC simulation events
    satisfying the \Zmm{} selection for the (a)
    CST and (b) TST \met~algorithms. The solid band shows the combined
    MC statistical and systematic uncertainties, and the insets at the
    bottom of the figures show the ratios of the data to the MC
    predictions. The far right bin includes the integral of all events above 20.}
  \label{fig:metSig}
\end{figure}

The TST \met{} is shown as an example of a pileup-suppressing
algorithm. The \sumet~is not always an accurate reflection of
the resolution when there are significant contributions from tracking resolution, as discussed in Section~\ref{sec:Perf_Z}. 
In particular, the performance of the TST reconstruction algorithm is determined by the tracking
resolution, which is generally more precise than the calorimeter
energy measurements because of the reduced pileup
dependence, especially for charged particles with lower \pT{}.
Neutral particles are not included in the
\sumet{} for the Track \met{} and TST algorithms, but they do affect
the resolution. In addition, a very small number of tracks do have very large
over-estimated momentum measurements due to multiple scattering or
other effects in the
detector, and the momentum uncertainties of these tracks are not
appropriately accounted for in the \sumet{} methodology.

\subsection{Tails of \met~distributions}
\label{sec:tails}

Many analyses require large \met~to select
events with high-\pT{} weakly interacting particles. The selection efficiency,
defined as the number of events with \met{} above a given threshold divided by the total number of
events, is used to compare the performance of various
\metmag~reconstruction algorithms. 
As \Zll~events very rarely include high-\pt{} neutrinos, they can be
rejected by requiring substantial \met{}. For events with intrinsic
\met~such as \Wln, higher selection efficiencies than the \Zll~events 
are expected when requiring reconstructed \met{}.
For both cases, it is important to evaluate the performance of the reconstructed 
\metmag{}. 

The selection efficiencies with various \met~algorithms are compared
for simulated \Zmm~and 
\Wmn~processes as shown in 
Figure~\ref{fig:tails_zw} using the MC simulation. 
The event selections discussed in
Sections~\ref{sec:evtselZll} and~\ref{sec:evtselWlv} are applied
except the requirements on \metmag~and \mT{} for the \Wmn{} selection. 

As shown in Figure~\ref{fig:tails_zw}(a), the selection efficiency
for \Zmm~events is around 1\% for \met~$>$~50 \GeV{},
for all \met~algorithms. Thus a \met{} threshold requirement can be
used to reject a large number of events without intrinsic \met{}. 
However, the \mettrue{}, which does not include detector resolution 
effects, 
shows the selection efficiency under ideal conditions,
indicating there may be additional potential for improvement
of the reconstructed \met{}. 
Namely, the selection efficiency with \mettrue{} provides a benchmark
against which to 
evaluate the performance of different
\met~algorithms.  
The STVF, TST, and Track
\metmag~distributions have
narrow cores, so for \met~threshold $\lesssim$ 50~\GeV{} these three
\met~definitions have the lowest selection efficiencies for
\Zmm~events. Above 50 \GeV{}, the Track \metmag~performance is degraded as a
result of missing neutral particles,
which gives it a very high selection efficiency. The TST and STVF 
\met~algorithms continue to have 
the lowest selection efficiency up to \met~threshold $\approx$~110
\GeV{}. For 110--160 \GeV{}, the TST \met{} has a longer tail than the
CST \met{}, which is a result of mismeasured low-\pT~particles 
that scatter and are reconstructed as high-\pT{} tracks. Such
mismeasurements\footnote{For the TST and Track
\met, mismeasured high-\pt~tracks with \pT~$>$~120 (200) \GeV{} are removed using the track quality
requirements in high (low) $|\eta|$ as defined in
Section~\ref{sec:tst}.} are rare but significant in the \met~tail. The TST, STVF, CST, and EJAF \met~algorithms provide
similar selection efficiencies for \met~$>$~160 \GeV{}. Above this
threshold, the \met{} is dominated by mismeasured high-\pT physics objects which are
identical in all object-based \met~definitions. 
Hence, the events with \met~$\gtrsim$~160
\GeV~are correlated among the TST, STVF, CST, and EJAF \met~distributions.

Figure~\ref{fig:tails_zw}(b) shows the selection efficiency for
the \Wmn~simulated events passing a \met~threshold for all
\metmag~algorithms. Requiring the \Wmn~events to pass the \met{} threshold
should ideally have a high selection efficiency similar to that of 
the \mettrue{}. The CST \met~algorithm gives 
the highest selection efficiency between 30--120 \GeV{}
but does not agree as well as that of the
other \met~algorithms with the \mettrue~selection efficiency for \met~threshold 
$\lesssim$ 110 \GeV{}. 
This comes from the positive-definite nature of the
\metmag{} and the worse resolution of the CST \met~relative to the other \met~definitions. 
The Track \met~has the efficiency closest to that of the \mettrue, but for
Track \met~$\gtrsim$~60 \GeV, the amount of jet activity increases, which
results in a lower selection efficiency
because of missing neutral particles. The EJAF,
STVF, and TST \met~distributions are closer than the CST to the \mettrue~selection efficiency 
for \met~threshold $\lesssim$ 100~\GeV, but the 
efficiencies for all the object-based algorithms and
\mettrue{} converge for \met~threshold $\gtrsim$ 110~\GeV{}. 
Hence, for large \met{} all object-based algorithms perform
similarly.

\begin{figure}[htbp]
 \subfigure[]{\includegraphics[height=60mm]{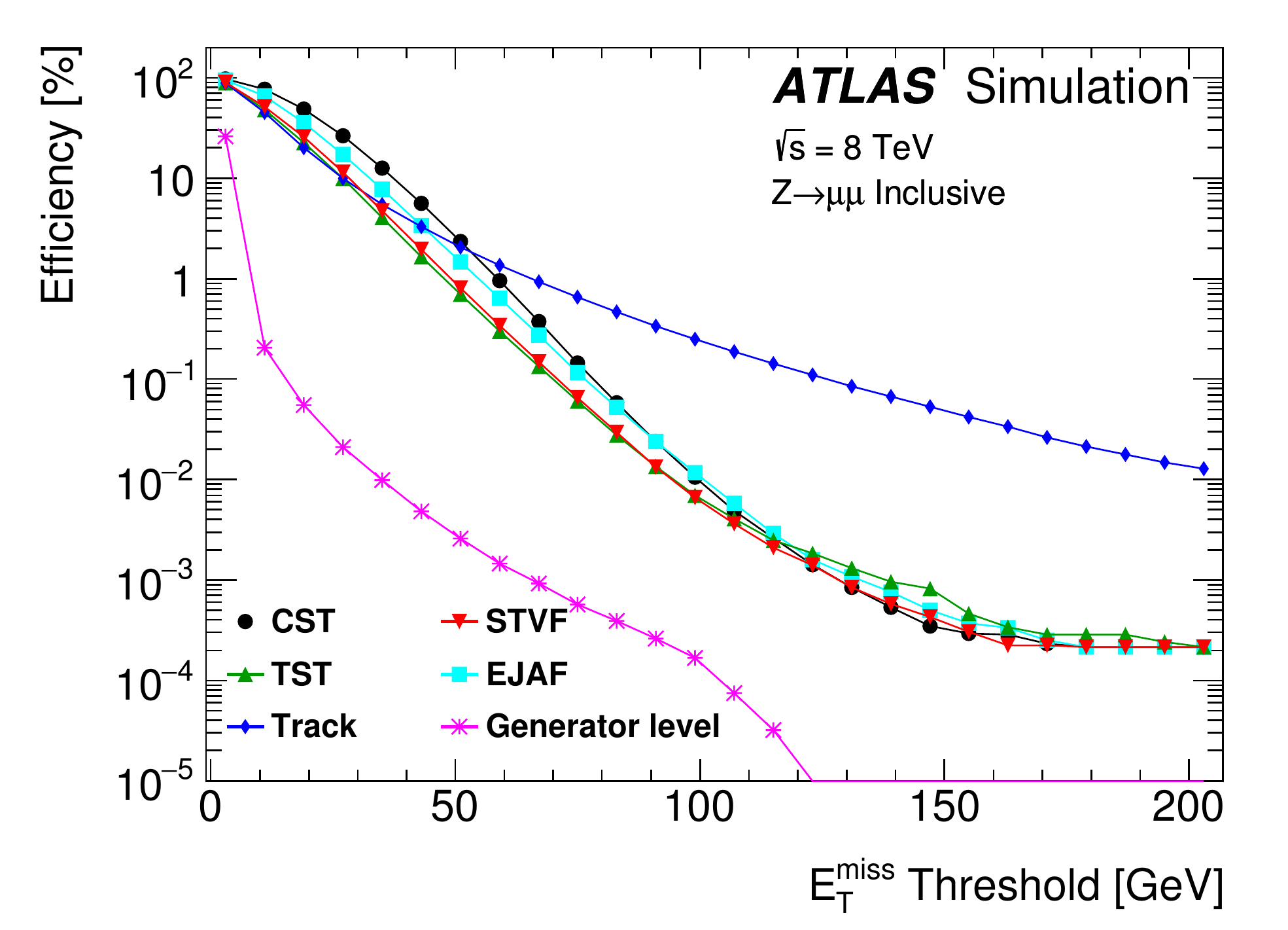}} 
\subfigure[]{\includegraphics[height=60mm]{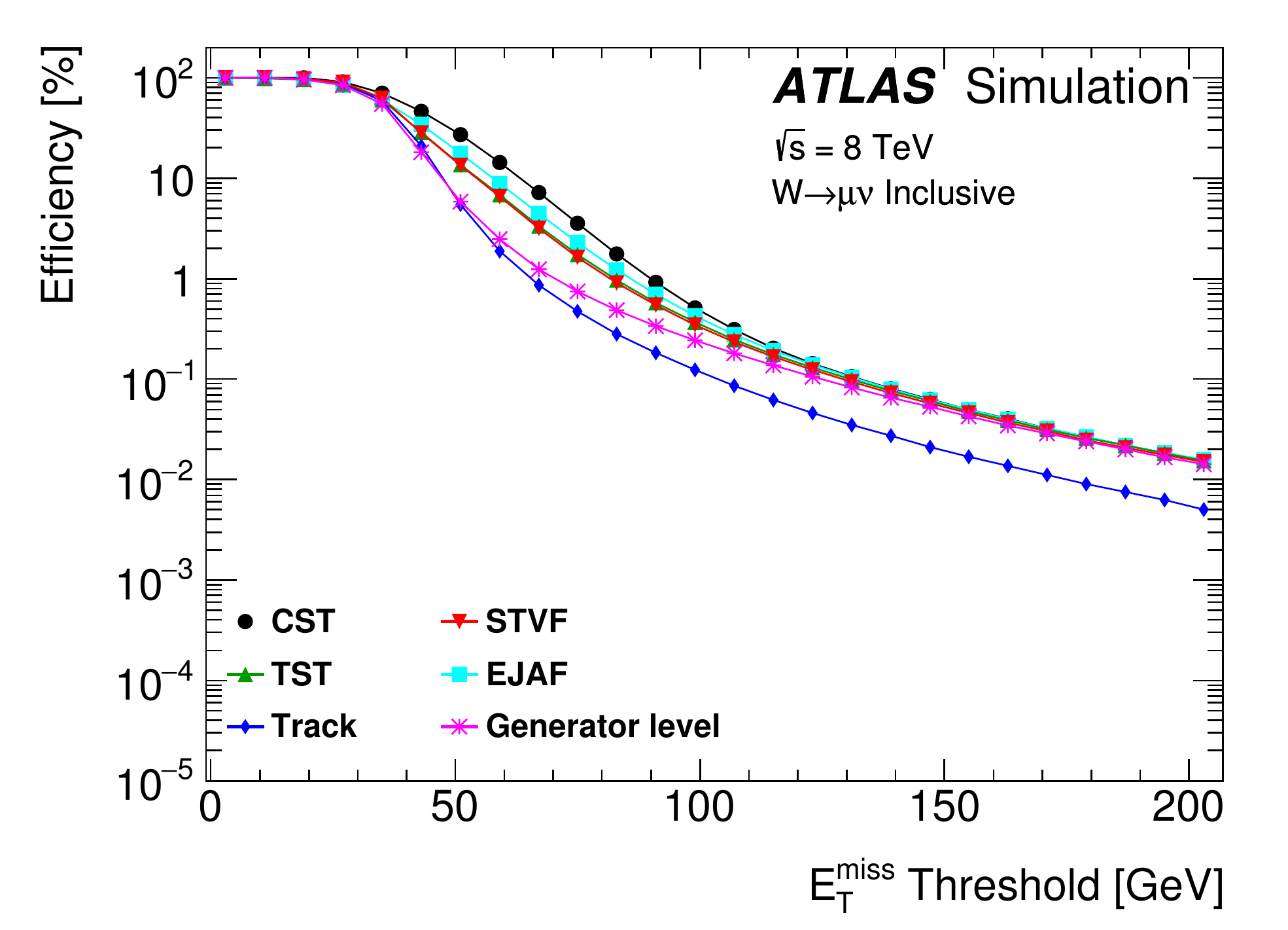}}
\caption{The selection efficiency is shown versus the
  \met~threshold for (a) \Zmm~and (b) \Wmn~inclusive MC simulation events.}
\label{fig:tails_zw}
\end{figure}

In Figure~\ref{fig:tails_cst_tst}, selection efficiencies are shown as
a function of the \met~threshold requirement for various
simulated physics processes defined
in Section~\ref{sec:MC} with no lepton, jet, or \mT~threshold
requirements. The physics object and event selection criteria are not
applied in order to show the selection efficiency resulting from the
\met~threshold requirement without biases in the event topology from the ATLAS detector acceptance for leptons or jets. Only the efficiencies
for the CST and TST \met~distributions are compared for brevity. 
In Figure~\ref{fig:tails_cst_tst}
(a), the efficiencies with the TST \met~selection are shown.
Comparing the physics processes while imposing a moderate \met~threshold requirement of $\sim$100 \GeV~results in a 
selection efficiency of 60\% for an ATLAS
search for gluino-pair production~\cite{Aad2015iea}, which is labelled as
``SUSY''. The VBF \Htau~and \ttbar~events are also selected with
high efficiencies of 14\% and 20\%, respectively. 
With the 100~\GeV{} \met~threshold
the selection efficiencies for these processes 
are more than an order of magnitude
higher than those for leptonically decaying $W$ bosons and more than
two orders of magnitude higher than for $Z$ boson events.

The \Zee~events
have a lower selection efficiency (around 20 times lower at
\met~$=$~100 \GeV{}) than the \Zmm~events. 
This is due to the
muon tracking coverage, which is limited to $|\eta|$~$<$~2.7, whereas the
calorimeter covers $|\eta|$~$<$~4.9. Muons 
behave as minimum-ionizing particles in the ATLAS calorimeters, 
so they are not included in the \met~outside the
muon spectrometer acceptance. The electrons on the other hand are measured by
the forward calorimeters. The electron and muon
decay modes of the $W$ boson have almost identical
selection efficiencies at \met~$=$~100~\GeV~because there is
\mettrue~from the neutrino. However, the differences in selection efficiency are around a
factor of four higher for \Wmn~than for \Wen~at \met~$=$~350~\GeV{}. 
Over the entire \met~spectrum, the differences between the electron
and muon final states for $W$ bosons are smaller than that for $Z$
bosons because there is a neutrino in \Wln{} events as opposed to none
in the \Zll{} final state.

In Figure~\ref{fig:tails_cst_tst}(b), the selection efficiencies for 
CST \met~threshold requirements are divided by those obtained 
using the TST \met. 
The selection efficiencies resulting from CST \met~thresholds for SUSY, 
\ttbar{}, 
and VBF \Htau{} are within 10\% of the efficiencies obtained using the TST \met{}. 
For \met~thresholds from 40--120~\GeV{}, the
selection efficiencies for $W$ and $Z$ boson events are higher by up to
60--160\% for CST \met~than
TST \met{}, which come from pileup contributions broadening the CST
\met~distribution. 
The \Zmm~and \Zee~events, which have no
\mettrue{}, show an even larger increase of 2.6 times as many
\Zee~events passing a \met~threshold of 50~\GeV{}. The increase is not
as large for \Zmm~as \Zee~events because neither \met~algorithm
accounts for forward muons ($|\eta|$~$>$~2.7) as discussed above.
Moving to a higher \met~threshold, 
mismeasured tracks in the TST algorithm cause it to select more 
\Zee~events with 120~$<$~\met~$<$~230~\GeV{}. 
In addition, the CST \met~also includes
electron energy contributions (\pT~$<$~20~\GeV{}) in the forward
calorimeters ($|\eta|$~$>$~3.1) that the TST does not.

The CST and TST \met~distributions agree within 10\% in selection
efficiency for \met~$>$~250~\GeV{} for all physics processes
shown. This demonstrates a strong correlation between the \met~distributions for
events with large \mettrue{}, or a strong correlation between 
the physics objects
that cause a large mismeasurement in \met~for $Z$ events. 


\begin{figure}[htbp]
 \subfigure[]{\includegraphics[height=60mm]{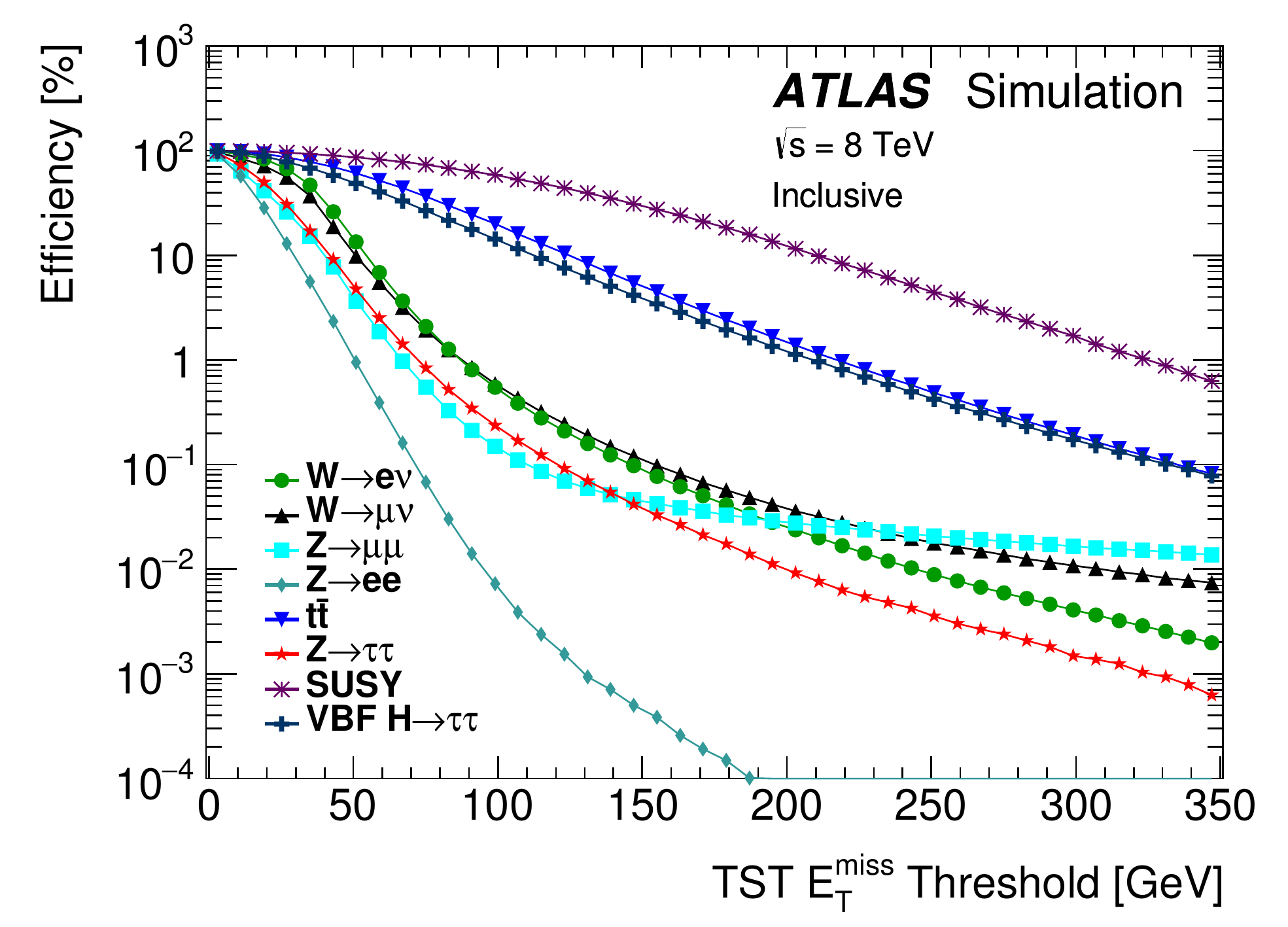}}
 \subfigure[]{\includegraphics[height=60mm]{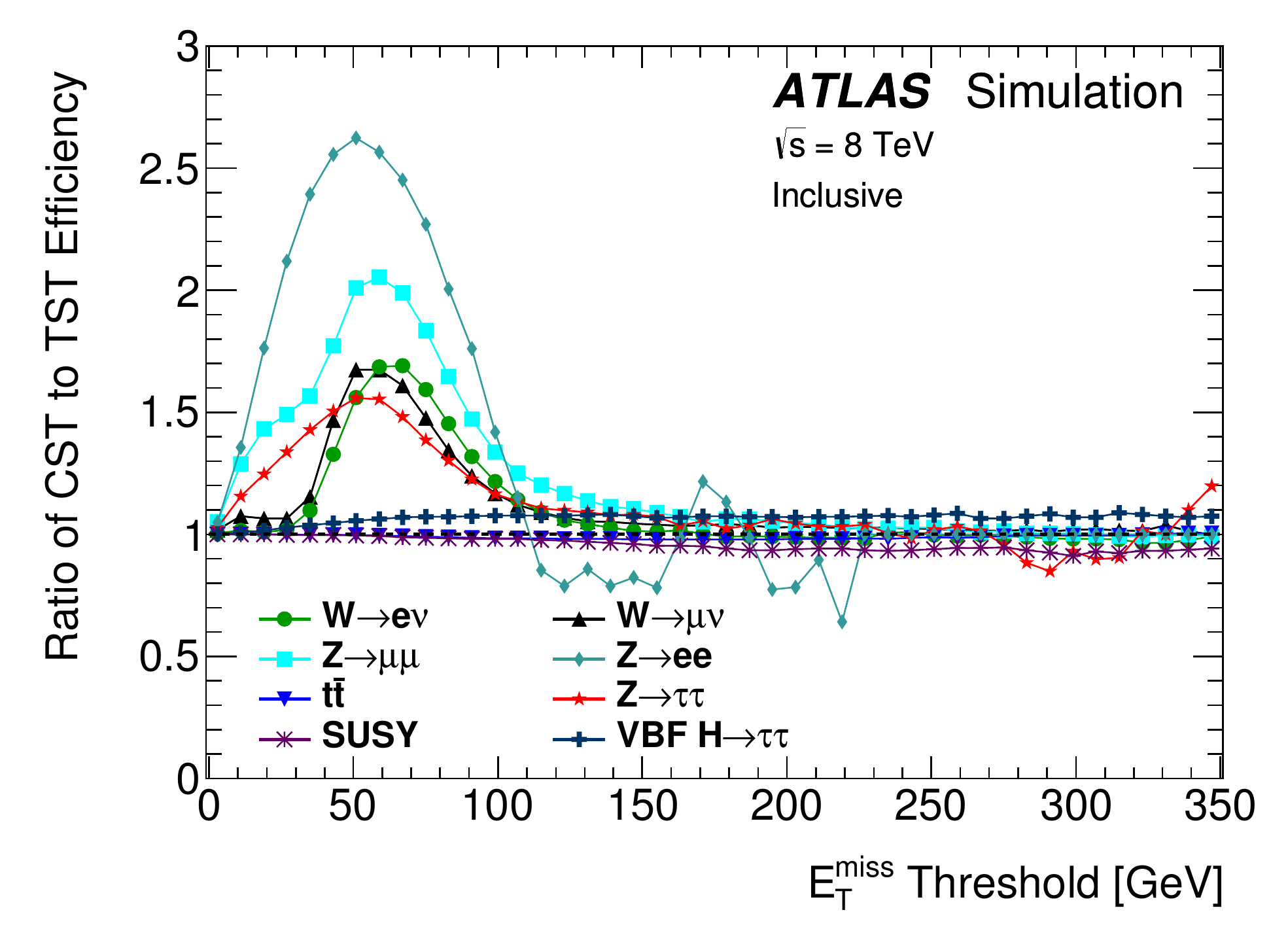}}
\caption{(a) The selection efficiency with TST \metmag{} versus the
  \metmag~threshold and (b) the ratio of CST to
  TST efficiencies versus \met{} threshold. In both cases, results are shown for several processes.}
\label{fig:tails_cst_tst}
\end{figure}

\subsection{Correlation of fake \met~between algorithms}
\label{sec:tail_correlation}

The tracking and the calorimeters provide almost
completely independent
estimates of the \met{}. These two measurements complement each other,
and the \met~algorithms discussed in this paper combine that information
in different ways. The distribution of the TST \met~versus the CST \met~is shown for the simulated
0-jet \Zmm~sample in
Figure~\ref{fig:tails_corr}. This figure shows the
correlation of fake \met~between the two algorithms, 
which originates from many sources including incorrect vertex association and miscalibration of high-\pT{} physics objects.

\begin{figure}[htbp]
\begin{centering}
\includegraphics[height=60mm]{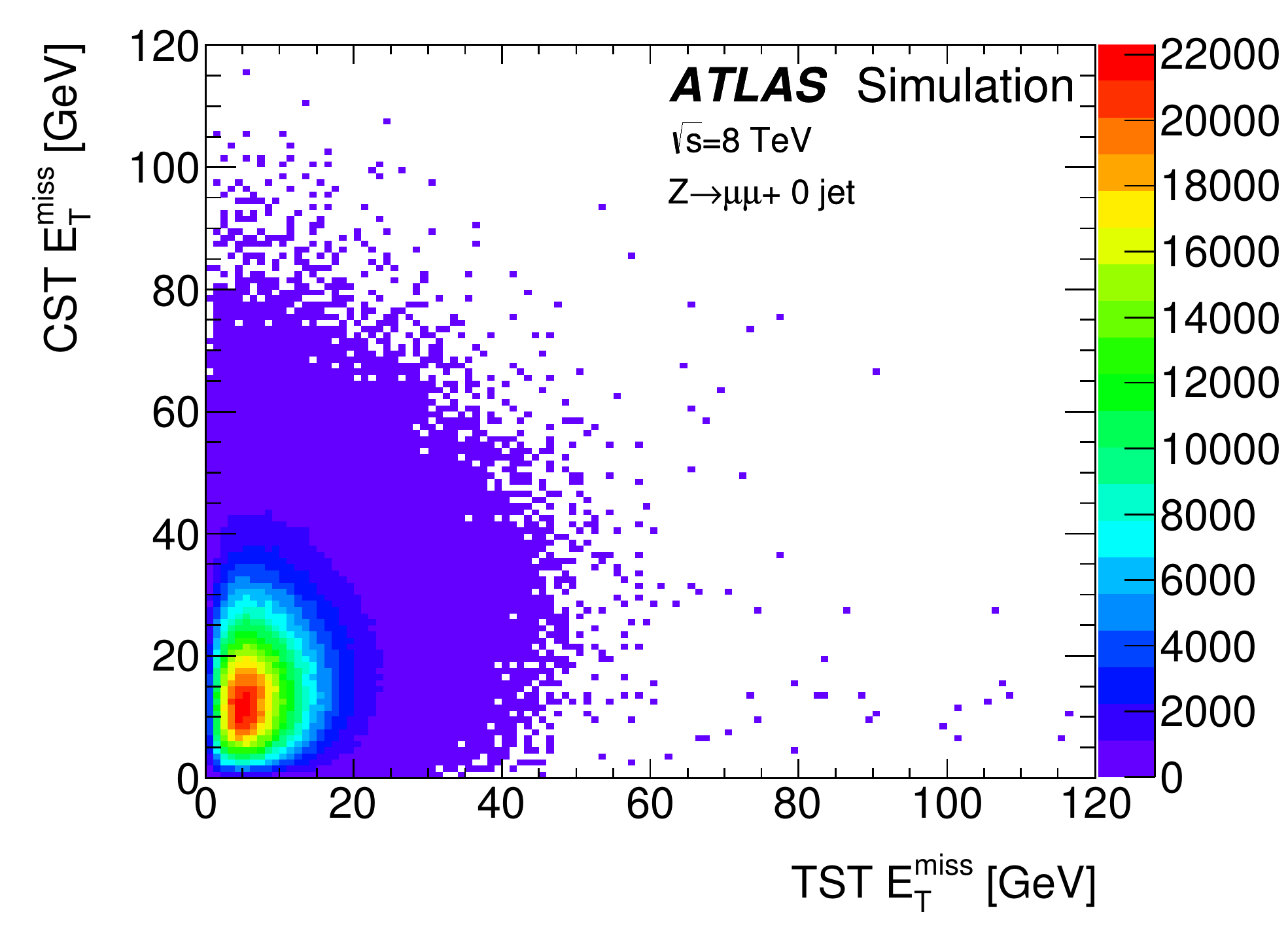}
\caption{The CST \metmag~versus the TST \metmag{} in
  \Zmm~$+$~0-jet events from the MC simulation. The vector
  correlation coefficient is 0.177~\cite{correlation}.} 
\label{fig:tails_corr}
\end{centering}
\end{figure}

Vector correlation coefficients~\cite{correlation}, shown in Table~\ref{tab:corr2d_tails}, are used to estimate the correlation between
the \met~distributions resulting from different reconstruction algorithms. 
The value of the vector correlation coefficients ranges from 0
to 2, with 0 being the least correlated and 2 being the most
correlated. The coefficients shown are obtained using the simulated 
0-jet and inclusive \Zmm~MC samples.
The least-correlated \met~distributions are the CST and Track \met{},
which use mostly independent momenta measurements in their reconstructions. The
correlations of the other \met~distributions to the CST \met~decrease
as more tracking information is used to suppress the
pileup dependence of the soft term, with the TST \met{} distribution having the second smallest
vector correlation coefficient with respect to the CST \met{}
distribution. Placing requirements on a
combination of \met{} distributions or requiring the difference in
azimuthal direction between two \met{} vectors to be small can greatly
reduce fake \met~backgrounds, especially using the least-correlated \met~distributions. 
Such strategies are adopted in several Higgs boson 
analyses in ATLAS~\cite{hww_atlas,hbb_atlas,zhinv_atlas}. 

\begin{table}
 \begin{center}
\caption{Vector correlation coefficients are shown between
  \met~definitions in \Zmm~MC simulation. Below the diagonal are
  events in the 0-jet sample, and above
  the diagonal are inclusive events. }\label{tab:corr2d_tails}
$\newline$ 
\begin{tabular}{l||rrrrr} 
 \met{} & CST  & TST & Track & STVF & EJAF \\ \hline \hline
 CST    & 2 & 0.261 & 0.035 &0.525 & 0.705  \\
 TST    & 0.177  & 2 &  0.232 & 1.557 & 0.866  \\
 Track    & 0.153  & 1.712 & 2 & 0.170 & 0.065\\
 STVF    & 0.585 & 1.190 & 1.017 & 2& 1.256 \\
 EJAF    & 0.761  & 0.472 & 0.401 & 1.000 & 2\\
\end{tabular}
 \end{center}
\end{table}

\section{Jet-\pT~threshold and vertex association selection}
\label{sec:jet_opt}

Jets can originate from pileup interactions, so tracks matched
to the jets are extrapolated back to the beamline to ascertain 
whether they are 
consistent with originating from the hard scatter or a pileup collision. 
The JVF defined in Section~\ref{sec:hard_term}
is used to separate pileup jets and jets from the hard scatter.
The STVF, EJAF, and TST \met~algorithms improve their jet identification 
by removing jets associated with pileup
vertices or jets that have a large degradation in momentum resolution
due to pileup activity. Energy contributions from jets not associated
with the hard-scatter vertex are included in the soft term. For the TST,
this means that charged particles from jets 
not associated with the hard-scatter vertex may then 
enter the soft term if their position along the beamline is consistent
with the $z$-position of the hard-scatter vertex.

Applying a JVF cut is a trade-off between 
removing jets from pileup interactions
and losing jets from the hard scatter. 
Therefore, several values of the JVF selection criterion are
considered in \Zll{} events with jets having \pt~$>$~20 \GeV{}; their impact
on the \met~resolution and scale is investigated in
Figure~\ref{fig:jet_jvf_opt}. Larger JVF thresholds on
jets reduce the pileup dependence of the \met{} resolution, but they
simulataneously worsen the \met{} scale. Thus the best compromise
for the value of the JVT threshold is chosen.
Requiring JVF~$>$~0.25 greatly
improves 
the stability of the \met{} resolution with respect to pileup
by reducing the dependence of the
\met~resolution on the number of reconstructed vertices as shown
in Figure~\ref{fig:jet_jvf_opt}(a). The \metvec{} in \Zll{} events ideally has
a magnitude of zero, apart from some relatively infrequent neutrino
contributions in jets. So its magnitude should be consistently zero along any
direction. The \ptZvec{} remains unchanged for different JVF
requirements, which makes its direction a useful reference to check the calibration of the \metvec{}. The difference from
zero of the average value of
the reconstructed \met{} along \ptZvec{} increases as tighter
JVF selections are applied as shown in Figure~\ref{fig:jet_jvf_opt}(b).
Requiring a JVF threshold of 0.25 or higher slightly improves the stability of 
the resolution with respect to pileup, whereas it visibly degrades the
\met{} response by removing too many hard-scatter jets. 
Lastly, pileup jets with \pT~$>$~50 \GeV{} are very rare~\cite{ATLASPU8TeV}, so applying the JVF
requirement above this \pT~threshold is not useful.
Therefore, requiring JVF
to be larger than 0.25 for jets with \pt~$<$~50 \GeV~within 
the tracking volume ($|\eta|$ $<$ 2.4) is the 
preferred 
threshold for the \met~reconstruction. 

\begin{figure*}
\begin{center}
  \subfigure[]{\includegraphics[height=55mm]{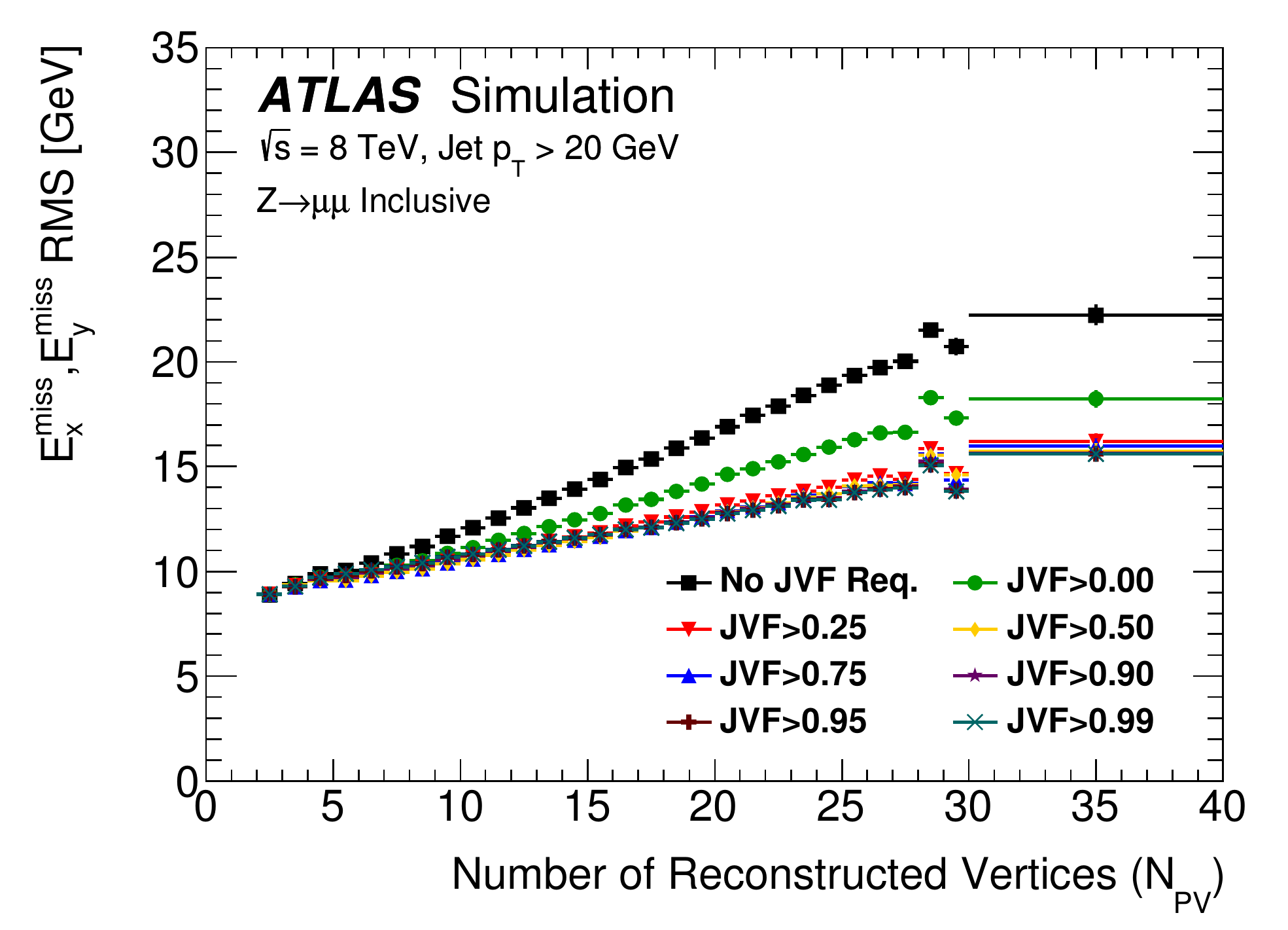}} 
  \subfigure[]{\includegraphics[height=55mm]{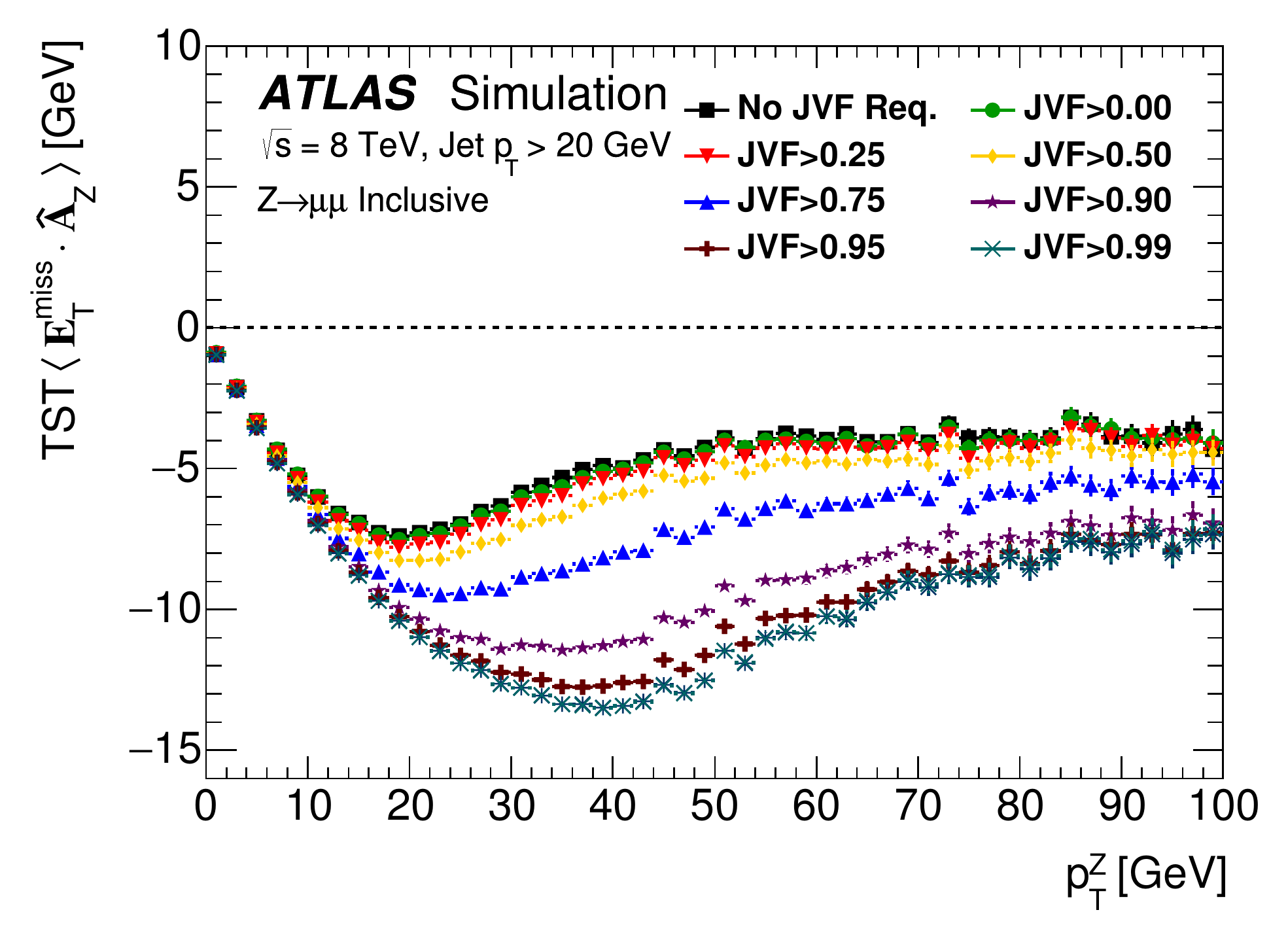}} 
\end{center}
\caption{The (a) TST \met{} resolution versus the number of
  reconstructed vertices per
  bunch crossing (\Npv) and the (b) TST \metvec~in the direction of the \ptZvec{} are shown 
  for the different JVF selection criterion values applied to jets with
  \pt~$>$~20 \GeV~and $|\eta|$ $<$ 2.4 using the \zmm\ simulation.}
\label{fig:jet_jvf_opt}
\end{figure*}

In addition, the \pT~threshold, which defines the boundary between the
jet and soft terms, is optimized. For these studies, the jets with \pt~$>$~20 \GeV~and $|\eta|$ $<$ 2.4 are
required to have JVF~$>$~0.25.
A procedure similar to that used for the JVF optimization is used for the jet-\pT~threshold 
using the same two metrics as shown in Figure~\ref{fig:jet_pt_opt}.
While applying a higher \pT~threshold improves the
\met~resolution versus the number of pileup vertices, by decreasing the
slope, the \metvec~becomes
strongly biased in the direction opposite to the \ptZvec{}. Therefore, 
the \pT~threshold of 20~\GeV{} is preferred.

\begin{figure*}
\begin{center}
  \subfigure[]{\includegraphics[height=55mm]{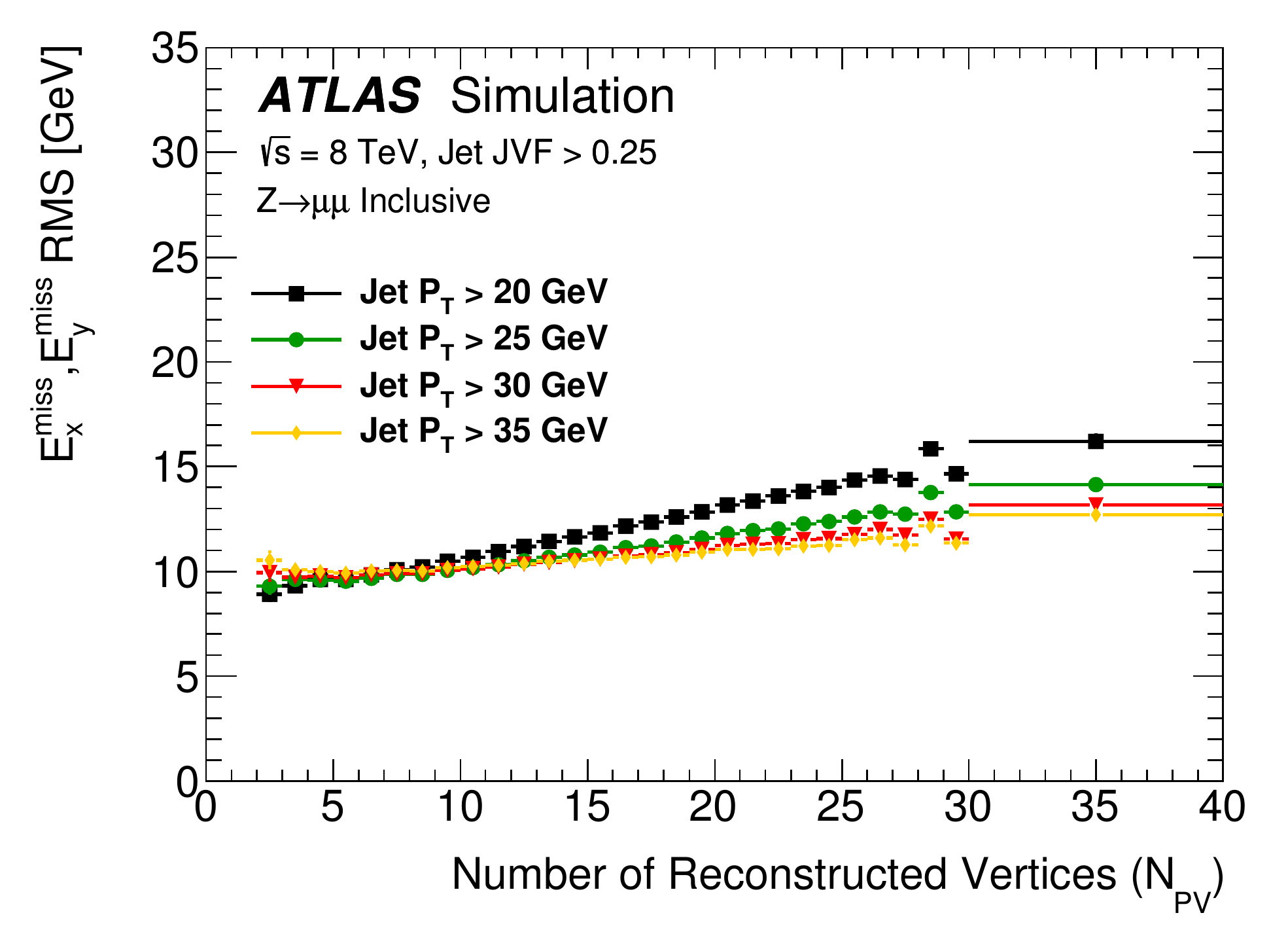}} 
  \subfigure[]{\includegraphics[height=55mm]{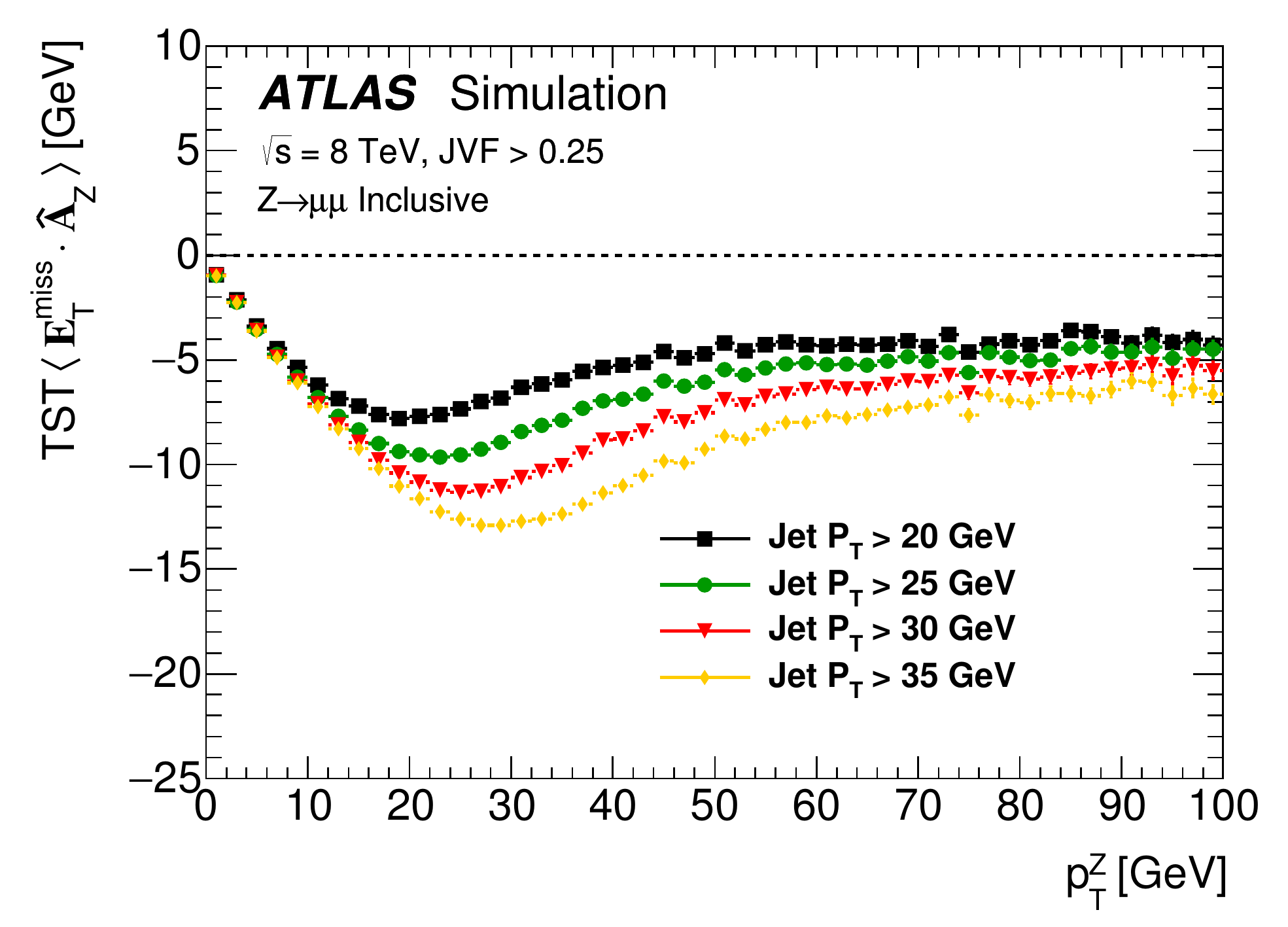}} 
\end{center}
\caption{The (a) TST \met{} resolution as a function of the number of
  reconstructed vertices per
  bunch crossing (\Npv)
  and the (b) TST \metvec~in the direction of the \ptZvec~are shown
  for different jet-\pT{} thresholds using the \zmm\
  simulation. JVF~$>$~0.25 is required for all jets with \pt~$>$~20 \GeV~and $|\eta|$ $<$ 2.4.}
\label{fig:jet_pt_opt}
\end{figure*}


\section{Systematic uncertainties of the soft term}
\label{sec:systematics}

The \metvec~is reconstructed from the vector sum of several terms 
corresponding to different types of contributions from reconstructed physics objects, as defined in 
Eq.~(\ref{eq7}).
The estimated uncertainties in the energy scale and momentum resolution for 
the electrons~\cite{Aad2014nim}, muons~\cite{ATLASMuonPerf12}, jets~\cite{ATLAS-CONF-2015-017}, \tauvis~\cite{Aad2014rga}, and photons~\cite{Aad2014nim} are propagated 
into the \met{}.
This section describes the estimation of the systematic uncertainties for the \met{} soft term.
These uncertainties take into account the impact of the 
generator and underlying-event modelling used by the ATLAS Collaboration, as well as effects from pileup.

The balance of the soft term with the calibrated physics objects
is used to estimate the soft-term systematic uncertainties in
\Zmm~events, which have very little \mettrue{}. The transverse momenta
of the calibrated physics objects, \pthardvec{}, is defined as 
\begin{eqnarray}
       \pthardvec =
       \sumve         +
       \sumvmu        +
       \sumvgamma   +
       \sumvtau    +
       \sumvjet{},
       \label{eqn:pthard}
\end{eqnarray}
which is the vector sum of the transverse momenta of the high-\pT{}
physics objects. 
It defines an axis (with unit vector \pthardhat{}) in the transverse plane of 
the ATLAS detector along which the \met~soft term is expected to 
balance \pthard{} in \Zmm~events. This balance is sensitive to the differences in
calibration and reconstruction of the \metsoft{} between data and MC
simulation and thus is sensitive to the uncertainty in the soft
term. This discussion is similar to the one in Section~\ref{sec:met_response};
however, here the soft term is compared to the hard term rather than
comparing the \metvec{} to the recoil of the $Z$.

\subsection{Methodology for CST}
\label{sec:calo_syst}

Two sets of systematic uncertainties are considered for the 
CST. 
The same approach is used for the STVF and EJAF algorithms
to evaluate their soft-term systematic uncertainties. 
The first approach decomposes the systematic uncertainties into the
longitudinal and transverse components along the direction of 
\pthardvec{}, whereas the 
second approach estimates the global scale and resolution
uncertainties. While both methods were recommended for analyses of the
8 \TeV{} dataset, the first method, described in
Section~\ref{sec:cstsys_balance}, gives smaller uncertainties. Therefore, the second method, which is discussed in Section~\ref{sec:cstsys_datamc}, is now treated as a cross-check.

Both methods 
consider a subset of \Zmm~events that do not have any jets with $\pt$~$>$~20
\GeV~and $|\eta|$~$<$~4.5. 
Such an event topology is optimal for
estimation of the soft-term systematic uncertainties
because only the muons and the soft term contribute to the \met.
In principle the methods are valid in event topologies with any jet multiplicity, 
but the \Zmm$+\geq$1-jet events are more susceptible to
jet-related systematic uncertainties.

\subsubsection{Evaluation of balance between the soft term and the hard 
  term}
\label{sec:cstsys_balance}

The primary or ``balance'' 
method exploits the momentum balance in the
transverse plane between the soft and hard terms in \Zll~events, and the level of disagreement between
data and simulation is assigned as a systematic uncertainty.

The \metsoftvec{} is decomposed along the \pthardhat{} direction.  
The direction orthogonal to \pthardhat{} is referred to as the 
perpendicular direction while the component parallel to \pthardhat~is
labelled as the longitudinal direction.
The projections of \metsoftvec{} along those directions are defined as: 
\begin{equation}
 \begin{array}{r@{}l}
 \metsoftpara &{}= \metsoft{} \cos\phi(\metsoftvec{},\pthardvec{}), \\
 \metsoftperp &{}= \metsoft{} \sin\phi(\metsoftvec{},\pthardvec{}), \label{eqn:metsyst}
 \end{array}
\end{equation}
The \metsoftpara~is sensitive to scale and resolution differences between the data and simulation 
because the soft term should balance the \pthardvec{} in
\Zmm~events. For a narrow range of \pthard{} values, the mean and width of the
\metsoftpara~are compared between data and MC simulation. On the other hand, the perpendicular 
component, \metsoftperp, is only sensitive to differences in resolution. 
A Gaussian function is fit to the \metvec{} projected onto
\pthardhat{} in bins of \pthard{}, and the resulting Gaussian mean and
width are shown 
in Figure~\ref{fig:syst_cst_EL_pthard}. The mean increases linearly
with \pthard{}, because the soft term is not calibrated to the correct
energy scale. On the other hand, the width is relatively independent of 
\pthard{}, because the width is mostly coming from pileup contributions.

The small discrepancies in mean and width between data and simulation are 
taken as the systematic uncertainties for the scale and 
resolution, respectively. A small dependence on
the average number of collisions per bunch crossing is
observed for the scale and resolution uncertainties for high
\pthard{}, so the uncertainties are computed in three ranges of
pileup and three ranges of \pthard{}. The scale uncertainty varies
from -0.4 to 0.3~\GeV{} depending
on the bin, which reduces the uncertainties from the 5\% shown in
Figure~\ref{fig:syst_cst_EL_pthard} for \pthard~$>$~10~\GeV{}. 
A small difference in the uncertainties for the resolution 
along the longitudinal and perpendicular directions is observed, so they are 
considered separately. The average uncertainty is about 
2.1\% (1.8\%) for the longitudinal (perpendicular) direction. 

\begin{figure}[t]
 \begin{center}
    \subfigure[]{\includegraphics[height=53mm]{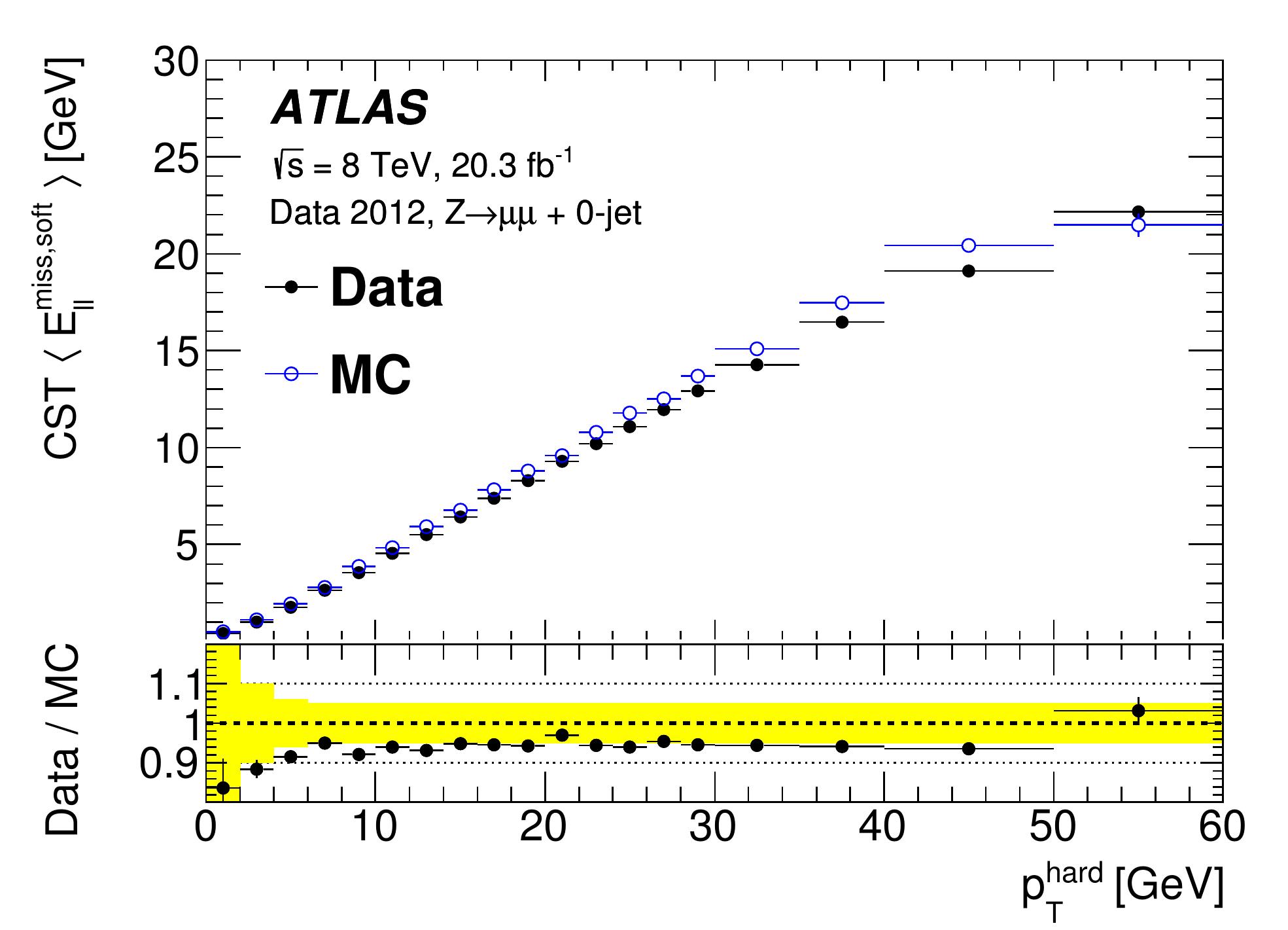}}
    \subfigure[]{\includegraphics[height=53mm]{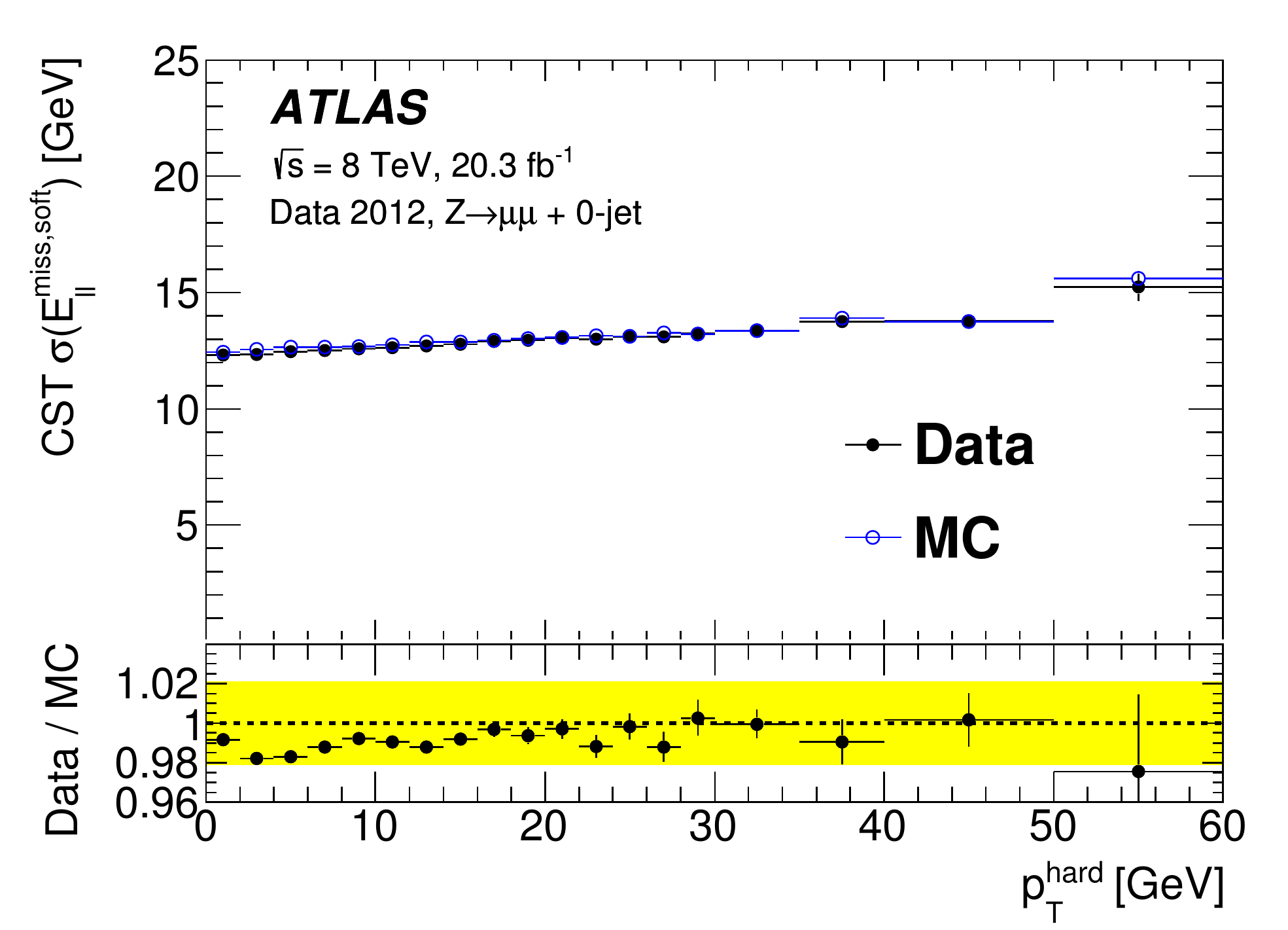}}
 \end{center}
\caption{The (a) mean and (b) Gaussian width of 
the CST \metvec~projected onto
\pthardhat~are each shown as
a function of \pthard~in \Zmm$+$0-jet events. The ratio of data to MC simulation
is shown in the lower portion of the plot with the band representing the
assigned systematic uncertainty.}
\label{fig:syst_cst_EL_pthard}
\end{figure}

\subsubsection{Cross-check method for the CST systematic uncertainties}
\label{sec:cstsys_datamc}

As a cross-check of the method used to estimate the CST uncertainties, the
sample of \Zmm$+$0-jet events is also used to evaluate the level of agreement between data and simulation. The projection of the \metvec~onto \pthardhat~provides a test for
potential biases in the \met~scale. The systematic uncertainty in the soft-term scale 
is estimated by comparing 
the ratio of data to MC simulation for $\langle\metvec\cdot \pthardhat \rangle$
versus \sumet(CST) as shown in Figure~\ref{fig:syst_cst_etmiss_sumet}(a). 
The average deviation from unity in the ratio of data to MC simulation is about 8\%, which is taken as a flat uncertainty in the absolute scale. 
The systematic uncertainty in the 
soft-term resolution is estimated by evaluating the level of agreement
between data and MC simulation in the \metx~and \mety~resolution as a function of the
\sumet(CST)~(Figure~\ref{fig:syst_cst_etmiss_sumet}(b)).
The uncertainty on the soft-term resolution is about 2.5\% and is
shown as the band in the data/MC ratio. 

\begin{figure}[h]
 \begin{center}
   \subfigure[]{\includegraphics[height=53mm]{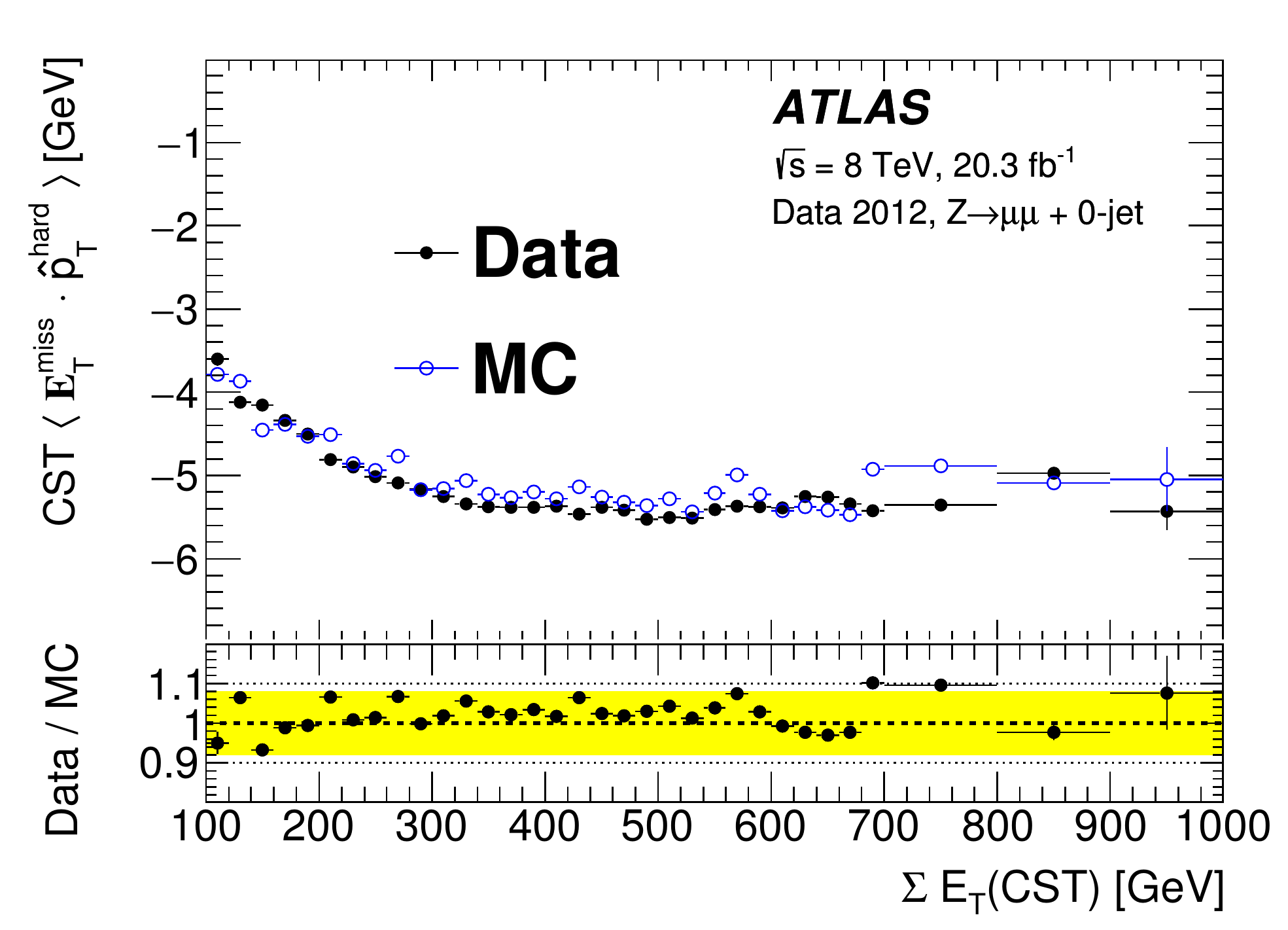}}
    \subfigure[]{\includegraphics[height=53mm]{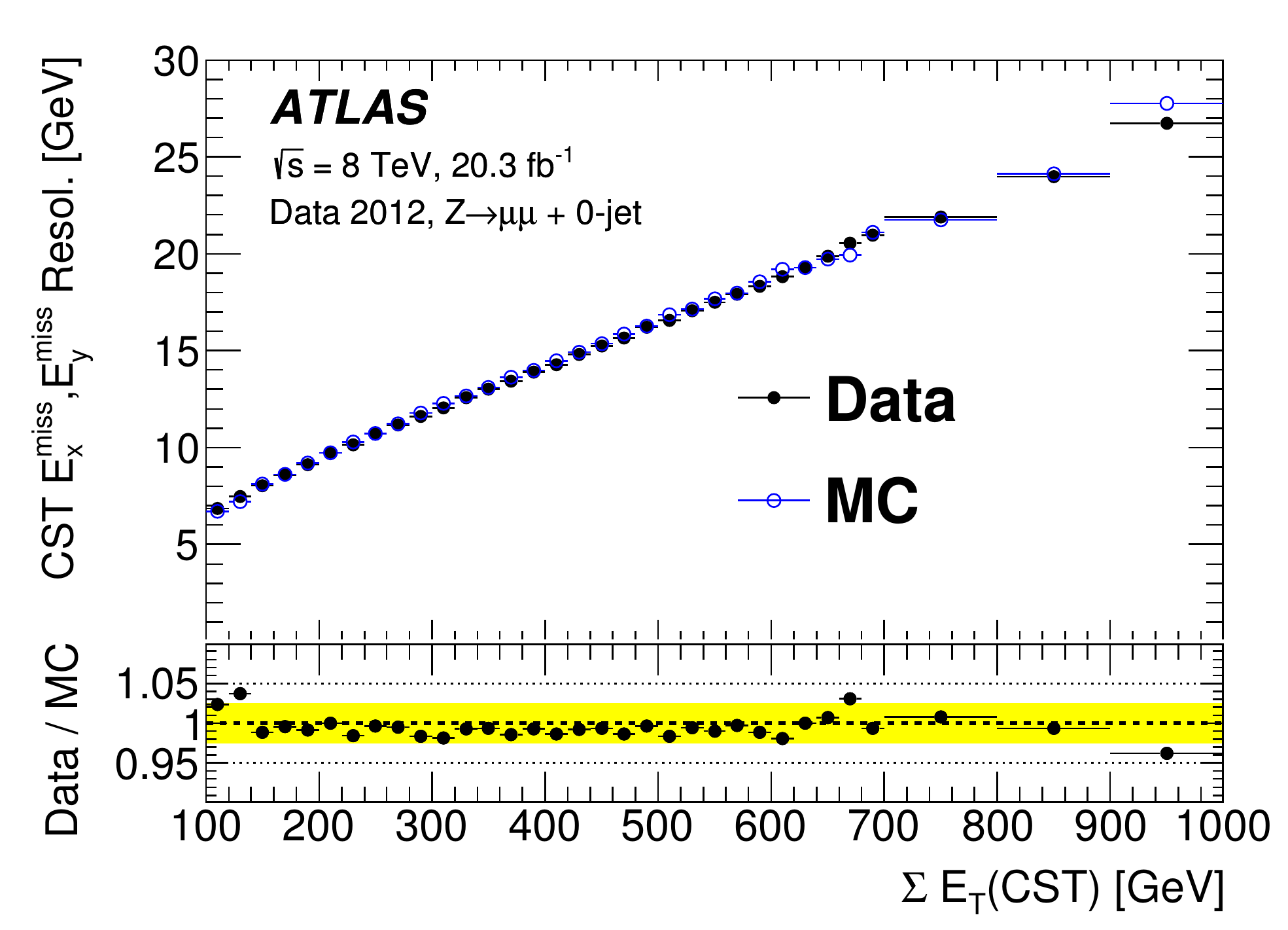}}
 \end{center}
\caption{The (a) projection of CST \metvec~onto \pthardhat{} and (b)
  the Gaussian width (resol.) of the combined distribution of CST \metx~and \mety~are shown versus \sumet(CST).
The ratio of data to MC simulation is shown in the lower portion of the plot with the solid band representing the
assigned systematic uncertainty.}
\label{fig:syst_cst_etmiss_sumet}
\end{figure}

Even though the
distributions appear similar, the results in this section 
are derived by projecting the full \met~onto the \pthardhat~in the 0-jet events,
and are not directly comparable to the ones in
Section~\ref{sec:cstsys_balance}, in which only the soft term is 
projected onto \pthardhat{}.

\subsection{Methodology for TST and Track \met}
\label{sec:track_syst}

A slightly different data-driven methodology is used to evaluate the 
systematic uncertainties in the TST and Track \met{}.
Tracks matched to jets that are included in the hard term are removed from the Track
\met~and are treated separately, as described in Section~\ref{sec:syst_rtrk}.

The method exploits the balance between the soft track term and
\pthardvec{} and is similar to the balance method for the CST.
The systematic uncertainties are split into two components: the longitudinal (\metsoftpara) and 
transverse (\metsoftperp) projections onto \pthardvec{} as defined in
Eq.~(\ref{eqn:metsyst}).

The \metsoftpara~in data is fit with the MC simulation convolved with
a Gaussian function, and the fitted Gaussian mean and width are used to extract the differences between simulation and data. 
The largest fit values of the Gaussian width and offset define the systematic uncertainties. 
For the perpendicular component, the simulation is only smeared by a
Gaussian function of width $\sigma_{\perp}$
to match the data. The mean, which is set to zero in the fit, is very
small in data and MC simulation because the hadronic recoil only affects \metsoftpara{}. The fitting is done in
5~\GeV\ or 10~\GeV\ bins of \pthard\ from 0--50~\GeV{}, and a single bin for \pthard~$>$~50~\GeV{}. 

An example fit is shown in Figure~\ref{fig:ptsoft_fits_example} for
illustration. The 1-jet selection with the JVF requirement is used to show that the differences between data and simulation, from the jet-related
systematic uncertainties, are 
small relative to the differences in the soft-term modelling.
The impact of the jet-related systematic uncertainties is less than 0.1\% in the Gaussian smearing
($\sigma$~$=$~1.61~\GeV{}), indicating that the jet-related systematic
uncertainties do not affect the extraction of the TST systematic uncertainties.

\begin{figure}[htb!]
 \begin{center}
\includegraphics[height=68mm]{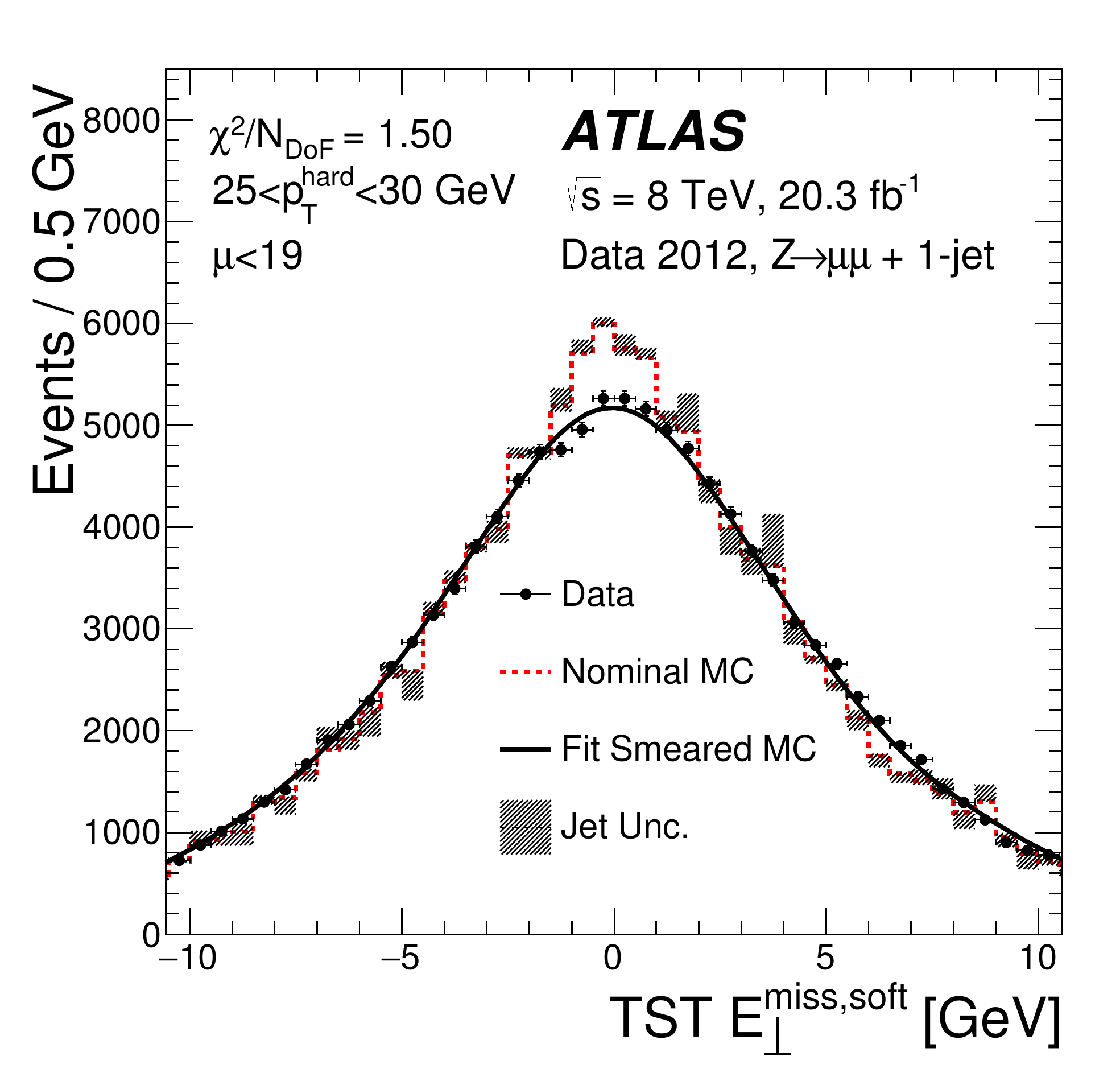}
 \end{center}
\caption{Fit to the TST \metsoftperp{} for $\mu$~$<$~$19$ and
  25~$<$~\pthard~$<$~30~\GeV\ in the 1-jet sample. The nominal MC
  simulation, the jet-related systematic uncertainties
  (hashed band), and the data are shown. The
  nominal MC simulation is convolved with a Gaussian function until it
  matches the data, and the resulting fit is shown with the solid
  curve. The jet counting for the 1-jet selection uses the same
      JVF criterion as the TST \met~reconstruction algorithm.
}
\label{fig:ptsoft_fits_example}
\end{figure}

The Gaussian width squared of \metsoftpara~and 
\metsoftperp~components 
and the fitted mean of \metsoftpara~for
data and MC simulation are shown versus
\pthard~in Figure~\ref{fig:syst_trkmet_summary}. The systematic uncertainty squared
of the convolved Gaussian width and the systematic uncertainty of the offset
for the longitudinal component are shown in
the bands. While the systematic uncertainties are applied to
the MC simulation, the band is shown centred around the data to show that 
all MC generators plus parton shower models agree with the data within the assigned uncertainties.
Similarly for the \metsoftperp{}, the width of the convolved Gaussian function for the perpendicular component 
is shown in the band. The \textsc{Alpgen+Herwig}~simulation
has the largest disagreement with data, so the Gaussian smearing
parameters and offsets applied to the simulation are used as the
systematic uncertainties in the soft term. The \pthard~$>$~50~\GeV~bin
has the smallest number of data entries; therefore, it has the largest
uncertainties in the fitted mean and width. In this
bin of the distribution shown in Figure~\ref{fig:syst_trkmet_summary}(a), the statistical
uncertainty from the \ALPGEN$+$\HERWIG{} simulation, which is not the most
discrepant from data, is added to the uncertainty
band, and this results in a systematic uncertainty band that spans the
differences in MC generators for $\sigma^2(\metsoftpara)$ for events with \pthard~$>$~50~\GeV{}.

\begin{figure}[h!]
 \begin{center}
    \subfigure[]{\includegraphics[height=53mm]{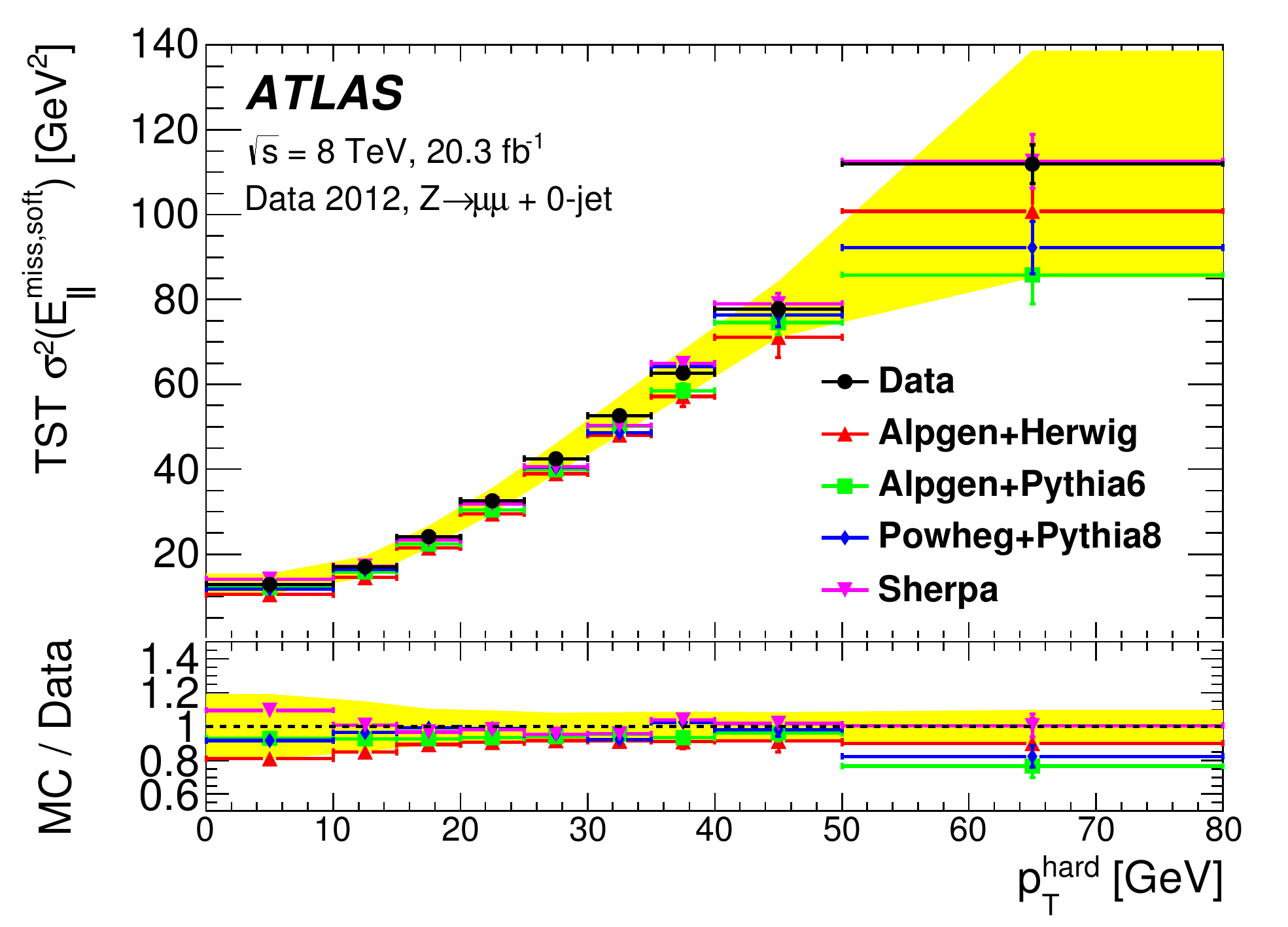}}
    \subfigure[]{\includegraphics[height=53mm]{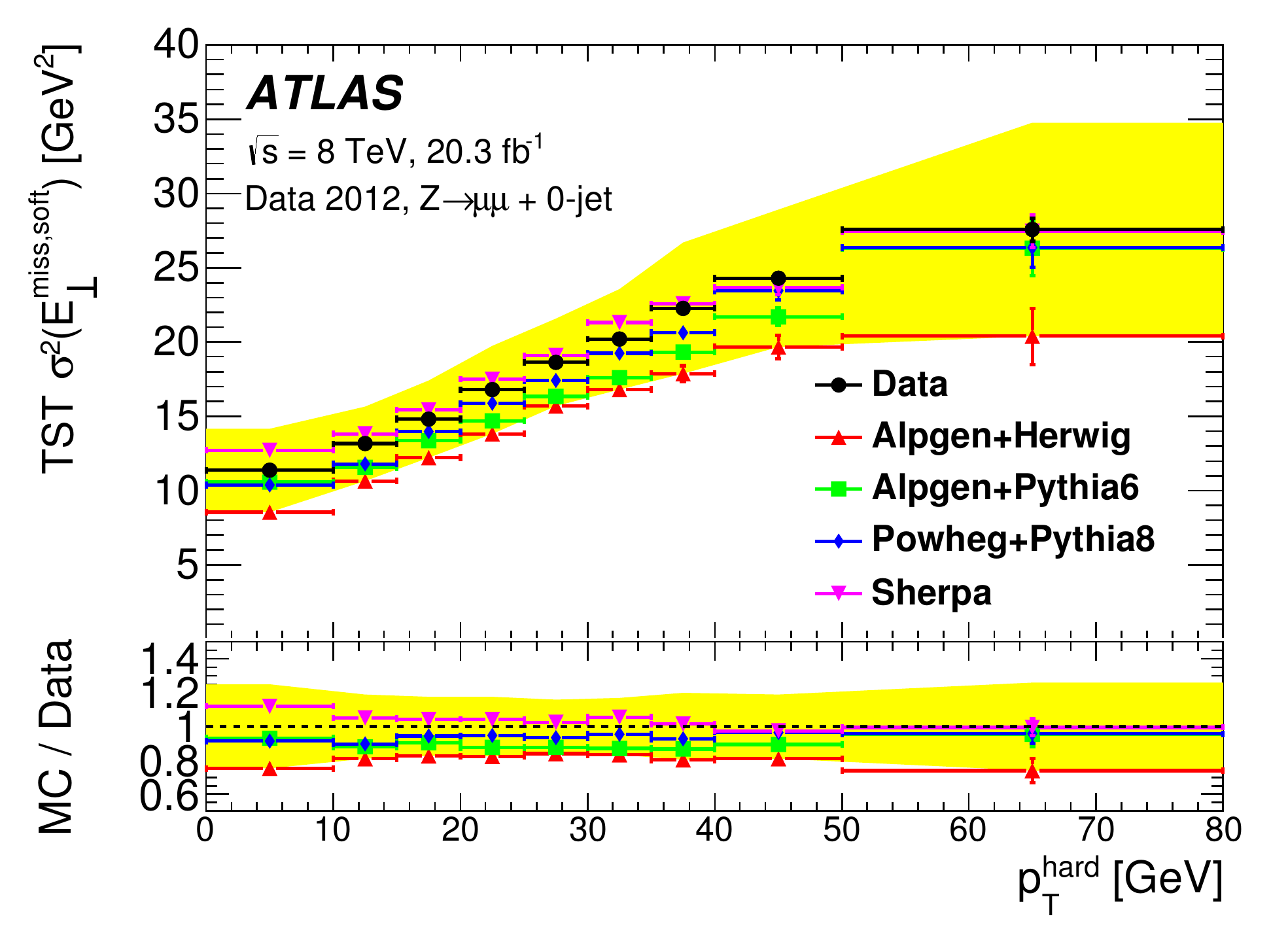}}\\
    \subfigure[]{\includegraphics[height=53mm]{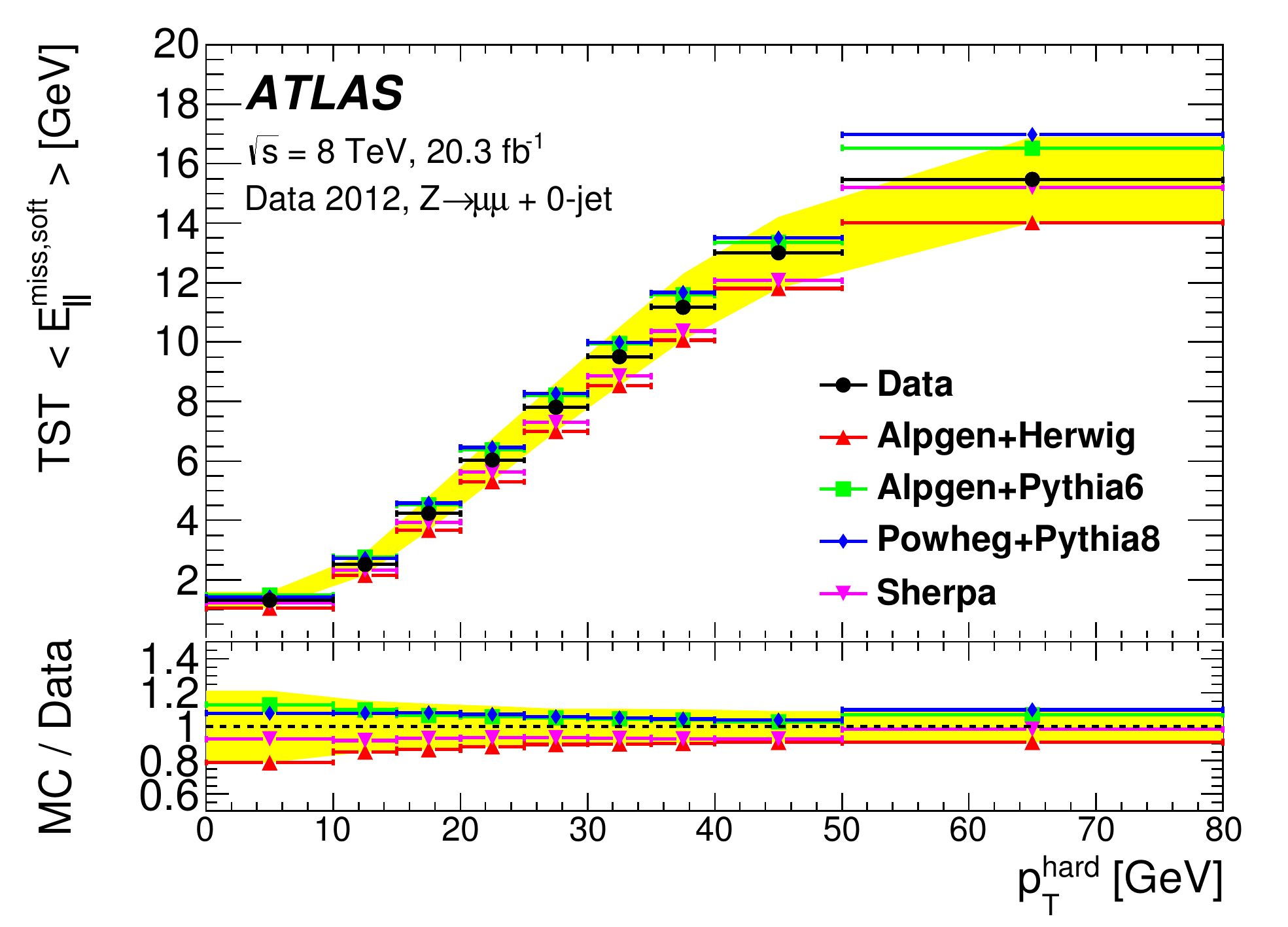}}\\

 \end{center}
\caption{The fitted TST
  (a) $\sigma^2(\metsoftpara)$, (b) 
  $\sigma^2(\metsoftperp)$, and (c) $\langle
  \metsoftpara\rangle$ in each case versus \pthard{} are shown in data
  and \ALPGEN$+$\HERWIG{}, \POWHEG{}$+$\PYTHIAEight{}, \SHERPA{}, and
  \ALPGEN$+$\PYTHIA~\Zmm\ simulation. The error bars on the data
  and MC simulation points are the errors from the Gaussian fits. The solid band,
  which is centred on the data,
  shows the parameter's systematic uncertainties from
  Table~\ref{tab:trkmetsyst2012}. The insets at the bottom of the figures show the ratios of the MC predictions to the data. }
\label{fig:syst_trkmet_summary}
\end{figure}

The impact of uncertainties coming from the parton shower model, the number of jets, $\mu$ dependence,
JER/JES uncertainties, and forward versus central jet differences was evaluated. 
Among the uncertainties, the differences between the generator and 
parton shower models have the most dominant effects. 
The total TST systematic uncertainty is summarized in 
Table~\ref{tab:trkmetsyst2012}.

\begin{table}
 \begin{center}
\caption{
  The TST scale ($\Delta_{\mathrm{TST}}$)
  and resolution uncertainties ($\sigma_{\parallel}$ and $\sigma_{\perp}$) are 
  shown in bins of \pthard{}.}\label{tab:trkmetsyst2012}
$\newline$ 
\begin{tabular}{l||rrr} 
  \pthard\ Range [\GeV{}] & $\Delta_{\mathrm{TST}}$~[\GeV{}]  & $\sigma_{\parallel}$~[\GeV{}] & $\sigma_{\perp}$~[\GeV{}]  \\ \hline \hline
  0--10     & 0.3 & 1.6 & 1.7 \\
  10--15   & 0.4 & 1.6 & 1.6 \\
  15--20 & 0.6 & 1.6 & 1.6 \\
  20--25 & 0.7 & 1.8 & 1.7 \\
  25--30 & 0.8 & 1.9 & 1.7 \\
  30--35 & 1.0 & 2.1 & 1.8 \\
  35--40 & 1.1 & 2.4 & 2.1 \\
  40--50 & 1.2 & 2.6 & 2.2 \\
  $>$~50     & 1.4 & 5.2 & 2.7 \\
\end{tabular}
 \end{center}
\end{table}

\subsubsection{Propagation of systematic uncertainties}

The CST systematic uncertainties from the balance method
defined in Section~\ref{sec:cstsys_balance} are propagated to the
nominal \metsoftvec~as follows:

\begin{subequations}
\begin{eqnarray}
\metsoftsysreso &=& (1 \pm R_{\parallel (\perp)})(\metsoftparaperp  - \langle\metsoftparaperp\rangle) + \langle\metsoftparaperp\rangle \label{eq:var2_cst_met} \\
\metsoftsysscale &=& \metsoftpara \pm \Delta_{\textrm{CST}} \label{eq:varscale2_cst_met}
\end{eqnarray}
\end{subequations}
where \metsoftsysreso~and \metsoftsysscale~are the values after
propagating the resolution and scale uncertainties, respectively, in
the longitudinal (perpendicular) directions. The mean values of
parameters are denoted using angled brackets. The $\Delta_{\textrm{CST}}$ is the
scale uncertainty, and the $R_{\parallel (\perp)}$ is the fractional resolution
uncertainty taken from the lower portion of Figure~\ref{fig:syst_cst_EL_pthard}(b). Both depend on the \pthard~and the average number of pileup
interactions per bunch crossing. Each propagation of the systematic
uncertainties in Eq.~(\ref{eq:varscale2_cst_met}) is called a
variation, and all of the variations are used in ATLAS analyses.

The systematic uncertainties in the resolution and scale for the CST
using the cross-check method
defined in Section~\ref{sec:cstsys_datamc} are propagated to the
nominal \metsoftvec~as follows:

\begin{subequations}
 \begin{eqnarray}
   \mexysoftsysreso &=& \metsoftxy \cdot \textrm{Gaus}(1,\hat{\sigma}_{\textrm{CST}}), \label{eq:var_cst_meta} \\
   \metsoftsysxyscale &=& \metsoftxy \cdot (1\pm \delta), \label{eq:varscale_cst1_met} 
 \end{eqnarray}
\end{subequations}
where \mexysoftsysreso~and \metsoftsysxyscale~are the values after
propagating the resolution and scale uncertainties, respectively, in
the $x$ ($y$) directions. Here, $\delta$ is the fractional scale
uncertainty, and
$\hat{\sigma}_{\textrm{CST}}$ corrects for the differences in resolution between the
data and simulation.

The systematic uncertainties in the resolution and scale for the TST
\metsoftvec{} are propagated to the nominal \metsoftvec{} as follows:

\begin{subequations}
\begin{eqnarray}
\metsoftsysreso &=& \metsoftparaperp  + \textrm{Gaus}(\Delta_{\textrm{TST}}, \sigma_{\parallel(\perp)}), \label{eq:var_csttrk_met} \\
\metsoftsysscale &=& \metsoftpara  \pm \Delta_{\textrm{TST}} \label{eq:varscale_csttrk_met}. 
\end{eqnarray}
\end{subequations}
The symbol
$\textrm{Gaus}(\Delta_{\textrm{TST}} , \sigma_{\parallel (\perp)})$ 
represents a random number sampled from a Gaussian distribution with 
mean $\Delta_{\textrm{TST}}$ and width $\sigma_{\parallel (\perp)}$.
The shift $\Delta_{\textrm{TST}}$ is zero for the perpendicular component. All of the TST systematic-uncertainty variations have a wider distribution
than the nominal MC simulation, when the Gaussian smearing is applied. To cover cases in which the data have a
smaller resolution (narrower distribution) than MC simulation, a
downward variation is computed using Eq.~(\ref{eq:log_normal}). To compute the yield of
predicted events in the variation,
$Y_{\textrm{down}}(X)$, for a given value $X$ of the \met{}, the yield is defined as the 
\begin{equation}
Y_{\textrm{down}}(X)  = \frac{[Y(X)]^2}{Y_{\textrm{smeared}}(X)},
\label{eq:log_normal}
\end{equation}
where the square of the yield of the nominal distribution, $Y(X)$, is
divided by the
yield of events after applying the variation with Gaussian smearing
to the kinematic variable, $Y_{\textrm{smeared}}(X)$. In practice, the yields are typically the content of
histogram bins before ($Y(X)$) and after ($Y_{\textrm{smeared}}(X)$)
the systematic uncertainty variations.
This procedure can be applied to any
kinematic observable by propagating only the smeared soft-term
variation to the calculation of the
kinematic observable $X$ and then computing the yield $Y_{\textrm{down}}(X)$ as defined in Eq.~(\ref{eq:log_normal}).

There are six total systematic uncertainties associated with the TST:
\begin{itemize}
\item Increase scale (\metsoftsysscaleplus{})
\item Decrease scale (\metsoftsysscaleminus{})
\item Gaussian smearing of \metsoftpara~(\metsoftsysparareso{})
\item The downward variation of the above
  \metsoftsysparareso{} computed using Eq.~(\ref{eq:log_normal})
\item Gaussian smearing of \metsoftperp~(\metsoftsysperpreso{})
\item The downward variation of the above
  \metsoftsysperpreso{} computed using Eq.~(\ref{eq:log_normal})
\end{itemize}

\subsubsection{Closure of systematic uncertainties}

The systematic uncertainties derived in this section for the CST and
TST \met{} are validated by
applying them to the \Zmm\ sample to confirm that the differences
between data and MC simulation are covered. 

The effects of these systematic uncertainty
variations on the CST \met{} are shown for the \Zmm~events in 
Figures~\ref{fig:syst_cst_closure1} and \ref{fig:syst_cst_closure2}
for the primary (Section~\ref{sec:cstsys_balance}) and the cross-check (Section~\ref{sec:cstsys_datamc})
methods, respectively. The uncertainties are larger for the
cross-check method, reaching around 50\% for \metsoft~$>$~60 \GeV{} in
Figure~\ref{fig:syst_cst_closure2}(a). 

The corresponding plots for the TST \met{} are shown in 
Figure~\ref{fig:syst_tst_closure2} using the \Zmm$+$0-jet control
sample, where the uncertainty band is the quadratic sum
of the variations with the MC statistical uncertainty. The systematic
uncertainty band for the TST is larger in
Figure~\ref{fig:syst_tst_closure2}(a) than the one for the primary
CST algorithm. In all the distributions, the systematic uncertainties in
the soft term alone cover the disagreement between data and MC simulation.

\begin{figure}[h!]
 \begin{center}
   \subfigure[]{\includegraphics[height=53mm]{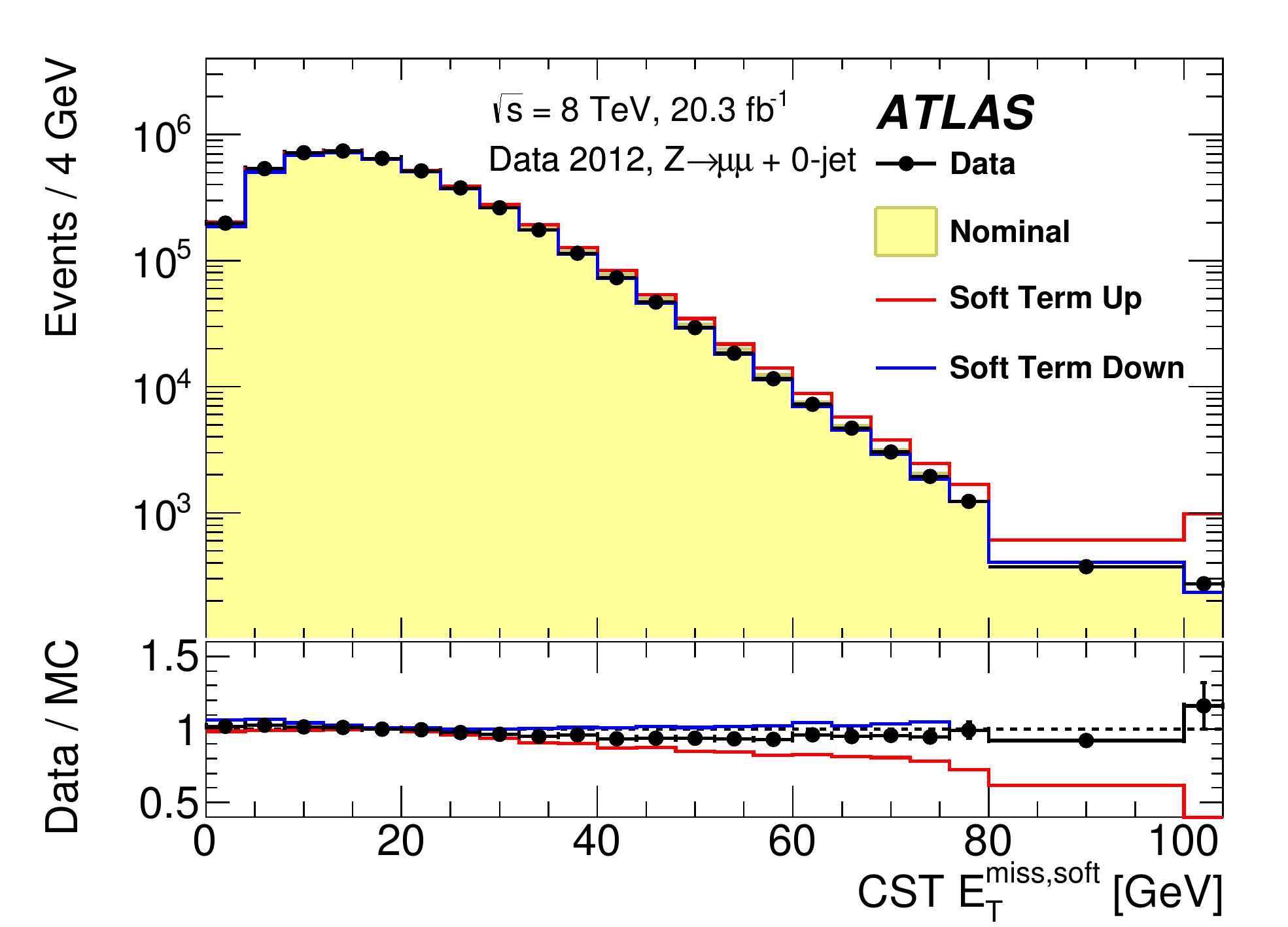}}
    \subfigure[]{\includegraphics[height=53mm]{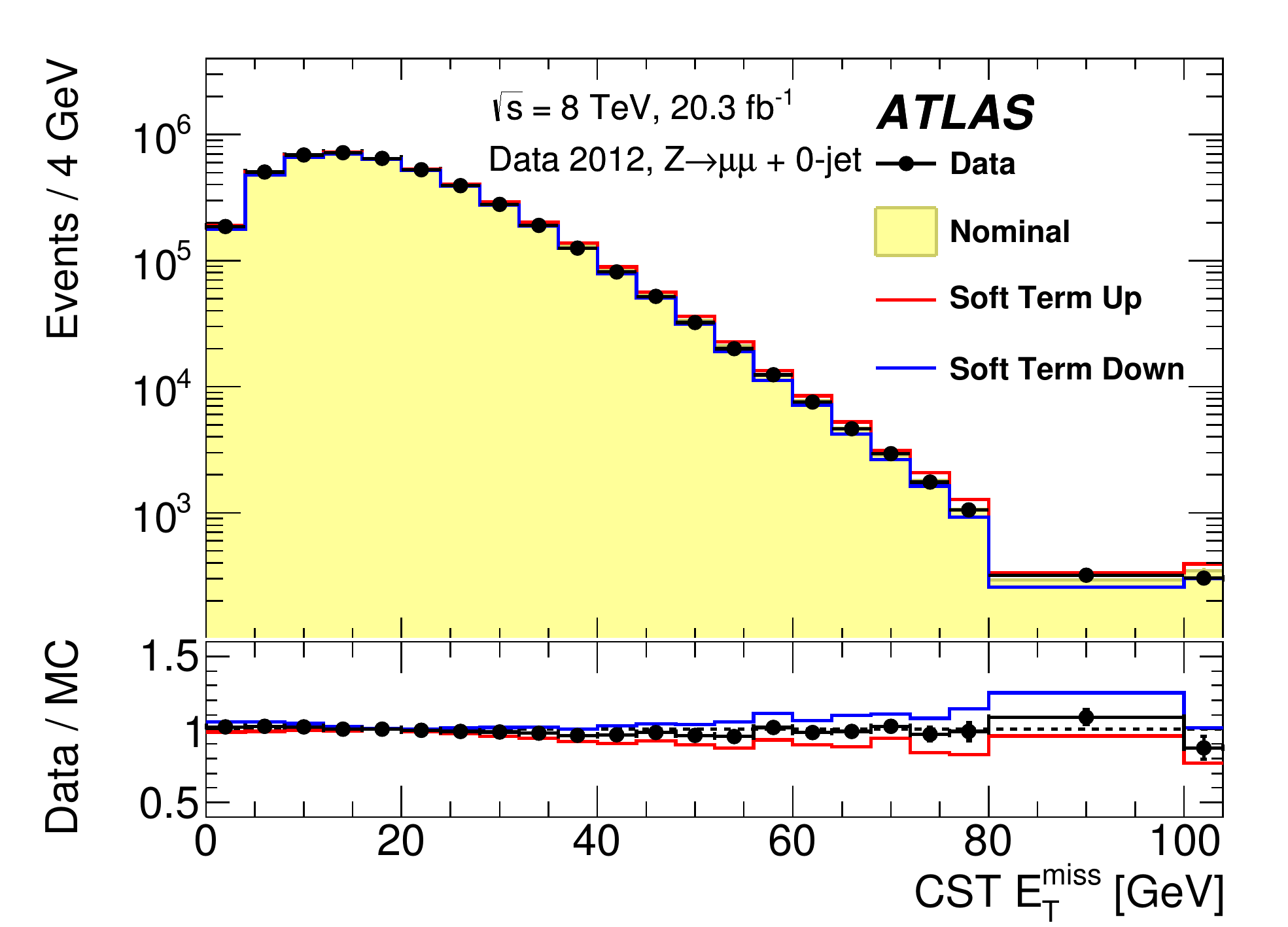}}
 \end{center}
\caption{Distributions of (a) \metsoft{} and (b) \metmag~with the
  CST algorithm. Data are
compared to the nominal simulation distribution as well as those
resulting from applying the shifts/smearing according to the
scale and resolution systematic uncertainties on the \metsoft{}. The
resulting changes from the variations are added in quadrature, and the
insets at the bottom of the figures show the ratios of the data to the
MC predictions. The
uncertainties are estimated using the balance method described 
in Section~\ref{sec:cstsys_balance}.}
\label{fig:syst_cst_closure1} 
\end{figure}

\begin{figure}[h!]
 \begin{center}
   \subfigure[]{\includegraphics[height=53mm]{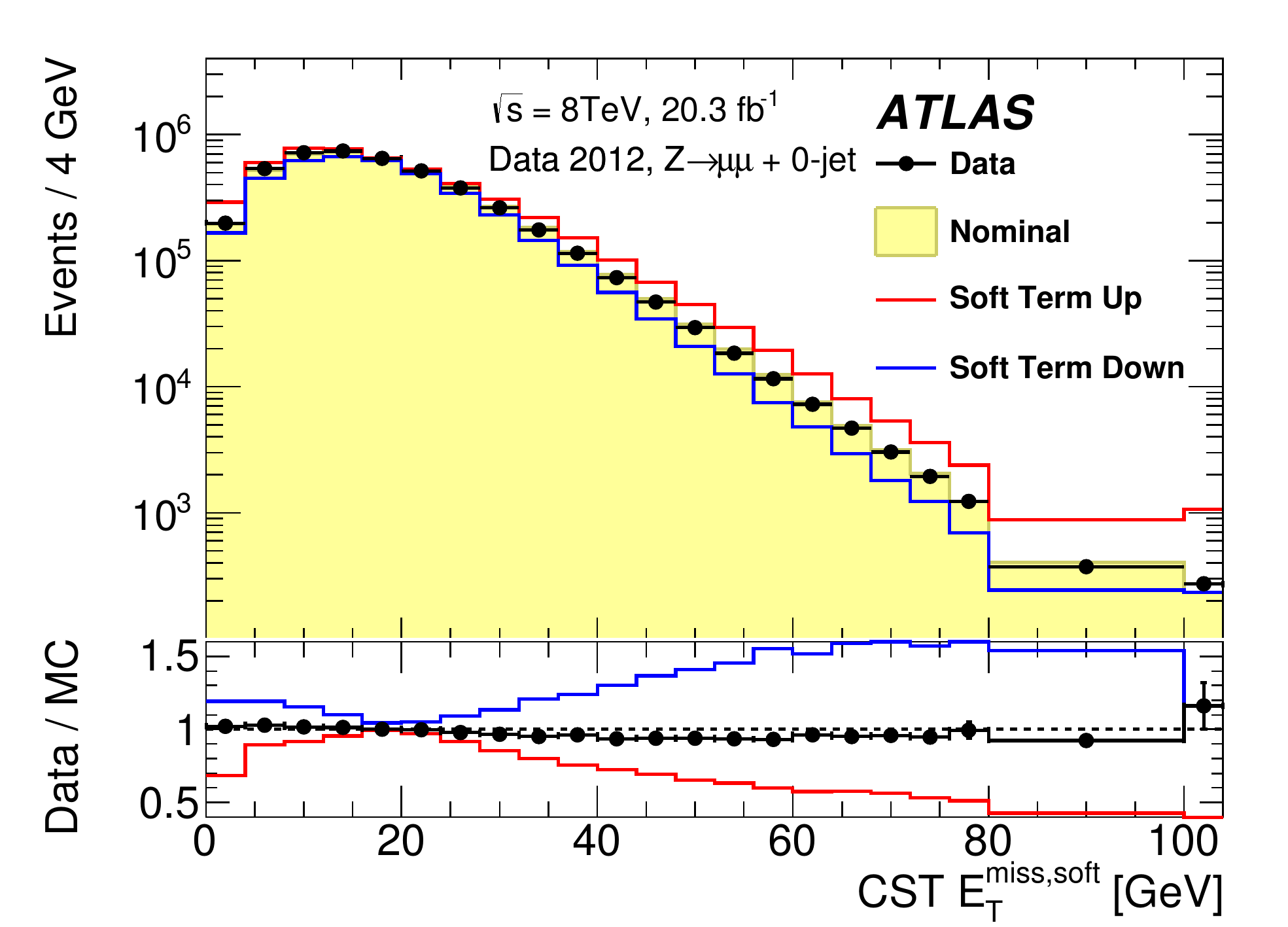}}\quad 
     \subfigure[]{\includegraphics[height=53mm]{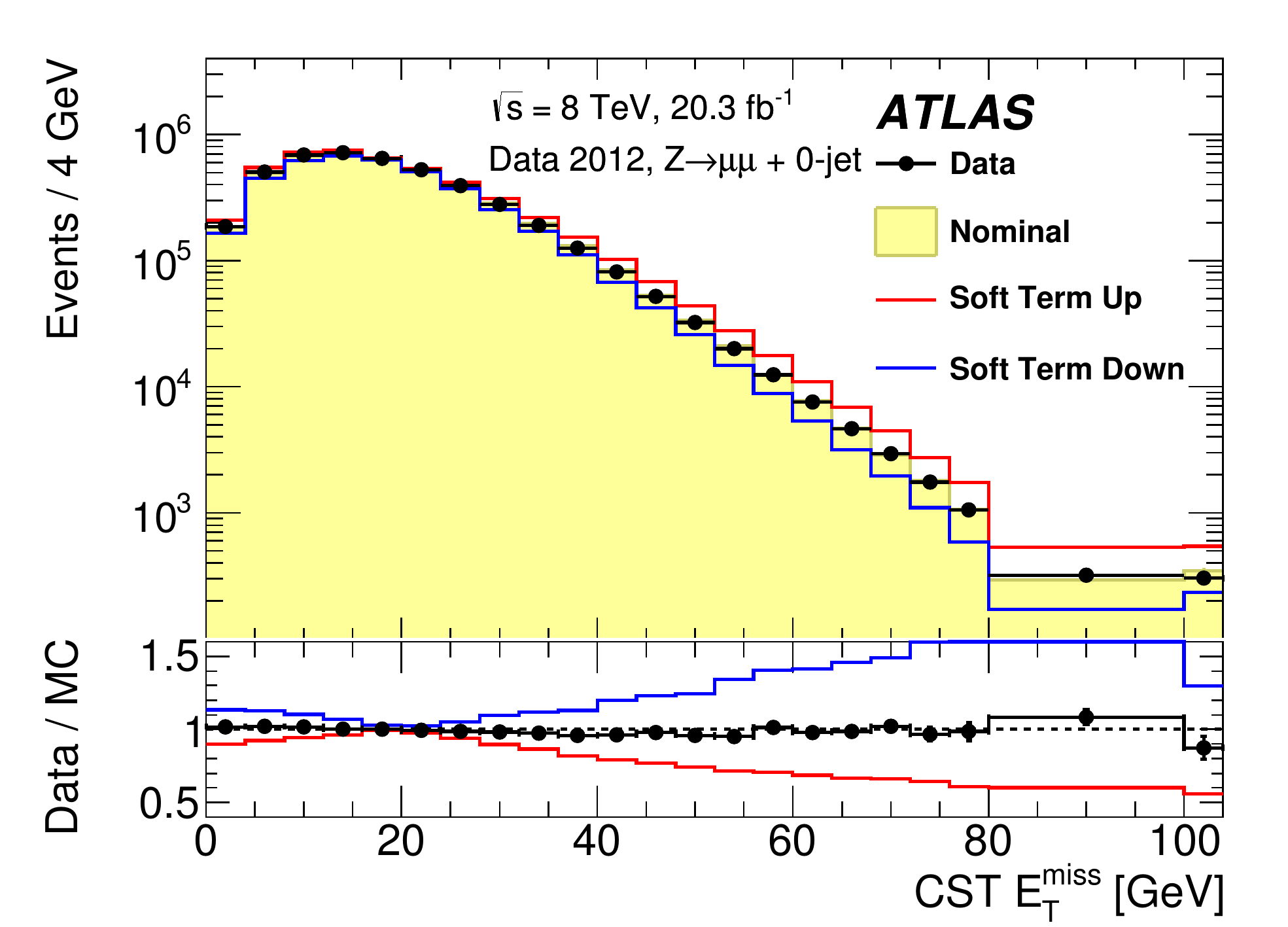}}
 \end{center}
\caption{Distributions of (a) \metsoft{} and (b) \metmag~with the CST
  algorithm. Data are compared to the nominal simulation distribution as well as those
resulting from applying the shifts/smearing according to the
scale and resolution systematic uncertainties on the \metsoft{}. The
resulting changes from the variations are added in quadrature, and the
insets at the bottom of the figures show the ratios of the data to the
MC predictions. The uncertainties are estimated from
the data/simulation ratio in Section~\ref{sec:cstsys_datamc}.}
\label{fig:syst_cst_closure2}
\end{figure}

\begin{figure}[h!]
 \begin{center}
    \subfigure[]{\includegraphics[height=53mm]{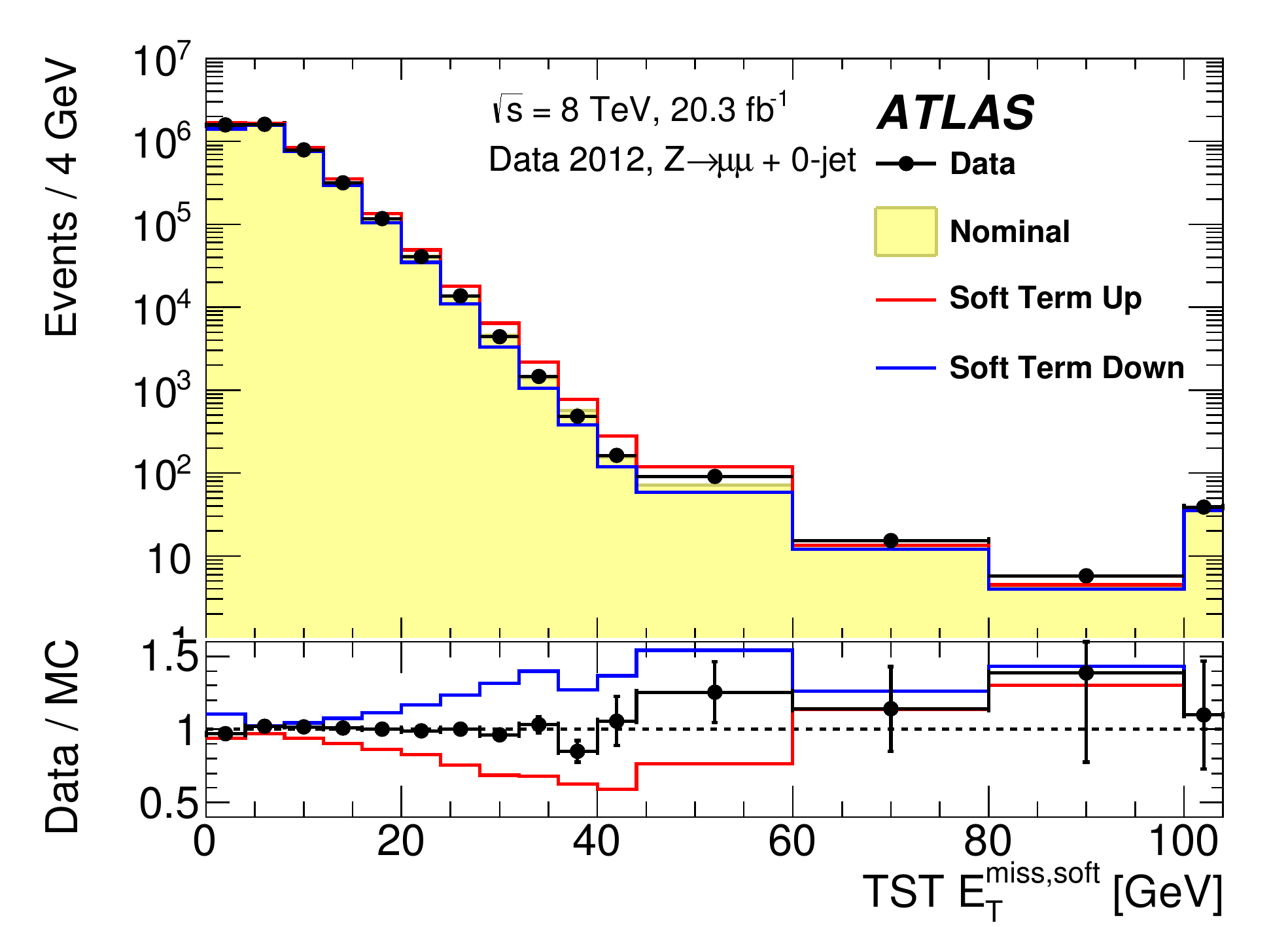}} \quad 
    \subfigure[]{\includegraphics[height=53mm]{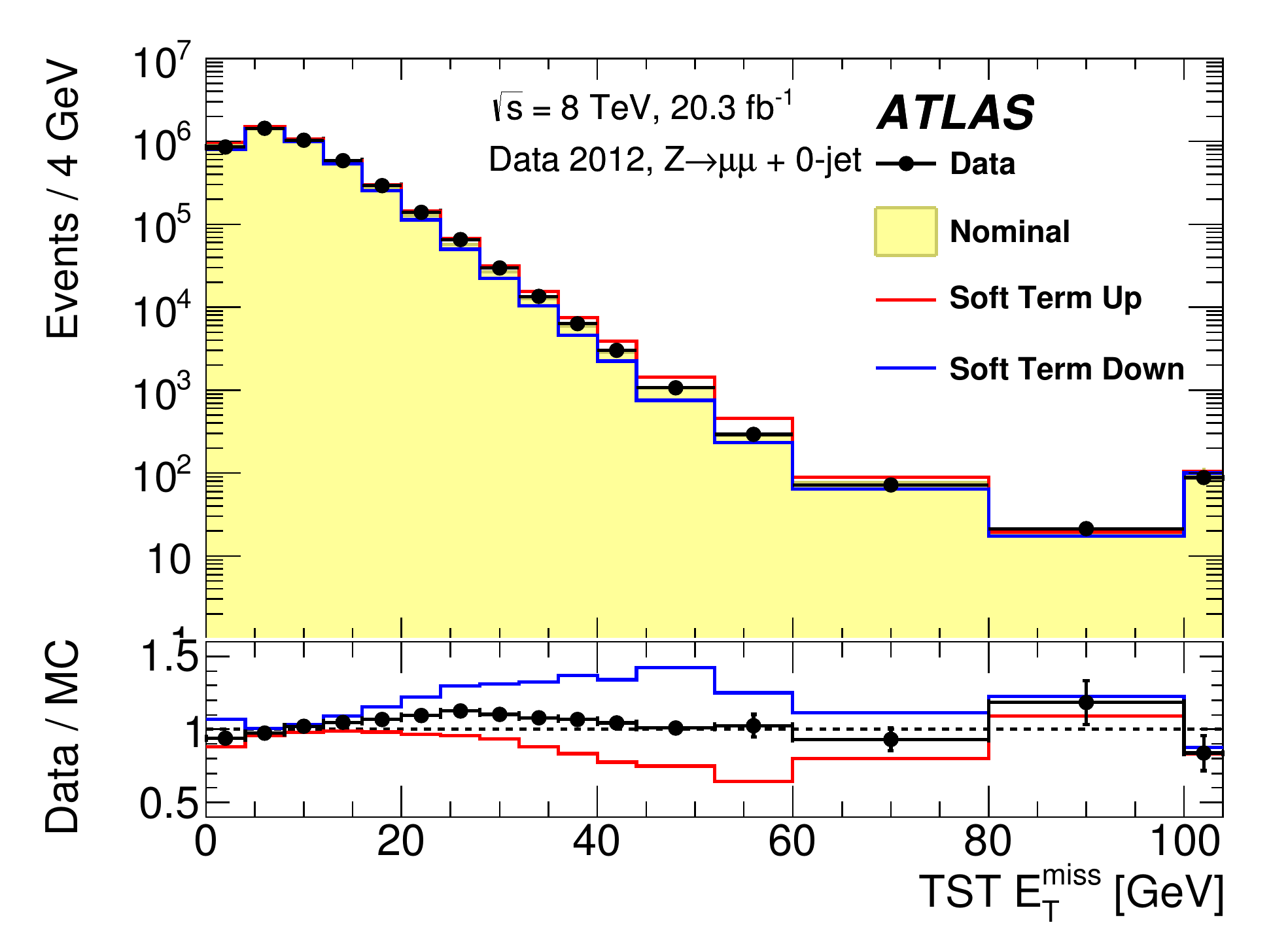}}
 \end{center}
\caption{Distributions of (a) \metsoft{} and (b) \metmag~with the TST algorithm. Data are
compared to the nominal simulation distribution as well as those
resulting from applying the
scale and resolution systematic uncertainties to the \metsoft{} and
adding the variations in quadrature, and the
insets at the bottom of the figures show the ratios of the data to the
MC predictions. The uncertainties are estimated from the method in Section~\ref{sec:track_syst}.}
\label{fig:syst_tst_closure2}
\end{figure}

\subsubsection{Systematic uncertainties from tracks inside jets}
\label{sec:syst_rtrk}

A separate systematic uncertainty is applied to the scalar summed \pT{} of tracks associated with high-\pT~jets in the Track \met{} because these tracks are
not included in the TST. 
The fraction of the momentum carried by charged particles within jets
was studied in ATLAS~\cite{ATLASJetEnergySys2011}, and its uncertainty
varies from 3\% to 5\% depending on the jet $\eta$ and \pT{}. 
These uncertainties affect the azimuthal angle
between the Track \met~and the TST \met{}, so the modelling is checked
with \Zmm~events produced with one jet.
The azimuthal angle between the Track \met{} and the TST \met{}
directions is well modelled, 
and the differences between data and MC simulation are within the systematic uncertainties.

\FloatBarrier
\section{Conclusions}
\label{sec:conclusions}

Weakly interacting particles, which leave the ATLAS detector
undetected, give rise to a momentum imbalance in the plane transverse to the beamline. An accurate measurement of the missing transverse momentum (\met{}) is thus important in 
many physics analyses to infer the momentum of these particles. 
However, additional interactions occurring in a given bunch crossing
as well as residual signatures from nearby bunch crossings make it difficult to reconstruct the \met{} from the hard-scattering process alone.

The \metvec~is computed as the
negative vector sum of the reconstructed physics objects including
electrons, photons, muons, $\tau$-leptons, and jets. The remaining energy
deposits not associated with those high-\pT{} physics
objects are also considered in the \metvec. 
They collectively
form the so-called soft term, which is the \met~component most affected by
pileup. The
calorimeter and the tracker in the ATLAS detector provide
complementary information to the reconstruction of
the high-\pT{} physics objects as well as the \met{} soft term. Charged 
particles are matched to a particular collision point or
vertex, and this information is used to determine which charged
particles originated from the hard-scatter collision. Thus tracking
information can be used to greatly reduce the pileup dependence of the
\met~reconstruction. This has resulted in the development of
\met~reconstruction algorithms that combine the information from the
tracker and the calorimeter. The performance of these reconstruction
algorithms is evaluated using data from 8~\TeV{} proton--proton collisions collected with the ATLAS detector at the LHC corresponding to an integrated luminosity
of 20.3~fb$^{-1}$.

The Calorimeter Soft Term (CST) is 
computed from the sum of calorimeter topological clusters
not associated with any hard object. 
No distinction can be made
between energy contributions from pileup and hard-scatter
interactions, which makes the resolution on the \metvec{} magnitude and direction very dependent on the number of pileup interactions. The 
pileup-suppressed \met~definitions clearly reduce the dependence on
the number of pileup interactions but also introduce a larger
under-estimation of the soft term than the CST. 

The Track Soft Term (TST) algorithm does not use calorimeter
energy deposits in the soft term and uses only the
inner detector (ID) tracks. It has stable \met~resolution with
respect to the amount of pileup; however, it does not have as good
a response as the CST \met, due mainly to missing neutral particles in the soft term. 
Nevertheless, its response is better than that of the other reconstruction algorithms 
that aim to
combine the tracking and calorimeter information. For large values of
\mettrue{}, the CST and TST \met~algorithms all perform similarly.
This is because contributions from jets dominate the \met~performance, 
making the differences in soft-term reconstruction less important.

The Extrapolated Jet Area with Filter (EJAF) and Soft-Term Vertex-Fraction (STVF) \met{} reconstruction algorithms correct for
pileup effects in the CST \met{} by utilizing a combination of the ATLAS
tracker and calorimeter measurements. Both apply a vertex association
to the jets used in the \met{} calculation. The EJAF soft-term reconstruction subtracts the pileup contributions to the soft term
using a procedure similar to jet area-based pileup corrections, and
the EJAF \met~resolution has a reduced dependence on the
amount of pileup, relative to the CST algorithm. The STVF
reconstruction algorithm uses an event-level correction of the CST,
which is the scalar sum of charged-particle \pT{} from the
hard-scatter vertex divided by the scalar sum of all charged-particle
\pT{}. The STVF
correction to the soft term greatly decreases the dependence of the \met{} resolution on the amount of pileup but causes the
largest under-estimation of all the soft-term algorithms. 

Finally, the Track \met~reconstruction uses only the inner detector tracks with
the exception of the reconstructed electron objects, which use the
calorimeter \eT~measurement. The resolutions on the Track \met{} magnitude and direction are very stable against pileup, but the
limited $|\eta|$ coverage of the tracker degrades the \met{} response, as does not accounting for high-\pT{} neutral particles, especially in events with many jets.

The different \met~algorithms have their own advantages and
disadvantages, which need to be considered in the context of each analysis. For
example, removing large backgrounds with low \met{}, such as
Drell--Yan events, may require the use of more than one \met~definition. 
The tails of the track and calorimeter \met~distributions remain 
uncorrelated, and exploiting both definitions in parallel allows one to 
suppress such backgrounds even under increasing pileup conditions. 

The systematic uncertainties in the \met~are estimated with
\Zmm~events for each reconstruction 
algorithm, and are found to be small.

\section*{Acknowledgements}


We thank CERN for the very successful operation of the LHC, as well as the
support staff from our institutions without whom ATLAS could not be
operated efficiently.

We acknowledge the support of ANPCyT, Argentina; YerPhI, Armenia; ARC, Australia; BMWFW and FWF, Austria; ANAS, Azerbaijan; SSTC, Belarus; CNPq and FAPESP, Brazil; NSERC, NRC and CFI, Canada; CERN; CONICYT, Chile; CAS, MOST and NSFC, China; COLCIENCIAS, Colombia; MSMT CR, MPO CR and VSC CR, Czech Republic; DNRF and DNSRC, Denmark; IN2P3-CNRS, CEA-DSM/IRFU, France; GNSF, Georgia; BMBF, HGF, and MPG, Germany; GSRT, Greece; RGC, Hong Kong SAR, China; ISF, I-CORE and Benoziyo Center, Israel; INFN, Italy; MEXT and JSPS, Japan; CNRST, Morocco; FOM and NWO, Netherlands; RCN, Norway; MNiSW and NCN, Poland; FCT, Portugal; MNE/IFA, Romania; MES of Russia and NRC KI, Russian Federation; JINR; MESTD, Serbia; MSSR, Slovakia; ARRS and MIZ\v{S}, Slovenia; DST/NRF, South Africa; MINECO, Spain; SRC and Wallenberg Foundation, Sweden; SERI, SNSF and Cantons of Bern and Geneva, Switzerland; MOST, Taiwan; TAEK, Turkey; STFC, United Kingdom; DOE and NSF, United States of America. In addition, individual groups and members have received support from BCKDF, the Canada Council, CANARIE, CRC, Compute Canada, FQRNT, and the Ontario Innovation Trust, Canada; EPLANET, ERC, FP7, Horizon 2020 and Marie Sk{\l}odowska-Curie Actions, European Union; Investissements d'Avenir Labex and Idex, ANR, R{\'e}gion Auvergne and Fondation Partager le Savoir, France; DFG and AvH Foundation, Germany; Herakleitos, Thales and Aristeia programmes co-financed by EU-ESF and the Greek NSRF; BSF, GIF and Minerva, Israel; BRF, Norway; Generalitat de Catalunya, Generalitat Valenciana, Spain; the Royal Society and Leverhulme Trust, United Kingdom.

The crucial computing support from all WLCG partners is acknowledged gratefully, in particular from CERN, the ATLAS Tier-1 facilities at TRIUMF (Canada), NDGF (Denmark, Norway, Sweden), CC-IN2P3 (France), KIT/GridKA (Germany), INFN-CNAF (Italy), NL-T1 (Netherlands), PIC (Spain), ASGC (Taiwan), RAL (UK) and BNL (USA), the Tier-2 facilities worldwide and large non-WLCG resource providers. Major contributors of computing resources are listed in Ref.~\cite{ATL-GEN-PUB-2016-002}.

%

\clearpage
\appendix
\part*{Appendix}
\addcontentsline{toc}{part}{Appendix}

\section{Calculation of EJAF}
\label{sec:ejaf}

A jet-level $\eta$-dependent pileup correction of the form 
\begin{equation}
\rhoetamed(\eta) = \rhoevtmed \cdot
P_{\textrm{fct}}^\rho(\eta,N_{\text{PV}},\muavno),
\label{eq:jet_pu_corr}
\end{equation}
\noindent{}is used, where the \Npv{} and \muavno{} are determined from the event
properties. This multiplies the median soft-term jet \pt-density,
\rhoevtmed{}, from Eq.~(\ref{eq:rhomedian}) by the functional form,
$P_{\textrm{fct}}^\rho(\eta,N_{\text{PV}},$\muavno$)$ as defined in Eq.~(\ref{eq:ejaf_pu_density}),
which was fit to the average transverse momentum density. The median transverse momentum density $\rhoevtmed$ is
determined from soft-term jets with $|\eta|$~$<$~2 and then extrapolated to
higher $|\eta|$ as discussed in Section~\ref{sec:PU_JetArea0} using
the fitted $P_{\textrm{fct}}^\rho(\eta,N_{\text{PV}},$\muavno$)$.

The pileup correction $\rhoetamed(\eta)$ from Eq.~(\ref{eq:jet_pu_corr}) is
applied to the transverse momenta of the soft-term jets passing a JVF
selection. The pileup-corrected jet \pT{} is labelled \ptjetcorri, and
it is computed as

\begin{equation}
\ptjetcorri= 
\left\{
\begin{array}{ll} 
0 &  \left(\ptjetfilteri\leq \rhoetamed(\etafilterjeti) \cdot \jetfilterareai \right) \\
\ptjetfilteri - \rhoetamed(\etafilterjeti) \cdot \jetfilterareai &
\left(\ptjetfilteri   > \rhoetamed(\etafilterjeti) \cdot \jetfilterareai \right).
\end{array} \right.
\label{eq:jet_pu_corr_complete}
\end{equation}  

The $x$ and $y$ components of \ptjetcorri~are used to compute the EJAF
soft term using Eq.~(\ref{eq:ejaf_met}), and only soft-term jets matched to the
PV with JVF~$>$~0.25 for $|\etafilterjeti|~<~2.4$ or jets with
$|\etafilterjeti|$~$\geq$~2.4 are used. Because of this JVF selection, the label of
``filter-jet'' is added to the catchment area (\jetfilterareai), to the
transverse momentum (\ptjetfilteri{}), and to the jet $\eta$ (\etafilterjeti{})
variables. 

While all other jets used in this paper use an $R$~$=$~0.4 reconstruction, the larger value of $R$~$=$~0.6 is used to reduce the number of
$k_{t}$ soft-term jets with \pT~$=$~0 (see Eq.~(\ref{eq:jet_pu_corr_complete})) in the central detector
region. While negative energy deposits are possible in the ATLAS
calorimeters, their contributions cannot be matched to the
soft-term jets by ghost-association. Studies that modify the
cluster-to-jet matching to
include negative-\pT~clusters indicate no change in the
\met~performance, so negative-\pT~clusters are excluded from the
soft-term jets. Finally, only
filter-jets with \ptjetfilteri~larger than the pileup
correction contribute to the EJAF soft term.

\printbibliography

\newpage 
\begin{flushleft}
{\Large The ATLAS Collaboration}

\bigskip

G.~Aad$^{\textrm 87}$,
B.~Abbott$^{\textrm 114}$,
J.~Abdallah$^{\textrm 65}$,
O.~Abdinov$^{\textrm 12}$,
B.~Abeloos$^{\textrm 118}$,
R.~Aben$^{\textrm 108}$,
M.~Abolins$^{\textrm 92}$,
O.S.~AbouZeid$^{\textrm 159}$,
H.~Abramowicz$^{\textrm 154}$,
H.~Abreu$^{\textrm 153}$
% ...
\bigskip
\\
% ...
$^{12}$ Institute of Physics, Azerbaijan Academy of Sciences, Baku, Azerbaijan\\
% ...
$^{65}$ University of Iowa, Iowa City IA, United States of America\\
% ...
$^{87}$ CPPM, Aix-Marseille Universit{\'e} and CNRS/IN2P3, Marseille, France\\
% ...
$^{92}$ Department of Physics and Astronomy, Michigan State University, East Lansing MI, United States of America\\
% ...
$^{108}$ Nikhef National Institute for Subatomic Physics and University of Amsterdam, Amsterdam, Netherlands\\
% ...
$^{114}$ Homer L. Dodge Department of Physics and Astronomy, University of Oklahoma, Norman OK, United States of America\\
% ...
$^{118}$ LAL, Univ. Paris-Sud, CNRS/IN2P3, Universit{\'e} Paris-Saclay, Orsay, France\\
% ...
$^{153}$ Department of Physics, Technion: Israel Institute of Technology, Haifa, Israel\\
$^{154}$ Raymond and Beverly Sackler School of Physics and Astronomy, Tel Aviv University, Tel Aviv, Israel\\
% ...
$^{159}$ Department of Physics, University of Toronto, Toronto ON, Canada\\
% ...
\end{flushleft}


\end{document}